\pdfoutput=1
\documentclass[apj,twocolappendix]{emulateapj}
\usepackage{apjfonts}
\usepackage{natbib}

\usepackage{color}

\newcommand{\HI}{\ion{H}{1}} 
\newcommand{\HII}{\ion{H}{2}}
\newcommand{\HJ}{{\rm HESS~J1841-055}}
\newcommand{\HJP}{{\rm HESS~J1837-069}}
\newcommand{\LP}{{\rm LAT\ PSR\ J1838-0537}}
\newcommand{\AN}{\rm G27}
\newcommand{\TSe}{{\rm TS}_{\rm ext}}

\newcommand{\XMM}{\emph{XMM-Newton}}
\newcommand{\Chandra}{\emph{Chandra}} 
\newcommand{\Ms}{M_{\odot}}
\newcommand{\NH}{{N_{\rm H}}}

\newcommand{\CC}{MA11}

\shorttitle{Extended Gamma-ray Emission from the G25.0+0.0 Region}
\shortauthors{J.\,Katsuta et al.}

\begin{document}


\title{Extended Gamma-ray Emission from the G25.0+0.0 Region:
A Star Forming Region Powered by the Newly Found OB Association?}


\author{
J.\,Katsuta\altaffilmark{1},
Y.\,Uchiyama\altaffilmark{2},
S.\,Funk\altaffilmark{3}
}
\altaffiltext{1}{Department of Physical Sciences, Hiroshima University, Higashi-Hiroshima, Hiroshima 739-8526, Japan; katsuta@hep01.hepl.hiroshima-u.ac.jp}
\altaffiltext{2}{Rikkyo University, Nishi-Ikebukuro, Toshima-ku, Tokyo 171-8501, Japan; y.uchiyama@rikkyo.ac.jp}
\altaffiltext{3}{Erlangen Centre for Astroparticle Physics, D-91058 Erlangen, Germany}

\begin{abstract}

We report a study of extended $\gamma$-ray emission with the Large Area Telescope (LAT) 
onboard the \emph{Fermi Gamma-ray Space Telescope}, which is likely to be the second case of 
a $\gamma$-ray detection from a star-forming region (SFR) in our Galaxy.
The LAT source is located in the G25 region, $1 \fdg 7 \times 2 \fdg 1$ around $(l, b) = (25\fdg0, 0\fdg0)$.
The $\gamma$-ray emission is found to be composed of two extended sources and one point-like source. 
The extended sources have a similar sizes of about $1 \fdg 4 \times 0 \fdg 6$.
An $\sim 0 \fdg 4$ diameter sub-region of one has a photon index of $\Gamma = 1.53 \pm 0.15$;
and is spatially coincident with HESS\,J1837$-$069, likely a 
pulsar wind nebula. 
The other parts of the extended sources have a photon index of $\Gamma = 2.1 \pm 0.2$ without significant spectral curvature.
Given their spatial and spectral properties, they have no clear associations with sources at other wavelengths.
Their $\gamma$-ray properties are similar to those of the Cygnus cocoon SFR, the only firmly established $\gamma$-ray detection of an SFR in the Galaxy.
Indeed, we find bubble-like structures of atomic and molecular gas in G25, which may be created by a putative OB association/cluster.
The $\gamma$-ray emitting regions appear confined in the bubble-like structure; similar properties are also found in the Cygnus cocoon.
In addition, using observations with the \XMM\ we find a candidate young massive OB association/cluster G25.18+0.26 in the G25 region.
We propose that the extended $\gamma$-ray emission in G25 is associated with an SFR driven by G25.18+0.26.
Based on this scenario, we discuss possible acceleration processes in the SFR and compare them with the Cygnus cocoon.

\end{abstract}

\section{Introduction}
\label{sec: intro}

Most massive stars are formed in clusters by the collapse of giant molecular clouds\,\citep[e.g.,][]{LadaLada}.
They produce strong radiation fields and stellar winds, which disrupt their natal molecular clouds and create bubble structures around them.
Such central OB associations/clusters and the accompanying bubbles constitute massive star-forming regions (SFRs).
SFRs could be sites of $\gamma$-ray emission, if the wind energy is efficiently transferred to the acceleration of particles.
Many models are proposed for acceleration processes in massive SFRs, such as diffusive shock acceleration (DSA) at the wind boundary and stochastic acceleration by magnetic turbulence\,\citep[see e.g,][]{CCJ83,Bykov01}. 
However, the limited number of $\gamma$-ray detected SFRs limits the possibility of studying the acceleration process in detail.

Observations with the {\it Fermi} Large Area Telescope (LAT) have recently revealed the extended $\gamma$-ray emission from an SFR in the Cygnus\,X\ region \citep[hereafter \CC]{Ackermann:2011eq}.
The $\gamma$-ray emission of this SFR (a.k.a. the Cygnus cocoon) spatially extends over a large diameter of $\sim 4^\circ$ and 
appears associated with the cavities probably created by a massive OB association, Cyg\,OB2.
The observed hard spectrum (photon index $\Gamma \sim 2.2$) suggests that the $\gamma$-ray emitting particles are accelerated in the region.
The accelerated particles are plausibly powered by Cyg\,OB2. 
However there is another possibility that the particles escaped from the nearby supernova remnant (SNR) $\gamma$ Cygni.

Recent TeV observations have found an emission whose morphology is consistent with that of the LAT observations\,\citep{2014ApJ...790..152B}.
The observed energy spectrum is smoothly connected to that of the LAT observations, and extends to a few TeV.
Prior TeV observations have also revealed the existence of relativistic particles of energies of up to a few tens of TeV in this region\,\citep{2007ApJ...658L..33A,2009ApJ...700L.127A,2012ApJ...745L..22B}. If all the observed $\gamma$ rays are attributed to the same object, its energy spectrum must steepen at TeV energies.

The $\gamma$-ray detection from the Cygnus cocoon is evidence that relativistic particles exist in the SFR.
This is the only firm case of a $\gamma$-ray detection from an SFR,
which may indicate that the $\gamma$-ray detection of the Cygnus cocoon is a special case, e.g., 
the relativistic particles are not accelerated at the SFR but are runaway particles from the SNR.
More studies are needed to answer the question of whether SFRs are sites of the particle acceleration.

In this paper, we report a study of LAT observations of extended $\gamma$-ray emission, which is likely to be the second example of a $\gamma$-ray detection of an SFR.
The LAT source is located in the $1 \fdg 7 \times 2 \fdg 1$ region around $(l, b) = (25 \fdg 0, 0 \fdg 0)$ (hereafter the G25 region).
The region was studied by \cite{Lande:2012jb} based on two years of LAT data.
They found an extended source which is spatially coincident with the TeV source HESS\,J1837$-$069.
In this paper we re-analyze this complex region using 57\,months of LAT data.
There is no known large ($\sim 1^\circ$) SFR associated with the G25 region.
However, in this direction, 
\cite{Rahman:2010in} recently found copious 8\,$\mu{\rm m}$ emission within a region of the brightest free-free emission reported by \cite{2010ApJ...709..424M}.
They claim that the emission is due to a star-forming complex with diameter $\lesssim 1^\circ$, 
which is created by hidden massive OB association(s).
The putative OB association might be difficult to detect optically due to heavy extinction, if it is distant from us.
In this paper, we investigate the G25 region using X-ray and radio data to check whether a massive SFR exists.
An optically hidden massive OB association is expected to be detectable in X-rays, because of the reduced extinction compared to that in the optical.
Here we analyze in particular an unidentified X-ray source AX\,J1836.3$-$0647 located in this region\,\citep{2001ApJS..134...77S} using \emph{XMM-Newton} data.

This paper is organized as follows. 
The observation and the analysis of the LAT sources in the G25 region is reported in Section\,\ref{sec: LAT}.
We describe the \XMM\ observations and analysis of AX\,J1836.3$-$0647 in Section\,\ref{sec: xmm}.
In Section\,\ref{sec: discussion}, we discuss the origin of the observed $\gamma$-ray and X-ray emissions.
Based on the discussion, we investigate possible acceleration mechanisms for the radiating particles.
In Section\,\ref{sec: conclusion}, we summarize our results and the outlook for advances with future observations.

\section{\emph{Fermi}-LAT Data}
\label{sec: LAT}

\subsection{Observation and Data Reduction}
\label{sec: obs}
The \emph{Fermi Gamma-ray Space Telescope} was launched on 2008 June 11. The LAT onboard  \emph{Fermi} is a pair-conversion telescope equipped with solid state silicon trackers and a cesium iodide calorimeter, sensitive to photons in the very broad energy band from $\sim 20$\,MeV to $> 300$\,GeV. The LAT has a large effective area ($\sim 8000$\,cm$^2$ above 1\,GeV for on-axis events), instantaneously viewing $\simeq 2.4$\,sr of the sky with a good angular resolution (per-photon 68\% containment radius better than $\sim 1^{\circ}$ above 1\,GeV). 
Details of the LAT instrument and data reduction are described in \citet{Atwood:2009ha}.

The LAT data used here were collected during $\sim 57$\,months from 2008 August 4 to 2013 May 1. We selected Pass 7 Reprocessed \textsf{SOURCE} class events and $\gamma$ rays with Earth zenith angles greater than $100^\circ$ were excluded to reduce the $\gamma$-ray background from the Earth limb, an intense source of $\gamma$ rays from cosmic-ray collisions with the upper atmosphere.
We also applied a cut that excludes time intervals during which any part of the region we analyzed (see Section~\ref{sec: ana}) was beyond the 100$\degr$ zenith angle limit.
We used the \textsc{P7REP\_SOURCE\_V15} instrument response functions for the analysis\,\citep{LATorbit}.
Note that we made a basic confirming check for the sources analyzed in this paper using the Pass 8 data with the surrounding sources in
the 3FGL catalog\,\citep{3FGL}. The results do not show strong quantitative differences from those in the following analyses.
\subsection{Analysis and Results}
\label{sec: ana}

The G25 region contains four sources in the {\it Fermi}-LAT Second Source Catalog\,(2FGL; \cite{2012ApJS..199...31N}): 
2FGL J1835.5$-$0649, J1836.8$-$0623c, J1837.3$-$0700c, and J1834.7$-$0705c.
Using the first two years of LAT data, \cite{Lande:2012jb} find an extended $\gamma$-ray source in this region.
The authors claim that two of the 2FGL sources are not distinct sources but an approximation for the extended source,
and the other two 2FGL sources are unrelated background sources.
They also find that the extended TeV source HESS\,J1837$-$069 is coincident with the LAT extended source.
However the LAT flux is about two times higher than what is expected from the TeV observation.
In addition, the LAT source is about two times larger than the H.E.S.S. source, with a different shape, and
the peak emission of the H.E.S.S. source is located on the edge of the LAT source reported by \cite{Lande:2012jb}.
This may indicate that other $\gamma$-ray sources contaminate the reported extended source in the LAT band.
The First {\it Fermi}-LAT Catalog above 10\,GeV\,(1FHL; \cite{2013ApJS..209...34A}), based on the first three years of data, reveals another new  source (1FHL\,J1839.4$-$0708) located south of the extended source.
This region also includes two other 1FHL sources, J1834.6$-$0703 and J1836.5$-$0655e, which correspond to 2FGL J1834.7$-$0705c and the extended source mentioned above, respectively.
The detection of the new source in 1FHL is consistent with the suggestion of other $\gamma$-ray sources in this region.
In this section we re-analyze this complex region using 57 months of LAT data, 
which is more than double the observation time considered by \cite{Lande:2012jb}.

To characterize the $\gamma$-ray sources in the region, we use {\sf gtlike}, part of the Science Tools analysis package\,(v9r32p5)\footnote{Available at the \emph{Fermi} Science Support Center, http://fermi.gsfc.nasa.gov/ssc}, and {\sf pointlike}\,\citep{Kerr:2011,Lande:2012jb}.
With {\sf gtlike} and {\sf pointlike}, we perform binned maximum likelihood fits on the observed $\gamma$ rays to optimize parameters of an input model taking into account the energy dependence of the point-spread function (PSF).
We first evaluate the spatial extent of the G25 region by using {\sf pointlike}; the {\sf pointlike} algorithm is optimized for speed to handle a large number of sources.
With the obtained spatial extent, we measure spectral energy distributions (SEDs) of the region by using {\sf gtlike}.
In this analysis, the test statistic (TS) is defined as $-2\Delta\ln ({\rm likelihood})$ between a model including a test source or parameter and a model without that source or parameter (the null hypothesis).  
For each model the likelihood is maximized over the free parameters before TS is evaluated.
The significance of the improvement of the maximum likelihood with the additional source or parameter can be evaluated from the TS value, 
because the value is expected to approximately follow the $\chi^2$ distribution with additional degrees of freedom\,\citep{1996ApJ...461..396M} in the case that the null hypothesis is a complete description of the region under study.

For the spatial analysis, we select a $9^\circ\times9^\circ$ region around G25, and
we include only energies above 3~GeV, because the LAT has better angular resolution at higher energies (68\% containment radius better than $\sim 0 \fdg 3$ above 3\,GeV).
In addition the spatial confusion between the G25 source and the Galactic diffuse component is reduced at higher energies, 
since the flux ratio of G25 to the Galactic diffuse emission increases with energy.
We use $0\fdg05$ spatial bins and divide the 3--500\,GeV energy range into 17 bins evenly on a log scale.
In the spectral analysis, we use a larger region, $21^\circ\times21^\circ$, around G25 in the energy range 0.2--500~GeV, given the relatively large LAT PSF at lower energies.
The size of the spatial bins is set at $0 \fdg 05$; the energy range is divided into 26 equal bins on a log scale.

\begin{figure*}[t] 
\begin{center}
\includegraphics[width=7in]{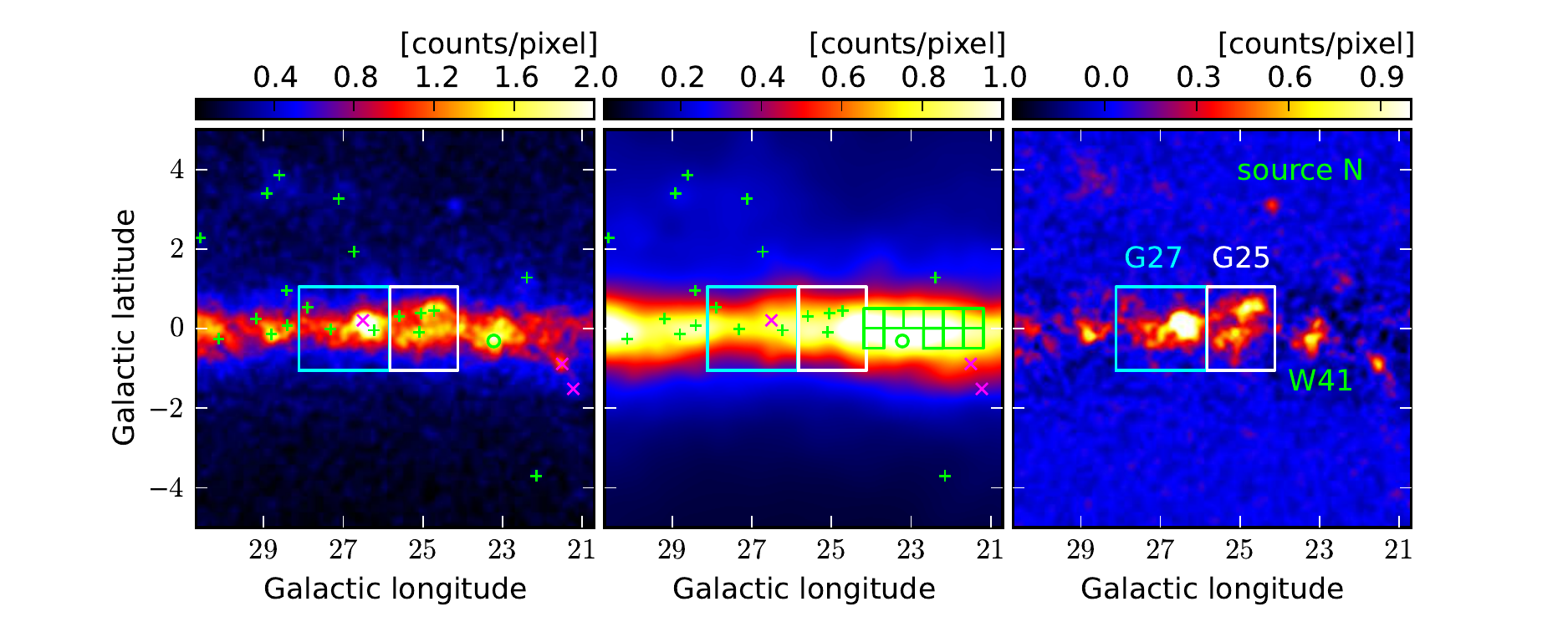}
\caption{\small
Left: {\it Fermi}-LAT counts map above 3\,GeV around the G25 region (in units of counts per pixel). The pixel size is $0\fdg025$. Smoothing with a Gaussian kernel of $\sigma = 0\fdg15$ is applied. The G25 and G27 regions are represented by white and cyan boxes, respectively. 
Green crosses and the magenta X represent positions of the 2FGL and the 2PC sources respectively. 
W41 is also shown with the position and the size of the corresponding source in our model (see text). 
Middle: background model map, which contains contributions from the Galactic diffuse emission and isotropic diffuse background. Green boxes represent the regions used for the evaluation of the accuracy of the Galactic diffuse model.
Right: background-subtracted count map. 
\label{fig: cmap_wide}}
\end{center}
\end{figure*}

The input model for the likelihood analysis includes cataloged sources in the G25 region, nearby sources outside the region, a model for the Galactic diffuse emission, and an isotropic component.
A $\sim 2^\circ$ square region in the vicinity of G25 (hereafter the G27 region; see Figure\,\ref{fig: cmap_wide}) appears to contain extended sources rather than a combination of 2FGL point sources. 
To perform precise measurements of G25, we also evaluate spatial and energy distributions of this nearby background region.
For the analysis we set positions and spectral parameters of the other background sources equal to those in the 2FGL catalog except for 2FGL J1834.3$-$0848 and LAT pulsars (PSRs) described in the LAT Second Pulsar (2PC) Catalog\,\citep{2013ApJS..208...17A}.
2FGL J1834.3$-$0848, which is associated with SNR~W41, was recently reported to be spatially extended~\citep{Castro:2013eo,Jeremie:2013}.
After \citet{Jeremie:2013}, we model its spectrum as a power law and its spatial distribution as a symmetrical 2D Gaussian centered at $(l, b) =$ $(23 \fdg 22, -0 \fdg 31)$ with $\sigma = 0 \fdg 15$.
When LAT PSRs are associated with 2FGL sources, we adopt positions and spectral functions described in the 2PC catalog instead of the 2FGL catalog.
Here 2FGL J1813.4$-$1246, J1826.1$-$1256, J1833.6$-$1032, and J1839.0$-$0539 are replaced by LAT PSRs J1813$-$1246, J1826$-$1256, J1833$-$1034, and J1838$-$0537 respectively.
We include LAT PSR J1835$-$1106, which has no counterpart in the 2FGL catalog, in the model.
Figure~\ref{fig: cmap_wide} shows that a point-like source (source\,N) is located $\sim 3^\circ$ off the Galactic plane.
Since there is no corresponding source in the 2FGL catalog, source\,N is also added into the model.
Note that there is a corresponding source (3FGL\,J1824.3-0620) in the 3FGL catalog.
Source\,N is modeled as a point source located at $(l, b) = (24 \fdg 18, 3 \fdg 09)$. 
The spectral shape is modeled with a simple power law with an exponentially cut-off function.
The Galactic diffuse emission is modeled using {\sf gll\_iem\_v05.fit} while an isotropic component (extragalactic diffuse backgrounds plus residual charged particles) is modeled using {\sf iso\_iem\_v05.txt}. Both background models are the standard diffuse emission models released by the LAT Collaboration\footnote{Available from the {\it Fermi} Science Support Center, http://fermi.gsfc.nasa.gov/ssc/data/access/lat/BackgroundModels.html. 
We note that an analysis using {\sf gll\_iem\_v05\_rev.fit} instead of {\sf gll\_iem\_v05.fit} does not significantly affect spectral parameters obtained in this analysis.  The former includes diffuse emission from molecular gas beyond the solar circle that was omitted from the latter.  At the low longitudes of the present analysis, the differences between the two models are unimportant.}

In all analyses, the normalizations of the diffuse components are set free unless otherwise mentioned.
In the $>3$\,GeV analysis, we set the normalizations free for sources within the fitting region. 
We also set the spectral parameters free for the sources of $>1 \times 10^{-9}$\,photon\,cm$^{-2}$\,s$^{-1}$ in the energy range of 3--100\,GeV in the 2FGL catalog.
In the $>0.2$\,GeV analysis, because of the large fitting region, there are many fitting parameters.
To reduce free parameters,
we first set the normalizations free for sources of $>2 \times 10^{-8}$\,photon\,cm$^{-2}$\,s$^{-1}$ in the energy range of 0.3--100\,GeV within the fitting region.
Spectral parameters of the sources with the fluxes $>1 \times 10^{-7}$\,photon\,cm$^{-2}$\,s$^{-1}$ are also set free.
Note that we use the $>0.3$\,GeV fluxes from 2FGL because the catalog does not provide the fluxes of $>0.2$\,GeV.
We fit the data using this model and then use the parameters from this first fit for the $>0.2$\,GeV analysis.
In the subsequent analysis, we determine the freedom of the spectral parameters
based on the fitting fluxes in the energy range of 0.2--500\,GeV (flux$_{0.2}$) and the angular distance from the center of the fitting region ($r$):
we set the normalizations free for the sources with flux$_{0.2}$ $>4 \times 10^{-8}$\,photon\,cm$^{-2}$\,s$^{-1}$ in the $5^{\circ} < r \leq 10^{\circ}$ region; in the $r \leq 5^{\circ}$ region, we set the normalization free for the sources with flux$_{0.2}$ $>2 \times 10^{-8}$\,photon\,cm$^{-2}$\,s$^{-1}$ and set the normalizations and spectral shapes free for the sources with flux$_{0.2}$ $>4 \times 10^{-8}$\,photon\,cm$^{-2}$\,s$^{-1}$.
We also include sources outside the fitting regions but within $15^\circ$ and $6^\circ$ for the analyses above 0.2\,GeV and 3\,GeV respectively, with their parameters fixed at those given in the 2FGL and 2PC catalogs.

\subsubsection{Analysis Procedure}
\label{sec: pd}

Here we describe how we evaluate the spatial and energy distributions of G25 as well as of nearby background region G27.
Modeling G27 is necessary for precise measurements of G25.
The $\AN$ region contains three 2FGL sources (2FGL J1839.3$-$0558c, J1840.3$-$0413c, and J1841.2$-$0459c) and a bright identified source, $\LP$.


In the first step, we construct an input model using these sources (hereafter ``$model1$").
We fit data above 3~GeV with the $model1$ by using {\it gtlike}. 
Figure~\ref{fig: cmap_wide} shows a smoothed count map of a $10^\circ \times 10^\circ$ region around G25 above 3~GeV, 
a corresponding background model map, and the background-subtracted counts map. 
The background map includes contributions from the modeled point sources, Galactic diffuse emission and isotropic diffuse background whose normalizations are set at the best-fit values obtained by the fit.
The $\gamma$-ray excesses associated with the G25 and $\AN$ regions are clearly visible in the background-subtracted map. 

Figure~\ref{fig: cmap_wide} might suggest that the $\gamma$-ray emissions of G25 and G27 are extended rather than a combination of 2FGL sources.
If the sources are extended, additional point sources (not in 2FGL) are expected to be needed to approximately model 
the observed $\gamma$-ray emission with a combination of point sources,
given that the accumulated time of this dataset (about 57\,months) is more than twice as long as that for the 2FGL catalog.
To evaluate in detail the spatial distributions of these regions with this dataset,
we added multiple point sources to explain the observed $\gamma$-ray emission instead of the 2FGL sources within the regions.

We focus first on G25, because the sum of TS values of the 2FGL sources within the G25 region is larger than that of the $\AN$ region.
To determine positions of the multiple point sources, we generate a TS map\footnote{A TS map is constructed by stepping a trial point source through a grid of positions, maximizing the likelihood and evaluating the likelihood test statistic of the trial source at each position.} of G25 using data above 3\,GeV. 
The input model for the first TS map is constructed by removing the four 2FGL sources within the G25 region from $model1$.
In the TS map, TS at each grid position is calculated by placing a point source with a power-law energy distribution with photon index set free. 
Based on the TS map obtained we modify the input model by adding a point source at a grid position whose TS is the largest in the TS map and then re-evaluate the TS map.
We iteratively add point sources into the modified model until the peak value of the TS in the map is less than 25 ($\sim 4\,\sigma$).
After the procedure, we obtain a model with the newly-detected point sources in G25.
Next, based on the obtained model, we iteratively add point sources to the $\AN$ region by the same procedure as for G25.
We exclude from the model the three 2FGL sources within the $\AN$ region. This modified model is used as an input model for the TS map of G27.
Note that $\LP$ in the $\AN$ region is not removed.
Since the G25 and $\AN$ regions are spatially close, positions of the multiple sources for each region might affect each other.
To minimize the effect, we relocalize the position of each source of G25 in descending order of TS by using {\sf pointlike}. 
Then we repeat the same procedure for the sources in G27. Note that the re-localizations do not significantly change the results.
Finally we obtain a model with the newly detected point sources in the G25 and G27 regions (``$model2$"; see also Table~\ref{tbl: mulpsc-sfc10}).

\begin{deluxetable}{ccr}
\tabletypesize{\scriptsize}
\tabletypesize{\footnotesize}
\tablecaption{Multiple Point-Source Model for G25 \label{tbl: mulpsc-sfc10}}
\tablewidth{0pt}
\tablehead{
\colhead{Point} & \colhead{Position} & TS\\
\colhead{source} & \colhead{($l$, $b$)} & 
}
\startdata
$p1$ & (24\fdg83, \,0\fdg52) & 121 \\
$p2$ & (24\fdg55, \,0\fdg60) & 83 \\
$p3$ & (25\fdg15, -0\fdg09) & 79 \\
$p4$ & (25\fdg50, -0\fdg30) & 55 \\
$p5$ & (25\fdg11, \,0\fdg47) & 48 \\
$p6$ & (24\fdg83, -0\fdg05) & 44 \\
$p7$ & (25\fdg16, -0\fdg41) & 29 \\
$p8$ & (25\fdg11, -0\fdg80) & 27 \\
\enddata
\tablecomments{Positions of the sources are illustrated in Figure\,\ref{fig: cmap_cls-sfc10}.}
\end{deluxetable}

\begin{deluxetable}{crrr}
\tabletypesize{\scriptsize}
\tabletypesize{\footnotesize}
\tablecaption{Multiple Point-Source Model for $\AN$ \label{tbl: mulpsc-g27}}
\tablewidth{0pt}
\tablehead{
\colhead{Point} & \colhead{Position} & TS\\
\colhead{source} & \colhead{($l$, $b$)} & 
}
\startdata
$q1$ & (26\fdg32, -0\fdg02) & 283 \\
$q2$ & (27\fdg34, -0\fdg01) & 64 \\
$q3$ & (26\fdg96, -0\fdg09) & 56 \\
$q4$ & (27\fdg18, -0\fdg39) & 47 \\
$q5$ & (25\fdg95, \,0\fdg18) & 41 \\
$q6$ & (27\fdg87, \,0\fdg60) & 28 \\
\enddata
\tablecomments{Positions of the sources are illustrated in Figure\,\ref{fig: cmap_cls-g27}.}
\end{deluxetable}

Tables~\ref{tbl: mulpsc-sfc10} and \ref{tbl: mulpsc-g27} list the eight and six point sources that are needed to
adequately model the $\gamma$-ray emission from the $\sim 1 \fdg 5$ square regions of G25 and G27, respectively (see also the left panels of Figures~\ref{fig: cmap_cls-sfc10} and \ref{fig: cmap_cls-g27}).
Note that the number of point sources has doubled from 2FGL.
The high density of point sources suggests possible extension of the observed $\gamma$-ray emission.
To check whether these sources are real point sources or an approximation of extended sources, 
we evaluate models with extended  $\gamma$-ray emission associated with the two regions. 
The method is described in detail in Section~\ref{sec: g25-sp} and Appendix~\ref{sec: g27}.

Finally, using the obtained spatial distributions, 
we measure spectral energy distributions (SEDs) for each source within the regions in the 0.2--500~GeV energy range.
We also check the significance of a possible steepening of the SEDs. To do this, we perform likelihood-ratio tests between a power-law function (the null hypothesis), and 
either an exponentially cut-off power-law or a smoothly broken power-law function (the alternative hypotheses).
The exponentially cut-off power-law function is described as
\begin{equation}
\frac{dN}{dE}=N_0\left(\frac{E}{E_0}\right)^{-\Gamma} \exp{\left(-\frac{E}{E_{\rm cut}} \right)},
\label{eq: g-cutoff}
\end{equation}
where $E_0$ is 1~GeV. The photon index $\Gamma$, cutoff energy $E_{\rm cut}$, and normalization factor $N_0$ are free parameters.
The smoothly broken power-law function is described as
\begin{equation}
\frac{dN}{dE}=N_0\left(\frac{E}{E_0}\right)^{-\Gamma_1}\left(1+\left(\frac{E}{E_{\rm br}} \right)^{\Gamma_2-\Gamma_1} \right)^{-1}, 
\label{eq: g-sbr}
\end{equation}
where $E_0$ is 1~GeV. The photon indices $\Gamma_1$ below the break, $\Gamma_2$ above the break, break energy ${E_{\rm br}}$, and normalization factor $N_0$ are free parameters. 
The resulting test statistic ${\rm TS_{Cut}} = -2\ln (L_{\rm PL}/L_{\rm Cut})$ and ${\rm TS_{BPL}} = -2\ln (L_{\rm PL}/L_{\rm BPL})$ provide information on the curvature of the SEDs.
The results are described in detail in Section~\ref{sec: g25-sed} and Appendix~\ref{sec: g27}.


\subsubsection{Spatial Distribution}
\label{sec: g25-sp}

\begin{figure*}[t] 
\begin{center}
\includegraphics[width=7.5in]{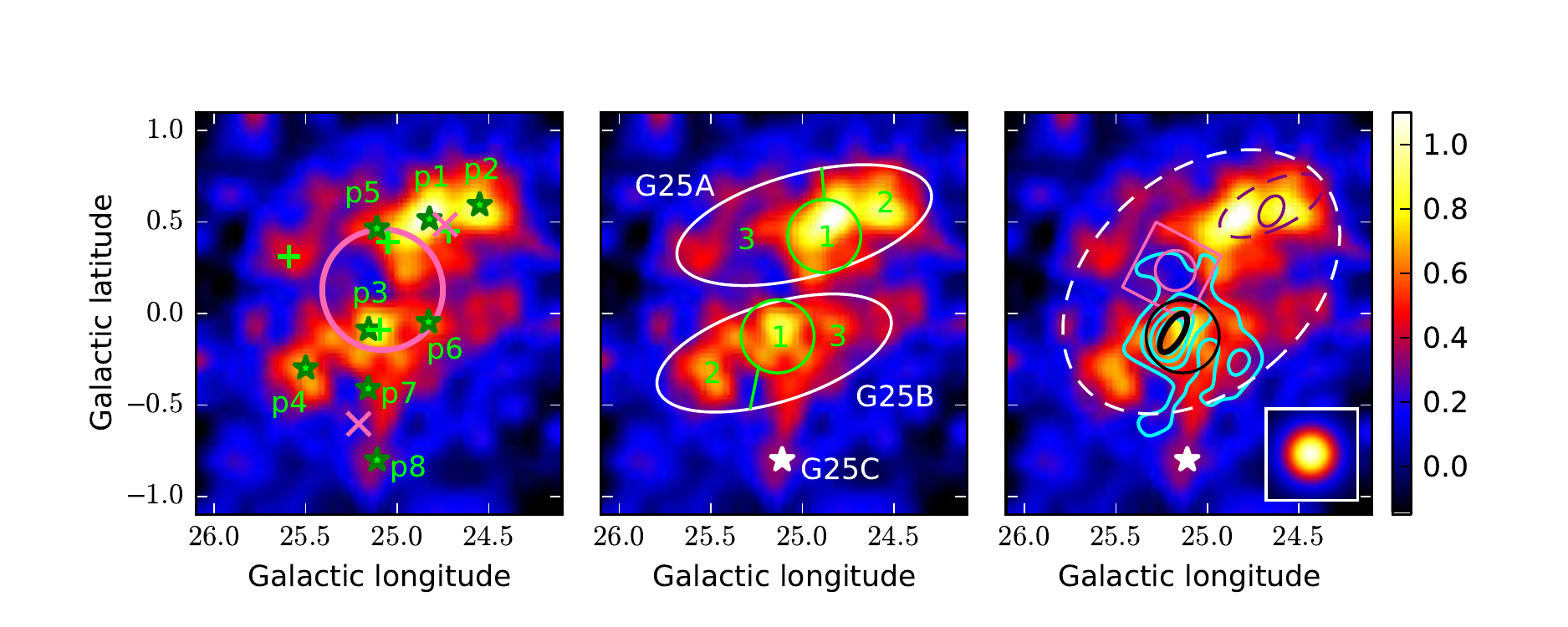}
\caption{\small
Left: {\it Fermi}-LAT background-subtracted map of the G25 region above 3~GeV (in units of counts per pixel). The pixel size is $0 \fdg 025$. Smoothing with a Gaussian kernel of $\sigma = 0 \fdg 15$ is applied.
Green stars represent the point sources described in Table~\ref{tbl: mulpsc-sfc10}.
Green crosses and the magenta X represent positions of the 2FGL and 1FHL sources respectively. The magenta circle shows an extended source 1FHL~J1836.5-0655e, which is based on the results of \cite{Lande:2012jb}.
Middle: same map as the left panel. The white ellipses delineate the G25A and G25B regions. Green circles and lines divide each region into three cells (see text). The point source G25C is represented by the white star.
Right: same map as the left panel.
The white dotted ellipse represents the spatial distribution obtained by fitting the $\gamma$ rays of G25A and G25B with one elliptical shape (see text).
The black circle and the white star represent G25B2 and G25C respectively.
$\HJP$ and its corresponding TeV image are displayed by the black ellipse and the cyan contours\,\citep{2006ApJ...636..777A}.
The contour levels are 15, 30, 45, and 60 counts per pixel, where the pixel size is 0 \fdg 005.
The magenta box and circle represent the field of view and target region of our X-ray analysis respectively (see Section\,\ref{sec: xmm} for details).
The purple dashed and solid ellipses represent sizes of a candidate PWN G24.7+0.6 based on \cite{1984A&A...133L...4R} and \citep{MAGPIS} respectively.
The simulated point source is shown in the inset (see text).
\label{fig: cmap_cls-sfc10}}
\end{center}
\end{figure*}

As described in Section~\ref{sec: pd}, eight point sources are needed if the observed $\gamma$-ray emission of G25 should be explained only by point sources. 
We modify $model2$ (see Section\,\ref{sec: pd}) and construct a new model (``$model3$") where we replace the sources in G27 by a spatial distribution estimated in Appendix\,\ref{sec: g27}.
The left panel in Figure~\ref{fig: cmap_cls-sfc10} shows a background-subtracted map of G25 using the best-fit values of $model3$.
Here the subtracted background model contains background sources outside the G25 region in addition to the contributions from the Galactic diffuse emission and isotropic diffuse background.
In the inset of the right panel, we show a point source simulated using a power-law of photon index 2.1, a typical spectral shape for sources in this region (see Table\,\ref{tbl: sed-sfc10}).
The high concentration of the point sources in the region rather might be explained by one or more extended sources.

To check for possible extension of the observed $\gamma$-ray emission, we perform the following steps:
(i) we adopt $model3$ as an input model and select the point source of the highest TS among the eight (i.e., $p1$ in Table~\ref{tbl: mulpsc-sfc10}).
(ii) we change the spatial distribution of the source from a point to an elliptical shape and fit the spatial parameters with {\sf pointlike}.
The spectral shape is not changed: a power-law function with its parameters free.
In the case that the TS of any of the other point sources decreases to less than 25 as a result, we remove them and refit the spatial shape of the target source.
(iii) we evaluate the significance of the spatial extension based on TS$_{\rm ext} = -2\ln (L_{\rm psc}/L_{\rm ell})$\,\citep{Lande:2012jb}.
If the value is larger than 14 ($\sim 3\,\sigma$ for three degrees of freedom; \cite{1996ApJ...461..396M}), we adopt the resulting elliptical shape.
In this case, we re-optimize the positions of the remaining point sources in descending order of TS by using {\sf pointlike},
since a change of the spatial shape might affect the other sources.
If ${\rm TS}_{\rm ext} < 14$, we keep the source as a point source in the model.
(iv) we continue to the source of the next highest TS among the remaining sources, and then repeat (ii)--(iv) until the source of the lowest TS is evaluated.

Here we describe details of the application of each step in the procedure.
We first fit $p1$ using an elliptical shape. The resulting elliptical size is so large ($\sim 1^\circ$) that the ellipse includes sources $p2$ and $p5$. Because the TS of the two sources decreases to less than 25, we remove both and refit the elliptical shape. This source is named ``G25A" as shown in Table\,\ref{tbl: sp-sfc10}. 
In Table\,\ref{tbl: sp-sfc10}, G25A has $\TSe$ of 158, which means that the source is significantly ($>5\,\sigma$) extended. This is not surprising given that the G25A region is explained by three point sources ($p1$, $p2$, and $p5$) in $model3$. 
We compare the likelihood of the model of the elliptical shape (G25A) and that of the three point sources. 
The resulting value is $-2\ln (L_{\rm 3psc}/L_{\rm G25A}) = 76$.
We note that this does not necessarily mean that the elliptical shape explains the observed $\gamma$-ray emission better than the three point sources
because the two models are not nested. Here we adopt the elliptical shape, given that it has higher likelihood even though it has 5 fewer degrees of freedom.
Next, we select source $p3$ and fit it using an elliptical shape; $p2$ was already removed in the previous procedure. 
Table\,\ref{tbl: sp-sfc10} shows that the resulting elliptical shape ``G25B" includes four point sources ($p3$, $p4$, $p6$, and $p7$) in the $model3$ ($\TSe$ of 126).
Since the resulting TS of each of those point sources is less than 25, we remove them and refit the elliptical shape as in the procedure of $p1$.
We also compare likelihood between models with the G25B region and the four point sources. 
The resulting value is $-2\ln (L_{\rm 4psc}/L_{\rm G25B}) = 14$.
Here we adopt the model with the ellipse, given that it has higher likelihood even though it has 9 fewer degrees of freedom.
The remaining point source is only $p8$. We fit $p8$ using an elliptical shape and find a $\TSe$ of 7, which means that this source is not significantly extended. 
We keep this source a point source and rename it ``G25C".
Finally we obtain ``$model4$ where we replace the point sources of G25 in $model3$ for G25A, B, and C.

\begin{deluxetable*}{lccccccrrc}
\tabletypesize{\scriptsize}
\tabletypesize{\footnotesize}
\tablecaption{Spatial Distributions of the Sources in G25 \label{tbl: sp-sfc10}}
\tablewidth{0pt}
\tablehead{
\colhead{Name} & \colhead{Spatial} & Center position & Positional & \colhead{Semi major} & \colhead{Semi minor} & \colhead{Angle\tablenotemark{b}} & $\TSe$ & TS & Sources\\
\colhead{} & \colhead{model} &  \colhead{($l$, $b$)}& error\tablenotemark{a} & axis (deg) & axis (deg) & (deg) &&& \colhead{ in Table\,\ref{tbl: mulpsc-sfc10} }
}
\startdata
G25A & ellipse & (24\fdg99, \,0\fdg48) & 0\fdg02 & $0.72\pm0.03$ & $0.28\pm0.02$ & $15\pm2$ & 158 & 412 & $p1, p2, p5$ \\
G25B & ellipse &  (25\fdg15, -0.22) & 0\fdg02 & $0.67\pm0.05$ & $0.27\pm0.02$ & $17\pm5$ & 126 & 283 & $p3, p4, p6, p7$  \\
G25C & point & (25\fdg11, -0\fdg80) & 0\fdg04 & -- &--&--& 7 & 27 & $p8$ \\
\enddata
\tablenotetext{a}{The positional error of the center position at 68\% confidence level}
\tablenotetext{b}{Measured counter-clockwise from the Galactic longitude axis to the major axis.}
\end{deluxetable*}

\begin{figure}[t] 
\begin{center}
\includegraphics[width=3.5in]{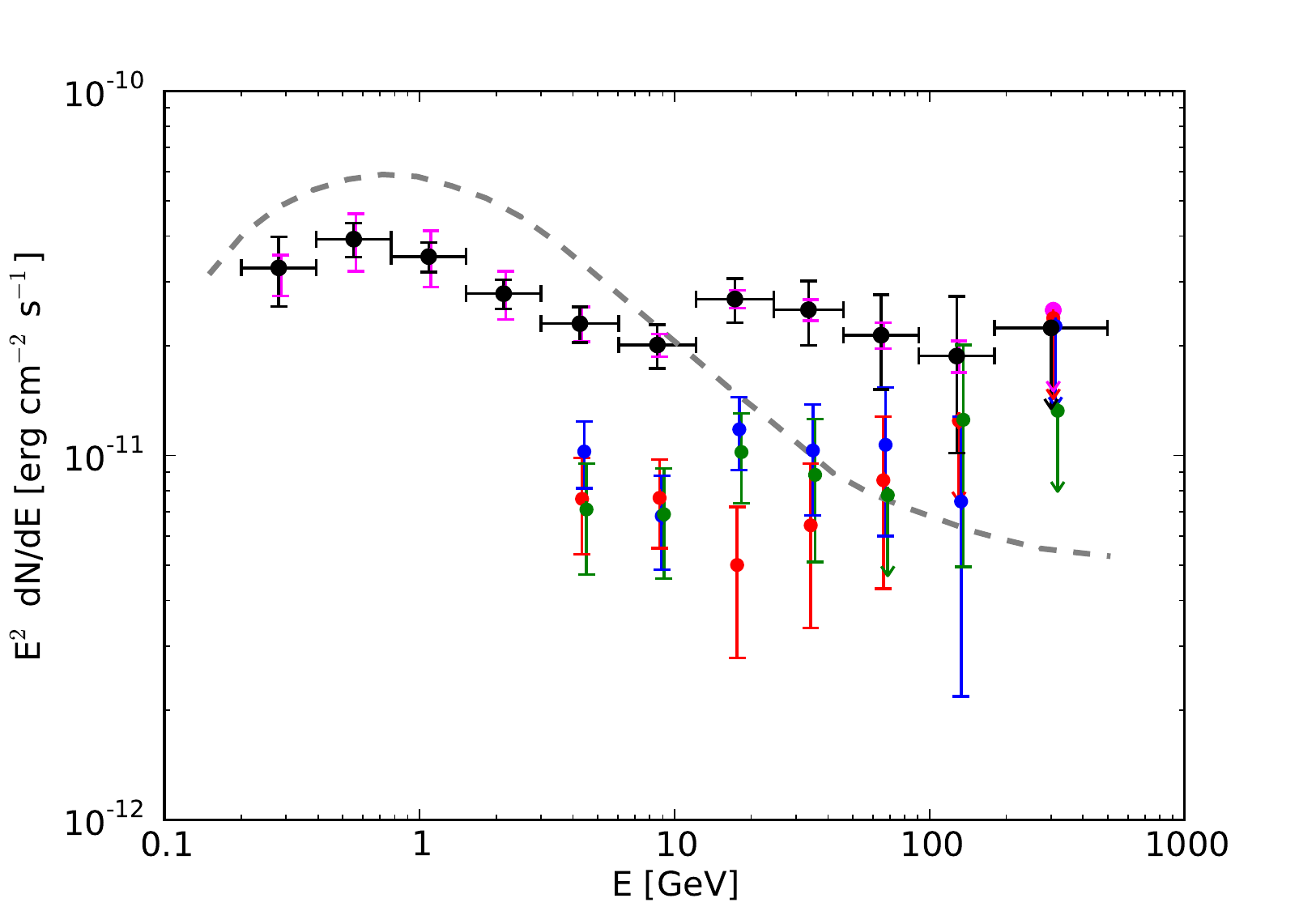}
\caption{\small
SED of the G25A region measured by the {\it Fermi} LAT. 
The arrows represent the 95\% confidence level upper limits. 
Upper limits are calculated for energy bins with ${\rm TS} < 4$.
The statistical errors ($1\,\sigma$) are indicated by the black error bars, while the total systematic errors (method\,1; see text) 
are indicated by the magenta error bars. 
The magenta points and arrows represent upper limits taking the systematic errors into consideration. 
SEDs of G25A1, A2, and A3 are also presented as red, blue, and green points with their statistical errors ($1\,\sigma$) respectively.
Note that the points and arrows are slightly off-set to clearly show the spectra.
The gray dashed line represents the Galactic diffuse emission from the G25A region, which is estimated by using the best-fit scaling of the Galactic diffuse model.
\label{fig: sed_x1}}
\end{center}
\end{figure}

Figure\,\ref{fig: cmap_cls-sfc10} (middle) and Table\,\ref{tbl: sp-sfc10} show the resulting spatial distributions and the corresponding parameters respectively.
Since the extended sources G25A and G25B are spatially close, we check whether one elliptical template is enough to explain $\gamma$ rays from the two sources.
We fit its spatial extent and spectral index. The resulting spatial shape is shown in Figure~\ref{fig: cmap_cls-sfc10} (right).
Here we perform a likelihood ratio test by setting this model (one ellipse) as the null hypothesis and the model of two ellipses (G25A and G25B) as the alternative hypothesis. The improvement of TS is 31, which corresponds to significance of 4.0\,$\sigma$ given that the alternative hypothesis has seven more free parameters\,\citep{1996ApJ...461..396M,Lande:2012jb}.
Note that the model with the two ellipses can take an independent spectral shape for each region, while the model with the one ellipse can take only one spectral shape.
The better likelihood might result from such a difference of spectral shapes rather than because of the difference of spatial shapes.
Actually, as we show in Section~\ref{sec: g25-sed}, a part of the G25B region (G25B1) has a much harder spectrum than the other regions. 
To check this possibility, we add the G25B1 template into the two models and re-fit them. 
If the above likelihood difference is mainly caused by the spectral discrepancy, the likelihood values of the two models will be similar.
The spectral shape of G25B1 is also assumed to be a simple power law.
We find that the improvement of TS between the two models is almost the same (${\rm TS} = 27$) as the models without G25B1. 
In addition, we test a model with a template of the HESS\,J1837$-$069 region (see Figure\,\ref{fig: cmap_cls-sfc10}) instead of G25B1. 
The resulting likelihood does not change, which indicates that the resulting TS is not sensitive to the spatial shape of the added template.

\subsubsection{Spectrum and Temporal Variability}
\label{sec: g25-sed}

\begin{figure}[t] 
\begin{center}
\includegraphics[width=3.5in]{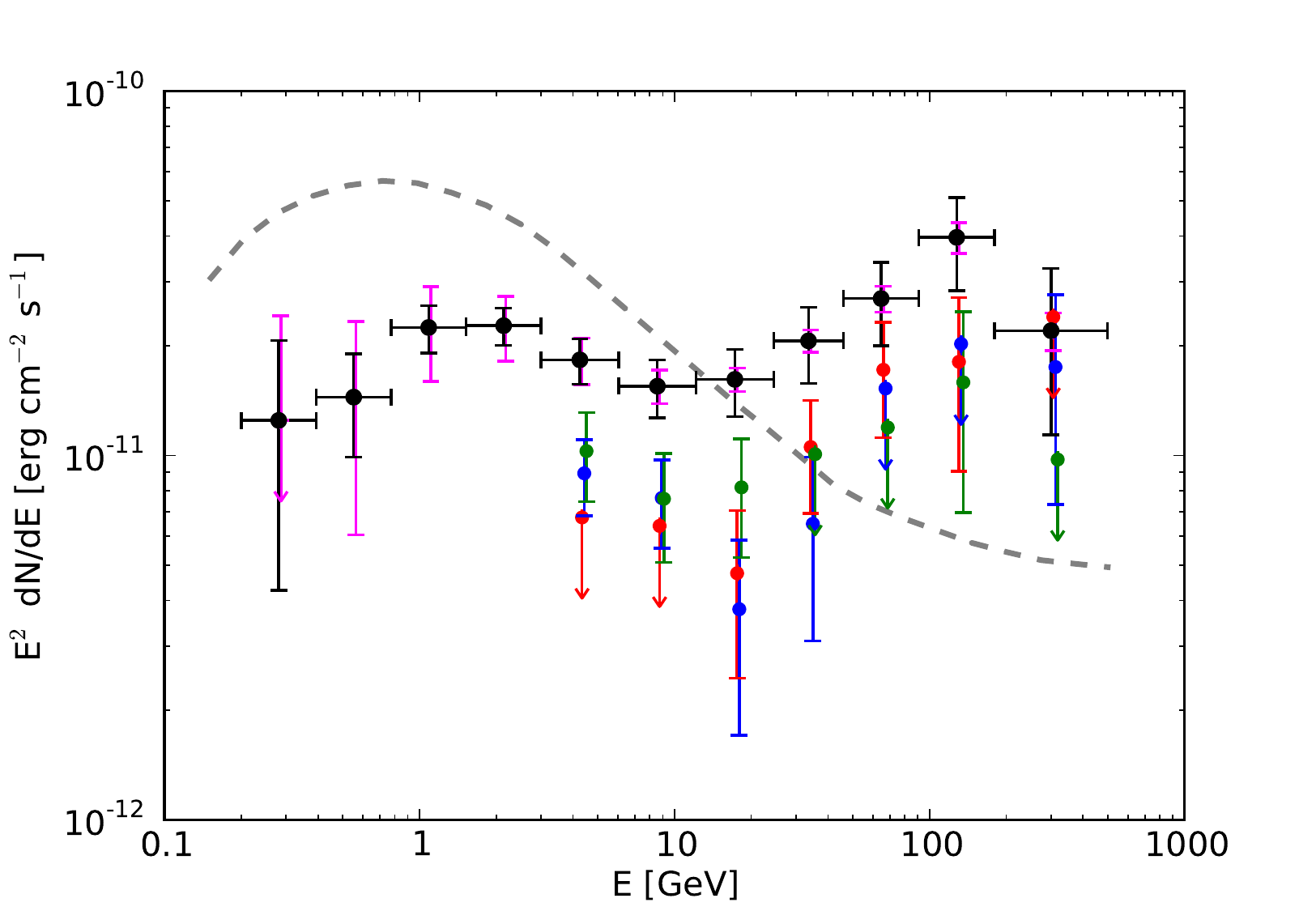}
\caption{\small
SED of the G25B region measured by the {\it Fermi} LAT.
The black and magenta SEDs have the same meaning as Figure\,\ref{fig: sed_x1}.
SEDs of G25B1, B2, and B3 are presented as red, blue, and green points with their statistical errors ($1\,\sigma$) respectively.
The gray dashed line represents the Galactic diffuse emission from the G25B region.
\label{fig: sed_x2}}
\end{center}
\end{figure}

\begin{figure}[h] 
\begin{center}
\includegraphics[width=3.5in]{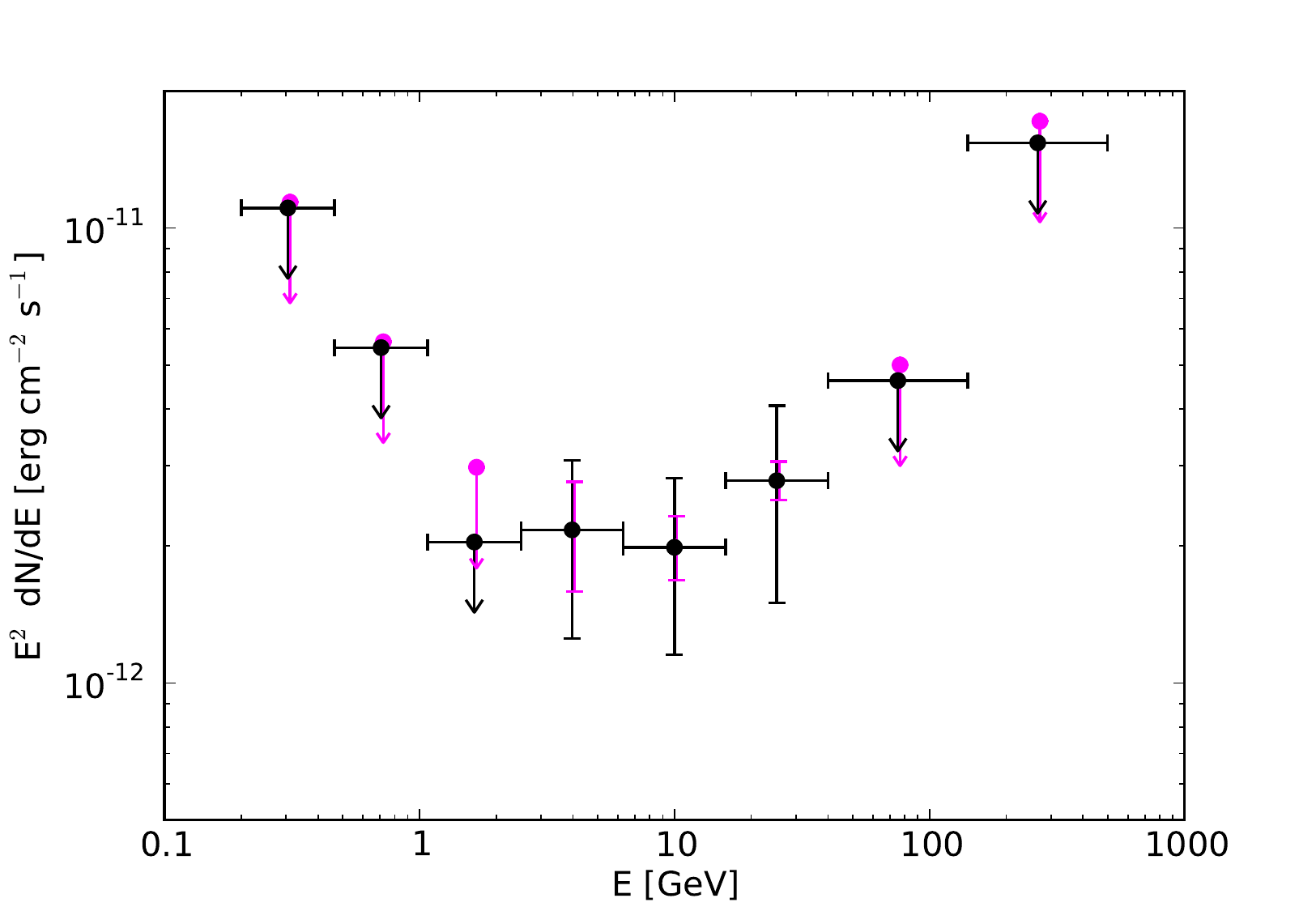}
\caption{\small
SED of the G25C source measured by the {\it Fermi} LAT.
The black and magenta SEDs have the same meaning as in Figure\,\ref{fig: sed_x1}.
\label{fig: sed_x3}}
\end{center}
\end{figure}

Figures\,\ref{fig: sed_x1}, \ref{fig: sed_x2}, and \ref{fig: sed_x3} show spectra of the sources in G25 (G25A, B, and C). The SEDs are obtained by dividing the 0.2--500\,GeV energy band into eleven, eleven, and eight logarithmically spaced energy bins respectively. All the sources are fitted with a simple power-law function in each energy bin with the photon index fixed at the value obtained from the broad-band fitting over the 0.2--500\,GeV energy range (see below). 
Since the LAT has better angular resolution at higher energy (Section\,\ref{sec: obs}), we divide the G25A and G25B regions, which are relatively large ($\gtrsim 1^\circ$) into three sections and evaluate the SED above 3\,GeV for each section (see Figure\,\ref{fig: cmap_cls-sfc10}). 
The regions are divided to meet the condition that each section contains a bright peak and has  TS $>25$.
The spectral shapes are fitted with a power-law function. The resulting TS values for G25A1, A2, and A3 are 71, 130, and 52 respectively, 
while those of G25B1, B2, B3, and C are 66, 61, 54 and 27 respectively.
The resulting spectral parameters are summarized in Table\,\ref{tbl: sed-sfc10}.


\begin{deluxetable*}{llccccc}
\tabletypesize{\scriptsize}
\tabletypesize{\footnotesize}
\tablecaption{Parameters of the Sub-Regions in G25 above 0.2\,GeV \label{tbl: sed-g25ab}}
\tablewidth{0pt}
\tablehead{
\colhead{Name} & Parameter &  \colhead{Value}& \colhead{Stat. error} & \multicolumn{2}{c}{Sys. error\,\tablenotemark{a}} 
 & \colhead{Sys. error\,\tablenotemark{b}}\\
\cline{5-6}
&&&& \colhead{method\,1} & \colhead{method\,2}}
\startdata
G25A &  Flux\,\tablenotemark{c} & 1.13 & $\pm 0.05$     & $\pm0.23$ & $\pm0.17$ & $\pm0.05$\\
            &  Photon Index               & 2.14 & $\pm 0.02$     & -0.04/+0.05 & $\pm0.07$ & $\pm0.02$\\
G25B$'$ &  Flux\,\tablenotemark{c} & 0.60 & $\pm 0.06$ & -0.24/+0.22 & $\pm0.31$ & $\pm0.03$\\
            &  Photon Index               & 2.11 & $\pm 0.04$ & -0.06/+0.13 & $\pm0.05$ & $\pm0.02$\\
\enddata
\tablenotetext{a}{Propagated uncertainties of the Galactic diffuse model. See text for details.}
\tablenotetext{b}{Propagated uncertainties of the LAT effective area. See text for details.}
\tablenotetext{c}{The flux is integrated over 0.2--500\,GeV in units of $10^{-7}$\,ph~cm$^{-2}$~s$^{-1}$.}
\end{deluxetable*}

\begin{deluxetable*}{llccccc}
\tabletypesize{\scriptsize}
\tabletypesize{\footnotesize}
\tablecaption{Parameters of the Sub-Regions of G25 above 3\,GeV \label{tbl: sed-sfc10}}
\tablewidth{0pt}
\tablehead{
\colhead{Name} & Parameter &  \colhead{Value}& \colhead{Stat. error} & \multicolumn{2}{c}{Sys. error\,\tablenotemark{a}}  & \colhead{Sys. error\,\tablenotemark{b}}\\
\cline{5-6}
&&&& \colhead{method\,1} & \colhead{method\,2}}
\startdata
G25A1 &  Flux\,\tablenotemark{c} & 1.5 & $\pm 0.3$   & -0.02/+0.01 & $\pm0.03$ & -0.07/+0.09\\
            &  Photon Index               & 2.08 & $\pm 0.16$ & -0.002/+0.004 & $\pm0.02$ & $\pm0.02$\\
G25A2 &  Flux\,\tablenotemark{c} & 2.0 & $\pm 0.3$    & $\pm0.2$ & $\pm0.2$ & $\pm0.1$\\
            &  Photon Index               & 2.03 & $\pm 0.12$ & $\pm0.03$ & $\pm0.04$ & -0.03/+0.02\\
G25A3 &  Flux\,\tablenotemark{c} & 1.6 & $\pm 0.3$     &$\pm0.18$ & $\pm0.3$ & -0.08/+0.09\\
            &  Photon Index               & 2.12 & $\pm 0.17$ & $\pm0.04$  &$\pm0.03$ & $\pm0.02$\\
\hline
G25B1 &  Flux\,\tablenotemark{c} & 1.0 & $\pm 0.2$   & $\pm0.001$ & $\pm0.05$ & $\pm0.06$ \\
            &  Photon Index               & 1.53 & $\pm 0.15$ & $\pm0.01$ & $\pm0.05$ & $\pm0.02$ \\ 
G25B2  &  Flux\,\tablenotemark{c} & 1.5 & $\pm 0.3$  & $\pm0.2$ & $\pm0.3$ & $\pm0.08$ \\
            &  Photon Index               & 2.06 & $\pm 0.18$ & $\pm0.05$ & $\pm0.05$ & $\pm0.02$ \\
G25B3 &  Flux\,\tablenotemark{c} & 1.7 & $\pm 0.3$   &  $\pm0.2$ & $\pm0.2$ & $\pm0.1$ \\
            &  Photon Index               & 2.14 & $\pm 0.18$ & $\pm0.06$ & $\pm0.05$ & $\pm0.02$\\
\hline
G25C &  Flux\,\tablenotemark{c} & 0.45 & $\pm 0.12$ & -0.03/+0.05 & $\pm0.02$ & -0.02/+0.04 \\
          &  Photon Index               & 2.1     & $\pm 0.3$   & -0.02/+0.05 & $\pm0.02$ & -0.01/+0.04 \\
\enddata
\tablenotetext{a}{Propagated uncertainties of the Galactic diffuse model. See text for details.}
\tablenotetext{b}{Propagated uncertainties of the LAT effective area. See text for details.}
\tablenotetext{c}{The flux is integrated over 3--500\,GeV in units of $10^{-9}$\,ph~cm$^{-2}$~s$^{-1}$.}
\end{deluxetable*}

Figure\,\ref{fig: sed_x1} suggests that the spectral shape of G25A is a simple power law in the 0.2--500\,GeV band. Quantitatively, the test for spectral curvature shows that ${\rm TS_{Cut}} = 0$ and ${\rm TS_{BPL}} = 0$, which means that no significant curvature is present. 
For the power-law model the photon index is $\Gamma = 2.14 \pm 0.02$ and the integrated 0.2--500\,GeV flux is $(1.13 \pm 0.05) \times 10^{-7}$\,photon\,cm$^{-2}$\,s$^{-1}$. 
Figure\,\ref{fig: sed_x2} shows clearly that the region G25B has a hard energy spectrum above $\sim 10\,{\rm GeV}$. This hard spectrum mainly comes from the region G25B1 (see Table\,\ref{tbl: sed-sfc10}). Therefore we separately measure SEDs of G25B1 and the other regions: G25B2 plus G25B3 (hereafter G25B$'$). 
We fit an SED of G25B1 using $>3\,{\rm GeV}$ data; in this energy range the LAT has the best angular resolution and so contamination from the Galactic diffuse emission component and the other sources is expected to be reduced relative to lower energies. The restricted energy range is sufficient for measuring the SED of G25B1 because of insignificant fluxes at lower energy due to the hard spectrum (see Figure\,\ref{fig: sed_x2}).
The spectral fitting results are summarized in Table\,\ref{tbl: sed-sfc10}.
We fit the spectrum of G25B$'$ in the 0.2--500\,GeV band with the SED of G25B1 fixed at the above values.
We find no significant curvature in the spectrum (${\rm TS_{Cut}} = 0$ and ${\rm TS_{BPL}} = 0$). 
The photon index is $\Gamma = 2.11 \pm 0.04$ and the integrated 0.2--500\,GeV flux is $(6.0 \pm 0.6) \times 10^{-8}$\,photon\,cm$^{-2}$\,s$^{-1}$. 
We fit the source G25C using a simple power law. 
We do not find any significant spectral curvature (${\rm TS_{Cut}} = 1$ and ${\rm TS_{BPL}} = 4$)                                                                                                       due to the low statistics.
The photon index is $\Gamma = 1.8 \pm 0.2$ and the integrated 0.2--500\,GeV flux is $(2.8 \pm 1.8) \times 10^{-9}$\,photon\,cm$^{-2}$\,s$^{-1}$. 

We also evaluate the systematic errors for the obtained SEDs.
The systematic errors in the spectral analysis are mainly due to uncertainties associated with the model of the underlying Galactic diffuse emission and uncertainties of the effective area of the LAT. The uncertainties of the Galactic diffuse emission are primarily due to imperfections in the Galactic diffuse model and/or the contributions from discrete sources not resolved from the diffuse background.
We evaluate the uncertainties of the Galactic diffuse emission by measuring the dispersion of the fractional residuals in ten regions near G25 as shown in Figure\,\ref{fig: cmap_wide} (middle), where the Galactic diffuse component dominates. The size of each box is $0\fdg5 \times 0\fdg5$.
The fractional residuals, namely (observed$-$model)/model, are calculated for each region using the energy range above 1\,GeV data. 
We adopt as the uncertainty of the Galactic diffuse model the second largest value among the ten residuals (90\% containment).
From the results, the uncertainty of the model is evaluated as 4\%. We apply this uncertainty over the entire energy range of 0.2--500\,GeV.
Systematic uncertainties of the effective area are 10\% at 0.1\,MeV, decreasing to 3\% at 0.3\,GeV, constant over 0.3--10\,GeV, and increasing to 13\% at 500\,GeV\,\citep{LATarea:2013}.
Total systematic errors are set by adding in quadrature the uncertainties due to the Galactic diffuse model and the effective area;
the systematic errors are dominated by the uncertainties of the Galactic diffuse model below 10\,GeV and by the uncertainty of the effective area above 100\,GeV.
We evaluate the systematic error due to the Galactic diffuse emission for the fitting parameters in the energy range 0.2--500\,GeV and 3--500\,GeV (see Table\,\ref{tbl: sed-g25ab} and \ref{tbl: sed-sfc10}) using the above method (method\,1). 
We also evaluate the systematic error using alternative Galactic diffuse models as in \cite{AltDiff}\,(method\,2). Eight models were tested: the parameters allowed to vary among the models are the radial distribution of the cosmic-ray sources (SNR-like or pulsar-like), the size of the cosmic-ray halo (4\,kpc or 10\,kpc) and the spin temperature of atomic hydrogen (150\,K or optically thin). 
In this analysis, a single scaling factor is allowed to vary for each alternative diffuse emission model.
The obtained systematic errors are marginally consistent with the ones estimated by method\,1 (see Figures\,\ref{fig: sed_x1}, \ref{fig: sed_x2}, and \ref{fig: sed_x3}).
We also estimate the systematic error due to the effective area for the fitting parameters by calculating bracketing instrument response functions as in \cite{LATorbit}. In most cases, the systematic errors due to uncertainties of the effective area are not important compared to those due to the Galactic diffuse model.

\begin{figure}[h] 
\begin{center}
\includegraphics[width=3.5in]{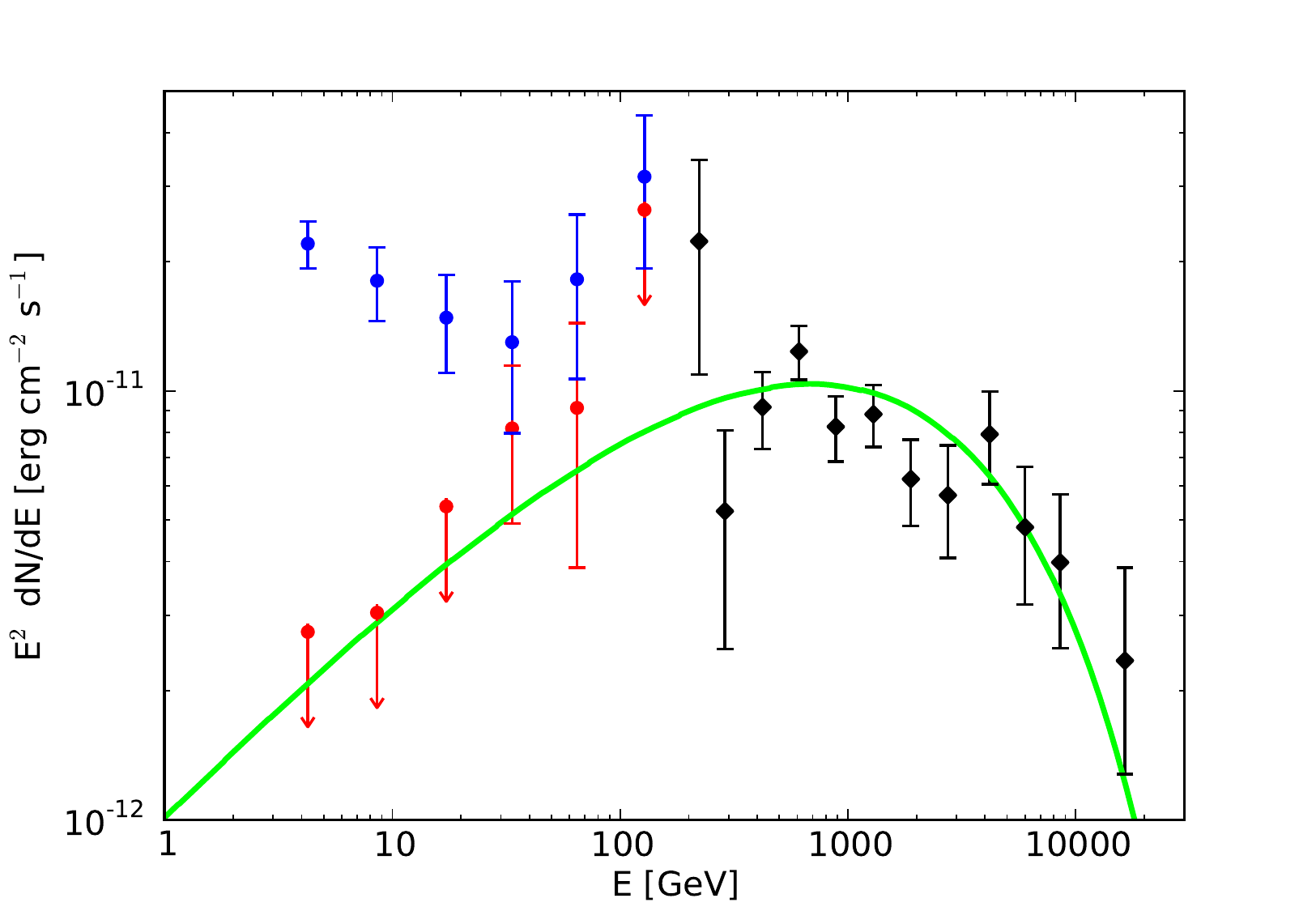}
\caption{\small
SED of the $\HJP$ region measured by the {\it Fermi} LAT (red points with statistical errors of $1\,\sigma$).
The arrows represent 95\% confidence level upper limits. 
Upper limits are calculated for energy bins with ${\rm TS} < 4$.
The TeV data of $\HJP$ (black diamond) are from \citet{2006ApJ...636..777A}.
Blue points represent the SED of the G25B region except for the $\HJP$ region (see text).
The green line shows the inverse Compton model for $\HJP$ (see Section\,\ref{sec: d_g25b} for details).
\label{fig: sed_TeV}}
\end{center}
\end{figure}

We also measure the SED of the region of \HJP\ above 3\,GeV.
Assuming that G25B is composed of the H.E.S.S. source and the other source, 
we divide G25B into an elliptical shape of the H.E.S.S. source ($0 \fdg 24 \times 0 \fdg 10$ in size; see Figure\,\ref{fig: cmap_cls-sfc10}) and the other region.
Figure~\ref{fig: sed_TeV} shows the resulting SEDs combined with the SED of the H.E.S.S. source\,\citep{2006ApJ...636..777A}.
The figure shows that the LAT SED of \HJP\ smoothly connects to the H.E.S.S. data.
The parameters obtained with the power-law model are photon index $\Gamma = 1.4 \pm 0.2$ and integrated 3--500\,GeV flux $(4.3 \pm 1.6) \times 10^{-10}$\,photon\,cm$^{-2}$\,s$^{-1}$. 
The result is consistent with what we obtained for G25B1 given that G25B1 contains the region of \HJP\ and is slightly larger than \HJP.
On the other hand, Figure~\ref{fig: sed_TeV} also shows that the G25B region except for \HJP\ has a hard spectral tail above $\sim 10$\,GeV,
while no clear hardening appears in the SEDs of G25B2 and G25B3 (see Figure\,\ref{fig: sed_x2}).
This suggests that the spatial region of the spectral hardness is larger than \HJP, but is mostly concentrated in the G25B1 region with its diameter of $0 \fdg 4$.
In addition, given that the hard tail remains up to a few hundred GeV, $\HJP$ may be larger than the previously reported size, even in the TeV energy range. 

To check for time variability in the 0.2--500\,GeV range, we divide the whole LAT observation period (about 57\,months) into 50, 30, and 4 bins for G25A, B, and C respectively.
The number of time bins is determined to meet the condition that the sources in each time bin have ${\rm TS} \gtrsim 4$.
In each time bin, the fluxes of the sources are evaluated by performing a {\sf gtlike} fit with the Galactic diffuse and isotropic components fixed to the best-fit values obtained for the whole time range in the energy range 0.2--500\,GeV. In addition, the photon indices of all sources are fixed at the values obtained from the broad-band fitting.
We do not find any indication of variability for any sources in the period spanned by the observations.
We also check the time variability of G25B1 in the 3--500\,GeV range by performing the same procedure as above and do not find any indication for variability.

\section{\XMM\ Data}
\label{sec: xmm}

\subsection{Observations and Data Reduction}

\begin{figure*}[t] 
\begin{center}
\includegraphics[width=7.in]{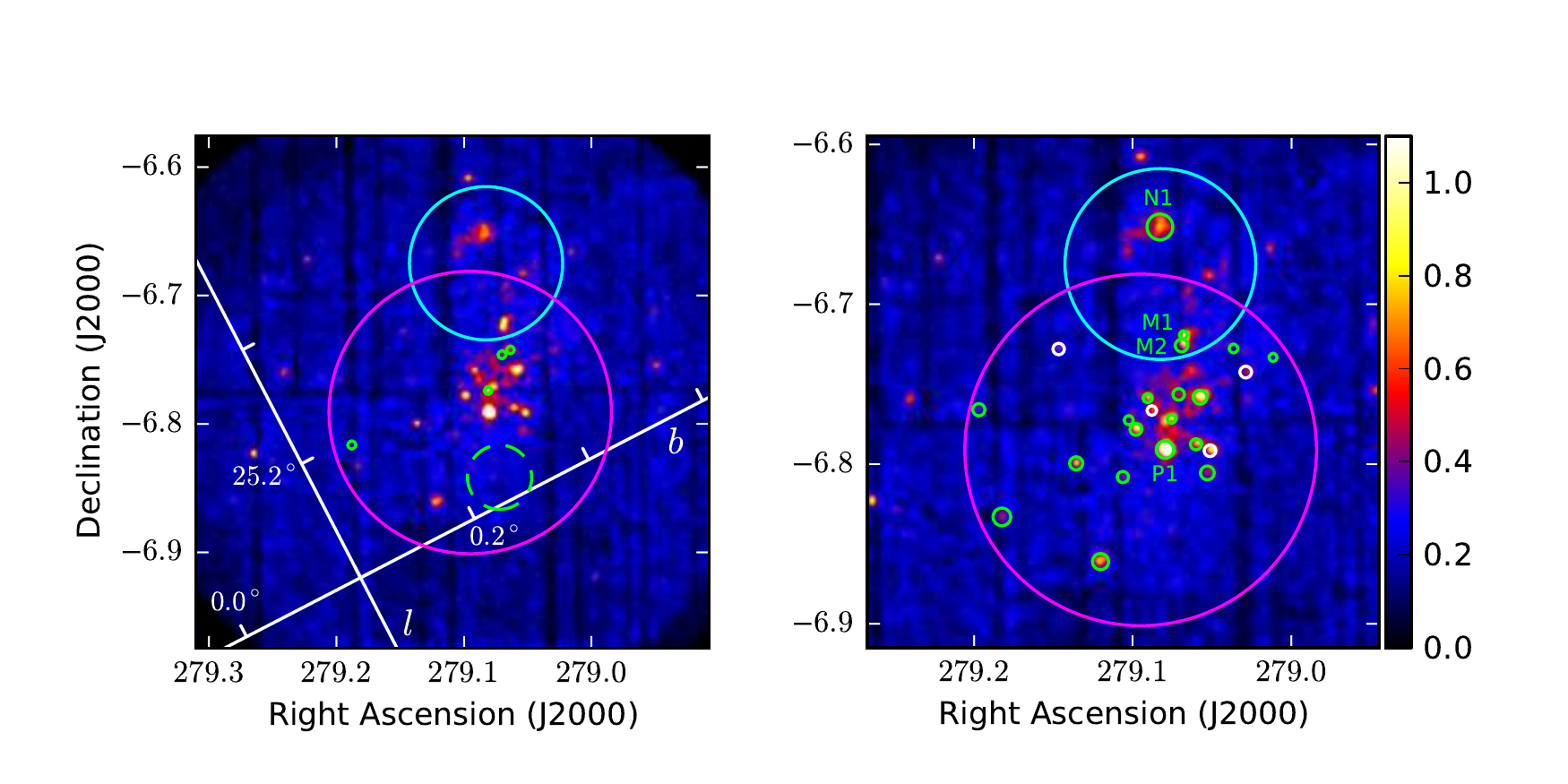}
\caption{
Left: 
combined MOS1+MOS2+pn count map of the AX\,J1836.3$-$0647 field in the 0.2--12\,keV band.
The magenta and the cyan circles represent the regions of group (i) and (ii) in Table\,\ref{tbl: xsrcs} respectively. 
The green dotted circle displays the background region.
The green solid circles denote the event-extraction regions for sources S1--S4.
Note that the magenta circle is the same as that in the right panel of Figure\,\ref{fig: cmap_cls-sfc10}.
Right: 
close-up image of the left panel. 
The small circles denote the event-extraction regions for sources M1--M4 (white) and sources P1, Q1--Q15, M1, M2 and N1 (green).
\label{fig: imgs-xray}}
\end{center}
\end{figure*}

The AX\,J1836.3$-$0647 field, centered at (RA, Dec) = (279.078, $-$6.787) (J2000), was observed with \XMM\,\citep{XMM} on 2007 September 18 for about 17\,ks.
The field is located between the $\gamma$-ray sources G25A and G25B (see the right panel of Figure\,\ref{fig: cmap_cls-sfc10}).
Data were acquired with the European Photon Imaging Camera (EPIC), which consists of three cameras: MOS1, MOS2, and pn\,\citep{pn,MOS}. The full-frame mode and the thick blocking filter were used for the observations.
The raw data are processed following standard procedures with the Science Analysis System (SAS) software\footnote{http://xmm.esac.esa.int/sas/} version 13.5.0:
event files of pn and MOS are extracted and cleaned with {\sc epproc} and {\sc emproc} respectively\footnote{http://xmm.esac.esa.int/sas/current/doc/epicproc/node3.html}. 
In addition, a count map of the X-ray data is created using {\sc emosaic}, which combines MOS1, MOS2, and pn images.
The spectral analyses are performed with {\sf XSPEC}\,(v12.8.1)\footnote{https://heasarc.gsfc.nasa.gov/xanadu/xspec}.

\subsection{Analysis and Results}
\label{sec: xray}

\begin{deluxetable}{llccr}
\tabletypesize{\scriptsize}
\tabletypesize{\footnotesize}
\tablecaption{Analyzed X-ray Sources \label{tbl: xsrcs}}
\tablewidth{0pt}
\tablehead{
\colhead{ID} &  \colhead{3XMM catalog}  & \colhead{RA} & \colhead{Dec} & \colhead{HR3}\\
\colhead{}     & \colhead{name}                 & \colhead{(deg)} & \colhead{(deg)} &  
}
\startdata
Group (i) \\
~P1     &  J183619.0-064728  &  279.080  &  -6.791  &  $-0.6 \pm 0.0$\\
~Q1     &  J183602.8-064358  &  279.012  &  -6.733  &  $0.1 \pm 0.2$\\
~Q2     &  J183608.7-064339  &  279.036  &  -6.728  &  $-0.0 \pm 0.2$\\
~Q3     &  J183612.8-064819  &  279.054  &  -6.806  &  $0.1 \pm 0.2$\\
~Q4     &  J183613.5-064528  &  279.056  &  -6.758  &  $-0.1 \pm 0.1$\\
~Q5     &  J183614.3-064714  &  279.060  &  -6.787  &  $-0.4 \pm 0.1$\\
~Q6     &  J183617.0-064521  &  279.071  &  -6.756  &  $-0.2 \pm 0.2$\\
~Q7     &  J183618.2-064623  &  279.076  &  -6.773  &  $-0.1 \pm 0.1$\\
~Q8    &  J183621.6-064530  &  279.090  &  -6.759  &  $-0.2 \pm 0.1$\\
~Q9    &  J183623.3-064639  &  279.097  &  -6.778  &  $-0.2 \pm 0.1$\\
~Q10    &  J183624.6-064626  &  279.103  &  -6.774  &  $0.1 \pm 0.2$\\
~Q11    &  J183625.6-064827  &  279.107  &  -6.808  &  $0.1 \pm 0.2$\\
~Q12    &  J183628.7-065138  &  279.120  &  -6.861  &  $-0.1 \pm 0.1$\\
~Q13    &  J183632.4-064759  &  279.135  &  -6.800  &  $-0.2 \pm 0.1$\\
~Q14    &  J183643.9-064959  &  279.183  &  -6.833  &  $-0.1 \pm 0.2$\\
~Q15    &  J183647.2-064556  &  279.197  &  -6.766  &  $-0.1 \pm 0.2$\\
~R1     &  J183606.8-064433  &  279.029  &  -6.743  &  $0.9 \pm 0.2$\\
~R2     &  J183612.2-064729  &  279.051  &  -6.792  &  $0.9 \pm 0.2$\\
~R3     &  J183621.1-064600  &  279.088  &  -6.767  &  $0.7 \pm 0.2$\\
~R4     &  J183635.0-064340  &  279.146  &  -6.728  &  $0.9 \pm 0.1$\\
~S1     &  J183615.2-064431  &  279.064  &  -6.742  &  $-0.6 \pm 0.2$\\
~S2     &  J183616.5-064444  &  279.069  &  -6.746  &  $-0.9 \pm 0.2$\\
~S3    &  J183619.3-064627  &  279.081  &  -6.774  &  $-0.8 \pm 0.1$\\
~S4    &  J183645.0-064900  &  279.188  &  -6.817  &  $-0.8 \pm 0.3$\\
Group (ii) \\
~M1     &  J183616.3-064312  &  279.068  &  -6.720  &  $0.4 \pm 0.1$\\
~M2     &  J183616.7-064330  &  279.070  &  -6.725  &  $0.0 \pm 0.1$\\
~N1     &  J183620.1-063904  &  279.084  &  -6.651  &  $-0.2 \pm 0.1$\\
\enddata
\end{deluxetable}

Figure\,\ref{fig: imgs-xray}\,(left) clearly shows that multiple ($\sim30$) point-like sources are concentrated near the center of the field.
Such a high concentration strongly suggests that these sources are collectively a stellar association/cluster. 
To study energy spectra of these sources, we divide the sources into three groups: 
(i) sources in the magenta circle and outside the cyan circle; (ii) those in the cyan circle; (iii) those outside of both circles (see Figure\,\ref{fig: imgs-xray}).
Information about sources in groups (i) and (ii) is summarized in Table\,\ref{tbl: xsrcs}.
The source positions and the values of HR3 we use are from the 3XMM-DR4 catalog\,\citep{3XMM}.
HR3 is a hardness ratio, which is defined as (H-M)/(H+M), 
where M and H are EPIC count rates in the ranges 1--2\,keV and 2--4.5\,keV respectively (see the 3XMM-DR4 catalog for more details). 
We do not analyze sources in group (iii), because most of them must be foreground/background sources unrelated to the putative association/cluster.
In group (ii), three bright sources are analyzed; the others are too dim.
We extract the events for each source in a circular region with radius displayed in Figure\,\ref{fig: imgs-xray}\,(right). 
In the spectral analyses, the background region is selected from blank sky within the field of view as shown in Figure\,\ref{fig: imgs-xray}\,(left). 

\begin{figure}[t] 
\begin{center}
\includegraphics[width=3.5in]{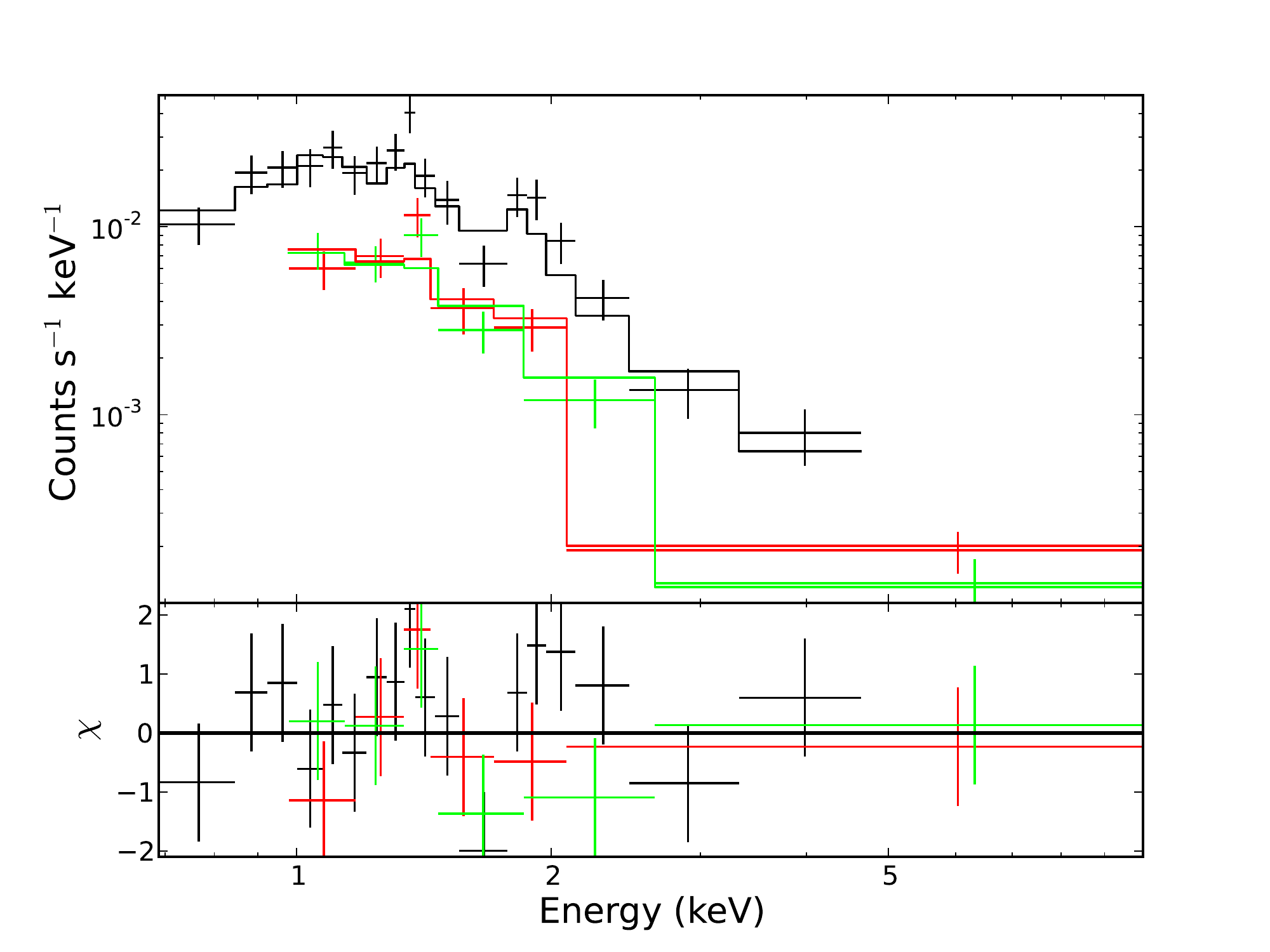}
\caption{
EPIC spectra of source P1 (black: pn; red: MOS1; green: MOS2).
The best-fit $2T$ model with abundance of 1.0 (Table\,\ref{tbl: xmodel-2T-PQ}) is imposed.
The bottom panel shows the distribution of $\chi$, which is defined as (observation -  model)/(statistical error)
at each energy point.
\label{fig: sed-P}}
\end{center}
\end{figure}

The brightest X-ray source (0.5--12\,keV) in this region is P1, and, 
as shown in Figure\,\ref{fig: sed-P}, clearly displays thermal features.
We fit the spectrum with an absorbed thermal model of {\sf wabs*apec} ($1T$ model) using {\sf XSPEC}\,\citep[the abundances of ][are used]{AE89}.
The resulting parameters are tabulated in Table\,\ref{tbl: xmodel-1T}. 
In the fit, we fix the abundances at 0.2 and 1, both of which can explain the observed data.
Since the X-ray image suggests the central sources are members of a stellar association, 
we also fit the observed spectrum with a typical spectral model for such sources.
Here we adopt an absorbed two-temperature thermal model of {\sf wabs*(apec$_1$ + apec$_2$)} ($2T$ model).
Typical temperatures are $0.1\,{\rm keV} \lesssim kT_1 \lesssim 1\,{\rm keV}$ and $1\,{\rm keV} \lesssim kT_2 \lesssim 5\,{\rm keV}$\,\citep[e.g.,][]{2009A&A...506.1055N,Townsley:2011jv}.
In the fit, the normalization of the second component is fixed at 20\% of the first component because of low statistics.
The ratio of the normalizations is chosen to be a typical value for O stars\,\citep{2009A&A...506.1055N}.
Note that the ratio does not strongly influence the resulting parameters (the column density and temperatures change 
by $\lesssim 10$\% when changing the ratio from 10 to 30\%).
Table\,\ref{tbl: xmodel-2T-PQ} shows that both the $1T$ and $2T$ models can explain the data.
Both models are dominated by the low-temperature ($< 1$\,keV) component and  
the obtained $\NH$ is almost the same for both models.

\begin{figure}[t] 
\begin{center}
\includegraphics[width=3.5in]{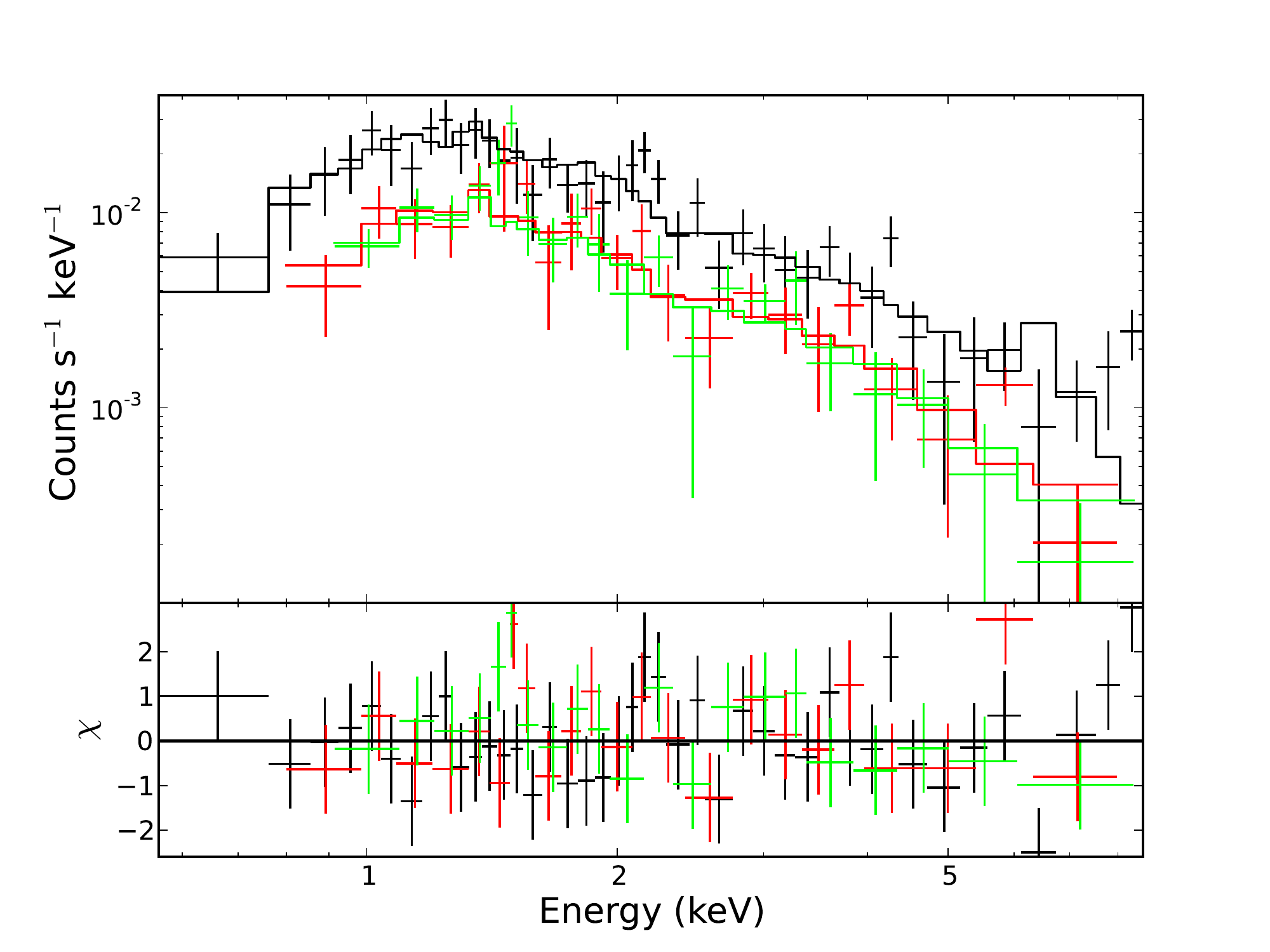}
\caption{
Same spectra as Figure\,\ref{fig: sed-P} but for Q$_{\rm sum}$.
\label{fig: sed-Q}}
\end{center}
\end{figure}

\begin{figure}[t] 
\begin{center}
\includegraphics[width=3.5in]{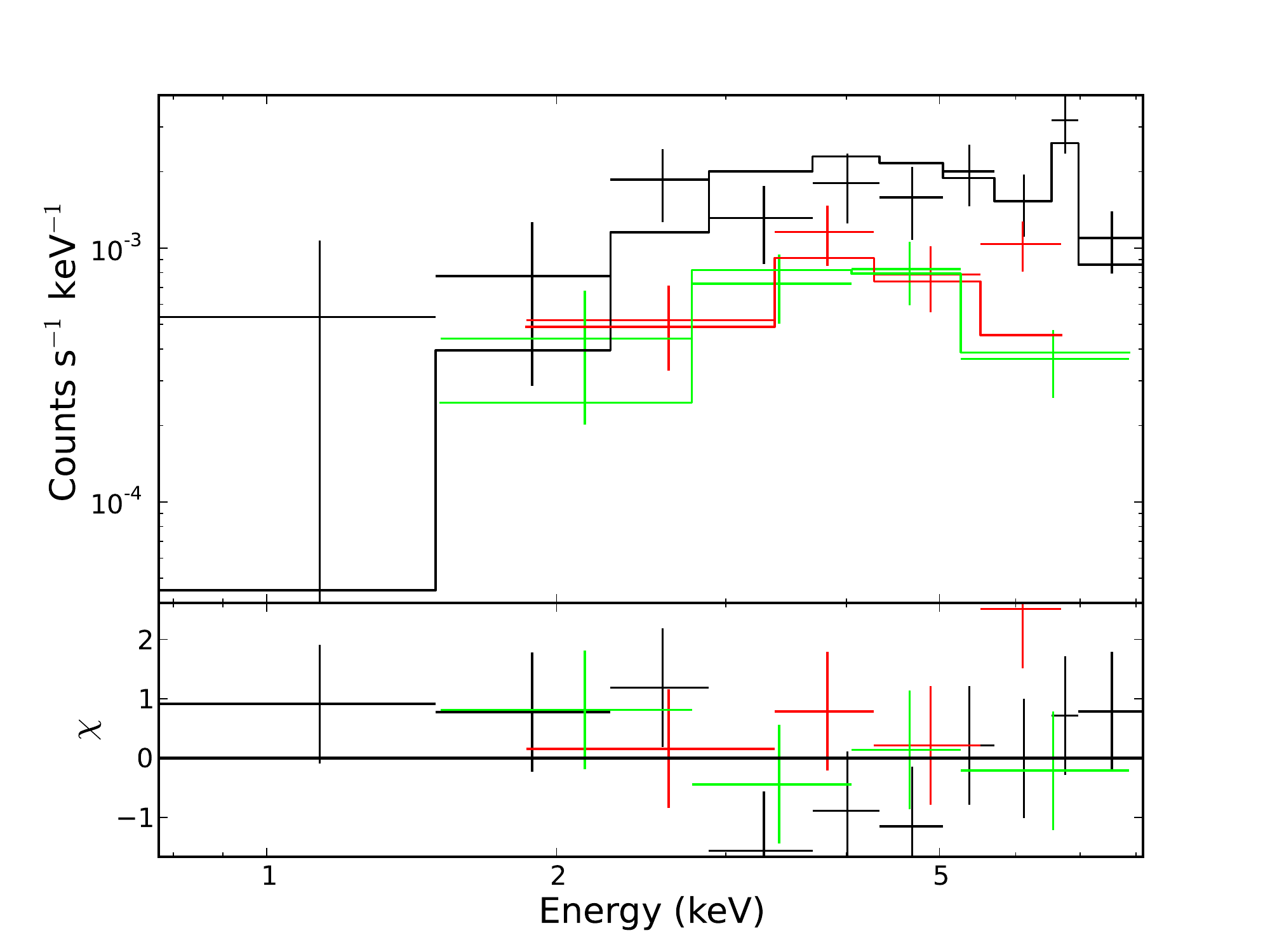}
\caption{
EPIC spectra of R$_{\rm sum}$ (black: pn; red: MOS1; green: MOS2).
The best-fit $1T$ model with abundance of 1.0 (Table\,\ref{tbl: xmodel-1T}) is imposed.
\label{fig: sed-R}}
\end{center}
\end{figure}

\begin{figure}[t] 
\begin{center}
\includegraphics[width=3.5in]{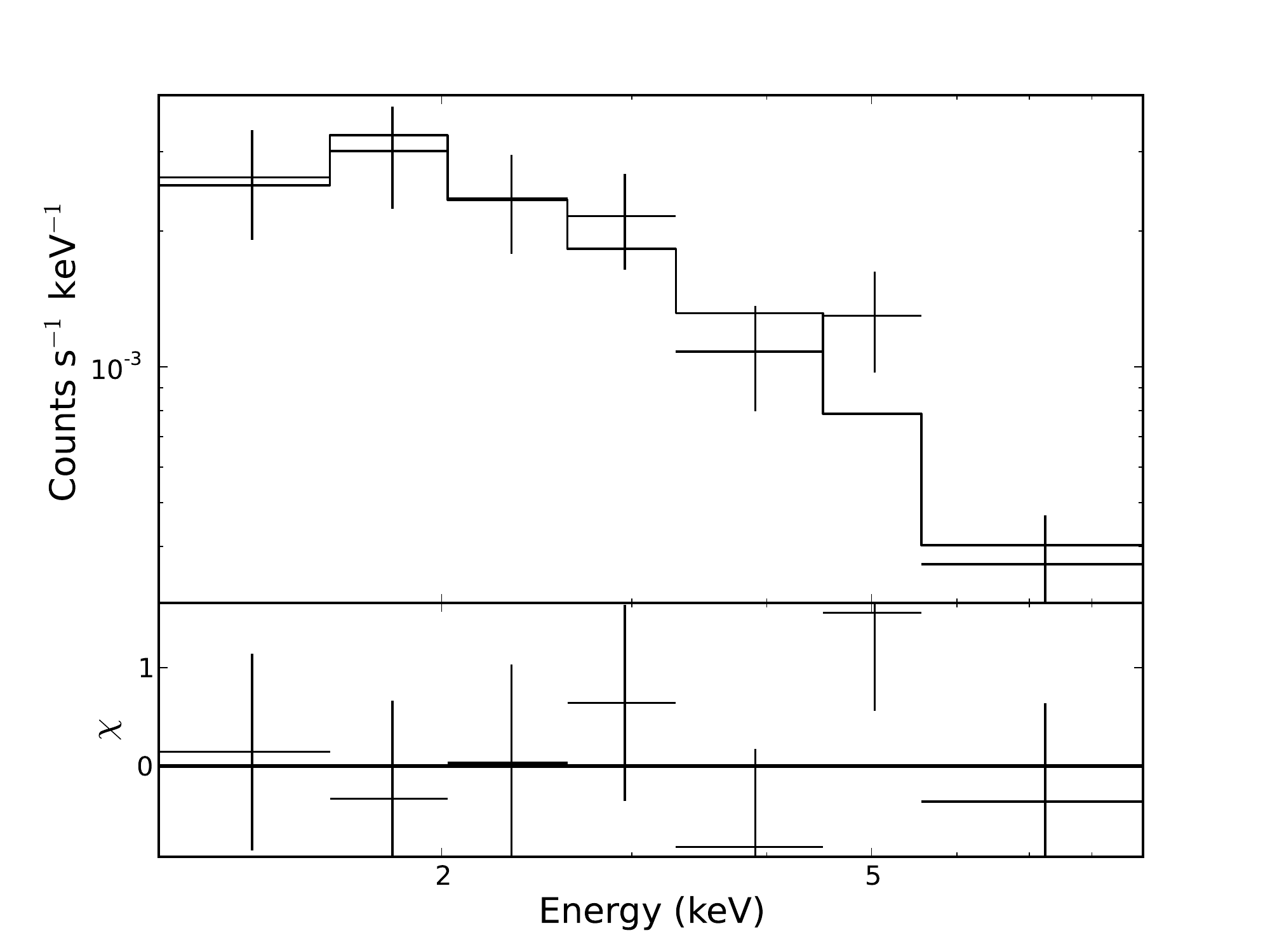}
\caption{
The pn spectra of M$_{\rm sum}$. Note that the spectra of MOS1 and MOS2 are not analyzed because of low statistics.
The best-fit $2T$ model with the abundance of 1.0 (Table\,\ref{tbl: xmodel-2T-PQ}) is imposed.
\label{fig: sed-M}}
\end{center}
\end{figure}

\begin{figure}[t] 
\begin{center}
\includegraphics[width=3.5in]{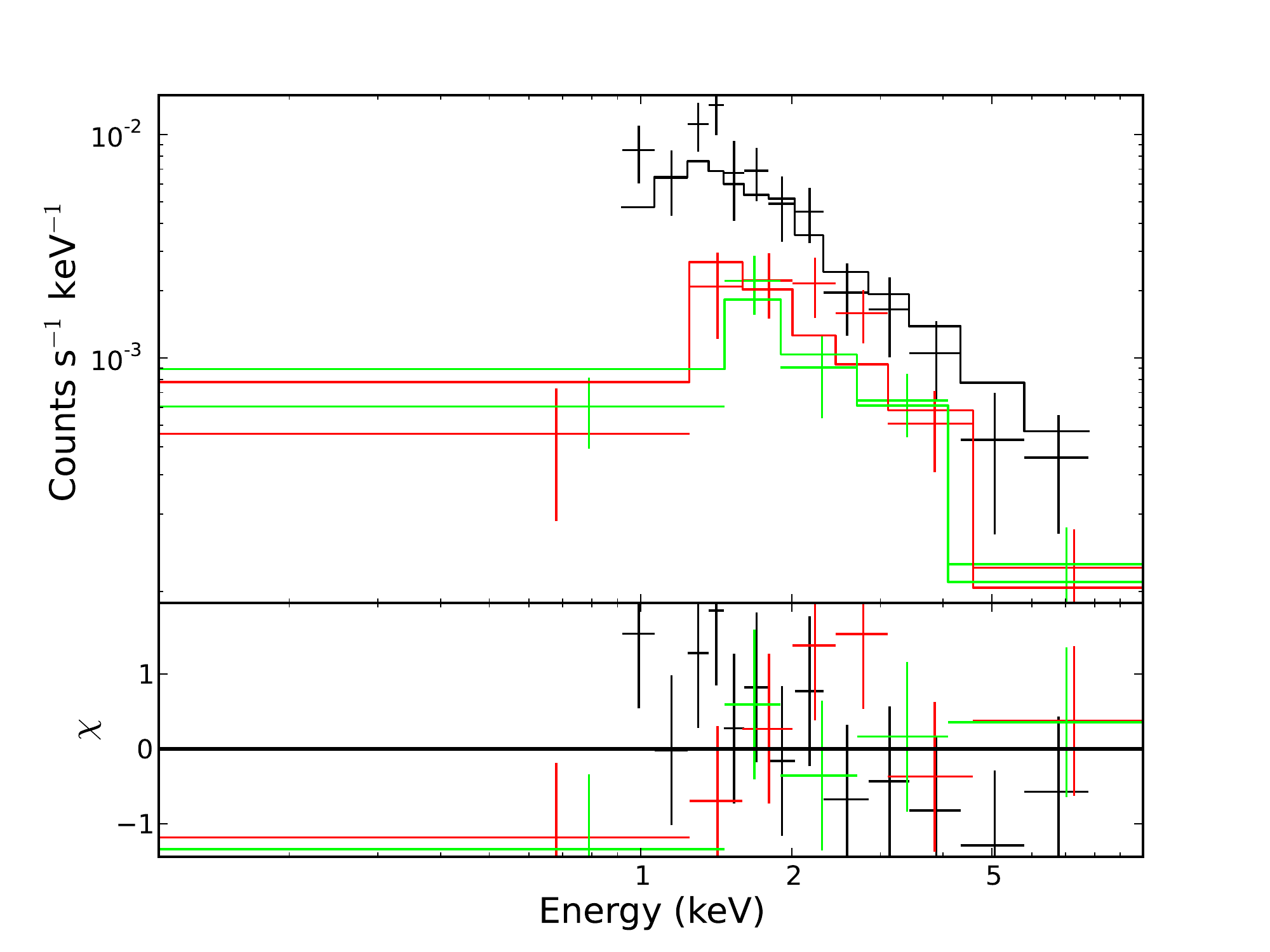}
\caption{
EPIC spectra of source N1 (black: pn; red: MOS1; green: MOS2).
The best-fit $2T$ model with the abundance of 1.0 (Table\,\ref{tbl: xmodel-2T-PQ}) is imposed.
\label{fig: sed-N}}
\end{center}
\end{figure}

Sources Q1--Q15 are relatively dim in the 0.5--12\,keV range.
Most have similar HR3 values around 0 as shown in Table\,\ref{tbl: xsrcs}, suggesting they have similar spectral shapes.
Given this and their proximities\,(see Figure \ref{fig: sed-Q}), 
we combine the 15 sources to study their general energy spectrum (hereafter Q$_{\rm sum}$).
First we fit the spectrum with an absorbed power-law function of {\sf wabs*pow} (PL model).
The resulting parameters are tabulated in Table\,\ref{tbl: xmodel-PL}.
We also fit the spectrum with $1T$ and $2T$ models (see Table\,\ref{tbl: xmodel-1T} and \ref{tbl: xmodel-2T-PQ}).
In the fit, we fix the abundance and the ratio of the normalization in the same manner as Source P1.
All models can explain the observed events.
The obtained $\NH$ is different between the $1T$ and $2T$ models,
because the temperature of the $1T$ model is very high ($\sim 6$\,keV) while the low-temperature component is dominant in the $2T$ model.

Here we evaluate a range of $\NH$ values for the $2T$ model with a solar abundance.
We fix the value of $\NH$ and then fit the $2T$ model; we re-fit for different values of $\NH$ to find the value where the resulting null hypothesis probability is 5\%.
This enables us to evaluate a range of $\NH$ where the null hypothesis probability is $>5$\%.
The resulting range of $\NH$ is $(1.0\mbox{--}1.4) \times 10^{22}\,{\rm cm}^{-2}$ and $(0.5\mbox{--}1.8) \times 10^{22}\,{\rm cm}^{-2}$ for P1 and Q$_{\rm sum}$ respectively.
Note that $\NH$ of Q$_{\rm sum}$ can be less if $kT1$ is allowed to be very low ($<0.1$\,keV).
When $\NH$ is set at $\gtrsim1.0 \times 10^{22}\,{\rm cm}^{-2} $, the resulting $kT1$ and $kT2$ are $\sim 0.3$\,keV and 2--4\,keV respectively, 
which are similar to those in Table\,\ref{tbl: xmodel-2T-PQ}.

\begin{deluxetable*}{lccccccc}
\tabletypesize{\scriptsize}
\tablecaption{Spectral Parameters of the $1T$ Models \label{tbl: xmodel-1T}}
\tablewidth{0pt}
\tablehead{
\colhead{ID} & \colhead{$N_{\rm H}$}  & \colhead{Abundance} & \colhead{$kT$} & \colhead{norm}  & \colhead{$f^{\rm obs}_{0.5\mbox{-}8\,{\rm keV}}$} & \colhead{$f^{\rm unabs}_{2\mbox{-}8\,{\rm keV}}$} & \colhead{$\chi_{\nu}^2$ (d.o.f.)}\\
\colhead{}     & \colhead{($10^{22}$\,cm$^{-2}$)}    & \colhead{(solar)} & \colhead{(keV)}  & \colhead{($10^{-4}$\,cm$^{-5}$)} &  \colhead{($10^{-13}$\,erg\,cm$^{-2}$\,s$^{-1}$)} &  \colhead{($10^{-13}$\,erg\,cm$^{-2}$\,s$^{-1}$)}  
}
\startdata
P1                       & $0.89 \pm 0.09$ & [0.2] & $0.89 \pm 0.08$ & $4.2 \pm 0.8$ & $0.72 \pm 0.03$ & $0.302 \pm 0.014$ & 1.40 (27) \\
                           & $1.18 \pm 0.06$ & [1.0] & $0.81 \pm 0.08$ & $2.3 \pm 0.4$ & $0.71 \pm 0.03$ & $0.300 \pm 0.014$ & 1.29 (27) \\
Q$_{\rm sum}$   & $0.40 \pm 0.06$ & [0.2] & $5.9 \pm 1.4$ & $ 3.7 \pm 0.3$ & $3.93 \pm 0.15$ & $3.11 \pm 0.12$ & 0.94 (83) \\
                           & $0.36 \pm 0.05$ & [1.0] & $6.8 \pm 1.1$ & $ 3.0 \pm 0.2$ & $4.23 \pm 0.16$ & $3.41 \pm 0.13$ & 1.04 (83) \\
R$_{\rm sum}$   & $9.4 \pm 1.6$ & [0.2] & [10.0] & $2.8 \pm 0.4$  & $1.51 \pm 0.12$  & $2.8 \pm 0.2$ & 1.38 (16) \\
                           & $8.5 \pm 1.4$ & [1.0] & [10.0] & $2.3 \pm 0.3$  & $1.62 \pm 0.12$ &  $2.8 \pm 0.2$ & 1.03 (16) \\
\enddata
\tablecomments{The parameters bracketed with [ ] are fixed in the fit. $f^{\rm obs}_{0.5\mbox{-}8\,{\rm keV}}$ and $f^{\rm unabs}_{2\mbox{-}8\,{\rm keV}}$ represent the observed and the absorption-corrected fluxes in the energy bands 0.5--8\,keV and 2--8\,keV respectively.}
\end{deluxetable*}

\begin{deluxetable*}{lccccccccc}
\tabletypesize{\scriptsize}
\tablecaption{Spectral Parameters of the $2T$ Models \label{tbl: xmodel-2T-PQ}}
\tablewidth{0pt}
\tablehead{
\colhead{ID} & \colhead{$N_{\rm H}$}  & \colhead{Abundance} & \colhead{$kT_1$} & \colhead{norm$_1$} & \colhead{$kT_2$} & \colhead{$f^{\rm obs}_{0.5\mbox{-}8\,{\rm keV}}$} & \colhead{$f^{\rm unabs}_{2\mbox{-}8\,{\rm keV}}$} & \colhead{$\chi_{\nu}^2$ (d.o.f.)}\\
\colhead{}     & \colhead{($10^{22}$\,cm$^{-2}$)}    & (solar) & \colhead{(keV)} & \colhead{($10^{-3}$\,cm$^{-5}$)} & \colhead{(keV)} & \colhead{($10^{-13}$\,erg\,cm$^{-2}$\,s$^{-1}$)} &  \colhead{($10^{-13}$\,erg\,cm$^{-2}$\,s$^{-1}$)}  
}
\startdata
P1       &  $0.98 \pm 0.07$ & [0.2] & $0.43 \pm 0.18$ & $0.8 \pm 0.4$ & $1.5 \pm 0.4$ & $0.77 \pm 0.04$ & $0.37 \pm 0.02$ & 1.32 (26)\\
            &  $1.15 \pm 0.06$ & [1.0] & $0.49 \pm 0.12$ & $0.31 \pm 0.11$ & $2.3 \pm 0.8$ & $0.85 \pm 0.04$ & $0.45 \pm 0.02$ & 1.13 (26)\\
Q$_{\rm sum}$       &  $1.11 \pm 0.08$ & [0.2] & $0.39 \pm 0.07$ & $2.0 \pm 0.3$ & $6 \pm 2$ & $ 3.98 \pm 0.15$ & $3.37 \pm 0.13$ & 0.95 (82)\\
                               &  $1.38 \pm 0.08$ & [1.0] & $0.29 \pm 0.03$ & $1.9 \pm 0.2$ & $4.0 \pm 0.7$ & $3.94 \pm 0.15$ & $3.54 \pm 0.14$ & 1.08 (82)\\
M$_{\rm sum}$      &  $2.4 \pm 0.3$ & [1.0] &           [0.29]             & $0.85 \pm 0.12$ &  [4.0] &$1.39 \pm 0.14$ & $1.57 \pm 0.15$ & 0.76 (5) \\
N1                    &  $1.52 \pm 0.11$ & [1.0] &        [0.29]            & $0.67 \pm 0.07$  & [4.0] & $1.33 \pm 0.09$ & $1.24 \pm 0.09$ & 0.91 (23) \\
\enddata
\tablecomments{norm$_2$ is set at $0.2 \times {\rm norm}_1$ in the fit. The other notes are the same as in Table\,\ref{tbl: xmodel-1T}.}
\end{deluxetable*}

\begin{deluxetable*}{lcccccc}
\tabletypesize{\scriptsize}
\tablecaption{Spectral Parameters of the PL Models \label{tbl: xmodel-PL}}
\tablewidth{0pt}
\tablehead{
\colhead{ID} & \colhead{$N_{\rm H}$}  & \colhead{$\Gamma$} & \colhead{norm} & \colhead{$f^{\rm obs}_{0.5\mbox{-}8\,{\rm keV}}$} & \colhead{$f^{\rm unabs}_{2\mbox{-}8\,{\rm keV}}$} & \colhead{$\chi_{\nu}^2$ (d.o.f.)}\\
\colhead{}     & \colhead{($10^{22}$\,cm$^{-2}$)}    & \colhead{} & \colhead{} & \colhead{($10^{-13}$\,erg\,cm$^{-2}$\,s$^{-1}$)} &  \colhead{($10^{-13}$\,erg\,cm$^{-2}$\,s$^{-1}$)}  
}
\startdata
Q$_{\rm sum}$   &  $0.51 \pm 0.07$ & $1.90 \pm 0.13$ & $(1.2 \pm 0.2) \times 10^{-4}$ & $3.94 \pm 0.15$ & $3.14 \pm 0.12$ & 0.92 (83) \\
R$_{\rm sum}$   &  $8.0 \pm 1.4$ & [1.5] & $(5.7 \pm 0.7) \times 10^{-5}$ & $1.52 \pm 0.12$ & $2.6 \pm 0.2$ & 1.33 (16) \\
\enddata
\tablecomments{The notes are the same as in Table\,\ref{tbl: xmodel-1T}.}
\end{deluxetable*}

Sources R1--R4 are also relatively dim but their spectral shapes must be different from Sources Q1--Q15
because of their high values of HR3 ($\sim0.9$; see Table\,\ref{tbl: xsrcs}).
Therefore we separately combine their events to study their general spectral shape (hereafter R$_{\rm sum}$).
As expected, the resulting spectrum is heavily obscured in the low-energy band ($\lesssim 2$\,keV; see Figure\,\ref{fig: sed-R}).
The spectrum is fitted with the $1T$ model with the temperature fixed at 10\,keV and 
the PL model with the photon index fixed at 1.5, a typical value for non-thermal extragalactic sources.
We fix these parameters at reasonable values due to low statistics of the data.
In addition we do not apply the $2T$ model because low-energy photons are too obscured to determine the low-temperature component.
The resulting parameters are tabulated in Table\,\ref{tbl: xmodel-1T} and \ref{tbl: xmodel-PL}.
Both models can explain the data with high column densities $N_{\rm H}$ of $(8\mbox{--}9) \times 10^{22}$\,cm$^{-2}$.
Note that we do not analyze sources S1--S4, which have soft spectral shapes (HR3 $\sim -0.8$), because of the limited statistics.

We also analyze sources in group (ii). Here we combine events of M1 and M2 because of their proximity and low statistics (hereafter M$_{\rm sum}$).
Since we cannot analyze their detailed spectra because of low statistics (see Figures\,\ref{fig: sed-M} and \ref{fig: sed-N}), 
we apply the $2T$ model fixed at the best-fitting spectral shape of $Q_{\rm sum}$.
Only $\NH$ and normalizations are set free in the fit. The resulting parameters are displayed in Table\,\ref{tbl: xmodel-2T-PQ}.
Note that the results are only representative since higher statistics are needed to determine the X-ray properties in detail.




\section{Discussion}
\label{sec: discussion}

\subsection{Counterparts to the Gamma-ray Emissions}
\label{sec: origin}

In our $\gamma$-ray analysis of Section\,\ref{sec: ana}, the G25 region is divided into two extended regions (regions G25A and G25B) and one point-like source (G25C).
For each region/source, we discuss what kind of astrophysical objects are responsible for the observed $\gamma$ rays based on the results of our analysis. 

\subsubsection{Region G25A}
\label{sec: d_g25a}

G25A is an elongated  $\gamma$-ray emitting region ($1 \fdg 44 \times 0 \fdg 56$; see Table\,\ref{tbl: sp-sfc10}) on the Galactic plane.
In Section\,\ref{sec: g25-sed}, we divided the region into three sections (G25A1, A2, and A3) and analyzed them.
Given their similar spectral shapes and proximity, they probably originate from the same celestial object (see Table\,\ref{tbl: sp-sfc10} and Figure\,\ref{fig: cmap_cls-sfc10}).
The observed hard spectral shape ($\Gamma = 2.14$) is different from the Galactic diffuse emission (see Figure\,\ref{fig: sed_x1}).
We infer that the $\gamma$ rays of G25A originate not from the Galactic diffuse emission but from a discrete source.

\begin{figure}[h] 
\begin{center}
\includegraphics[width=4in]{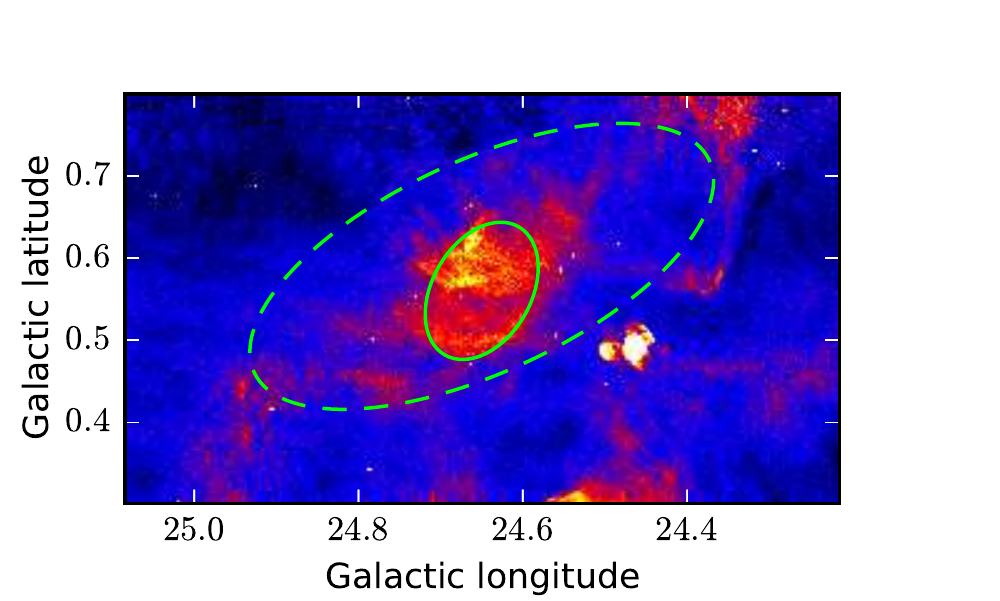}
\caption{\small
The 20\,cm data for the PWN candidate G24.7+0.6 (see text). 
The dashed ellipse indicates the size of G24.7+0.6 reported by \cite{1984A&A...133L...4R}, while the solid one is the size estimated from the 20\,cm data.
The ellipses are the same as those shown in purple in Figure\,\ref{fig: cmap_cls-sfc10}.
The bright western sources are \HII\ regions.
\label{fig: pwn}}
\end{center}
\end{figure}

Since PWNe and SNRs are the most prominent candidates for extended $\gamma$-ray sources, we first investigate the possibility that G25A is composed of such sources.
To date, no PSR with spin-down luminosity $>1\times10^{34}$\,erg\,s$^{-1}$ or PWN is known in the G25A region.
Although there is no established PWN, the G25A region contains a candidate SNR G24.7+0.6 (right panel in Figure\,\ref{fig: cmap_cls-sfc10}).
This source was found in the radio band by \cite{1984A&A...133L...4R}, 
who reported extended emission with a size of $30' \times 15'$ and a polarized filled central core with a hard spectrum (spectral index of $\alpha \sim 0.2$).
They concluded that the radio source is a candidate composite SNR whose PWN is powered by an undetected PSR.
Recent observations at 20\,cm with better angular resolution\,\citep{MAGPIS} confirm that there is a central source of $\sim 0 \fdg 1$ radius.
On the other hand, the brightness of the surrounding emission is no brighter than the background emission (see Figure\,\ref{fig: pwn}).
This suggests that the candidate PWN (G24.7+0.6) is the central source. 
The previously-reported surrounding emission is probably irrelevant to the candidate PWN; it may be diffuse emission from the Galactic plane.
No X-ray or TeV detection has been reported from this region.
If G24.7+0.6 is responsible for the $\gamma$ rays, the $\gamma$-ray spatial size should be much larger than that in the radio band 
and its spatial shape should be very asymmetric. 
This suggests that the $\gamma$-ray emission of G25 is unlikely to come from G24.7+0.6.

Next we consider the possibility that G25 is an unknown PWN.
In this case, the observed $\gamma$ rays are expected to come from the inverse Compton (IC) emission by relativistic electrons.
As calculated in Section\,\ref{sec: g25-sed}, the $\gamma$-ray spectrum is represented by a power-law function with a photon index of 2.14 over three decades (0.2--500\,GeV).
To reproduce such a spectral shape, a distribution of relativistic electrons is required to be $dN/dp \propto p^{-s}$ where $p$ is momentum of the electrons and $s$ is about 3.0. The index $s$ is much softer than a typical value of 2.0 indicated by multi-wavelength observations of other PWNe.
Given the absence of  PWNe or energetic PSRs in the regions, the $\gamma$ rays of G25A are unlikely to come from a PWN. 

\begin{figure*}[hp!]
\begin{center}
\includegraphics[width=7.5in]{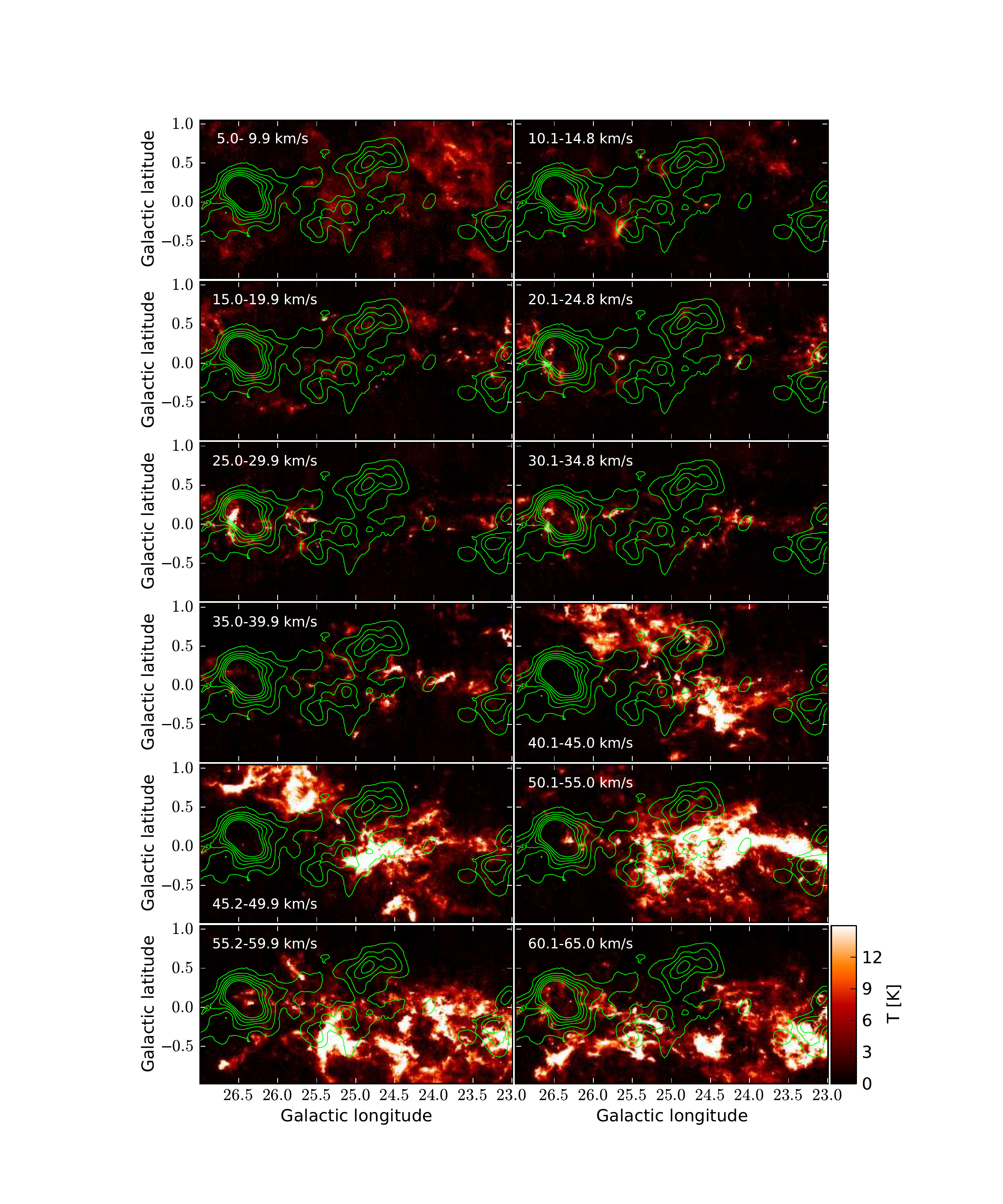}
\caption{\small
Integrated $^{13}{\rm CO}\,J=1\mbox{--}0$ emission every 5\,km\,s$^{-1}$ from 5 -- 65\,km\,s$^{-1}$. 
The green contours show the residual map of the LAT data above 3\,GeV in Figure\,\ref{fig: cmap_wide}.
\label{fig: co-all1}}
\end{center}

\end{figure*}

\begin{figure*}[hp!] 
\begin{center}
\includegraphics[width=7.5in]{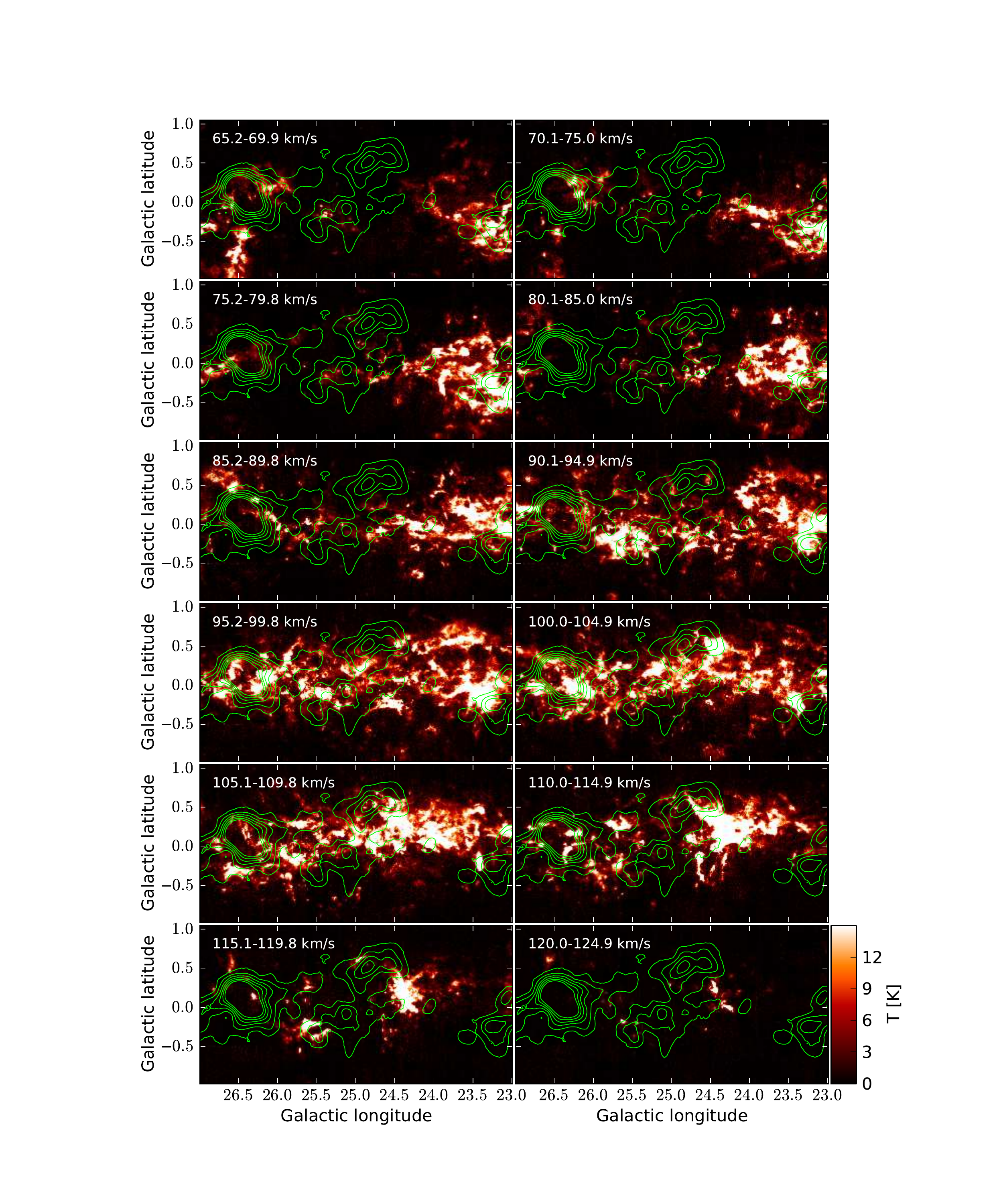}
\caption{\small
The same as Figure\,\ref{fig: co-all1} except for velocity intervals every 5\,km\,s$^{-1}$ from 65 -- 125\,km\,s$^{-1}$. 
\label{fig: co-all2}}
\end{center}
\end{figure*}

We do not find any evidence from multi-wavelength data that SNR shells exist in this region.
Molecular clouds are also candidates for discrete extended $\gamma$-ray sources.
If relativistic particles escape from nearby acceleration sites such as SNRs, 
enhanced $\gamma$ rays are expected from the molecular clouds illuminated by the accelerated particles\,\citep[e.g.,][]{Escape96,Escape08}.
To check this possibility, we examine whether molecular clouds exist toward the G25A region.
We use observations of the $^{13}$CO\,$J = 1\mbox{--}0$ line, which traces molecular clouds, from the Galactic Ring Survey (GRS; \cite{GRS}).
Figures\,\ref{fig: co-all1} and \ref{fig: co-all2} show the maps for G25, where we integrate them in 5\,km\,s$^{-1}$ steps from 5 to 125\,km\,s$^{-1}$.
The figure indicates no molecular cloud covering the entire region of G25A at any distance.
It is unlikely that the $\gamma$ rays of G25A come from molecular clouds.


There is another candidate for Galactic extended $\gamma$-ray sources: SFRs.
LAT observations revealed the extended $\gamma$-ray emissions from the massive SFR Cygnus cocoon\,(\CC).
In Section\,\ref{sec: sfr}, we will discuss in detail the possibility of an SFR origin for the observed $\gamma$ rays.


\subsubsection{Region G25B}
\label{sec: d_g25b}
 
G25B is also an elongated $\gamma$-ray region ($1 \fdg 34 \times 0 \fdg 54 $; see Table\,\ref{tbl: sp-sfc10}) on the Galactic plane.
We divided the region into three sections (G25B1, B2, and B3) and analyzed them.
As for G25A, the $\gamma$ rays of G25B1, B2, and B3 can be ascribed to discrete sources.
In addition, we found that G25B1 clearly displays a hard spectral shape ($\Gamma = 1.53 \pm 0.15$), while the regions G25B2 and B3 have softer spectra ($\Gamma \simeq 2.1 \pm 0.2$).
The consistency of spectral shapes and the proximity of G25B2 and B3 indicate that 
they probably originate from the same celestial object (see Table\,\ref{tbl: sp-sfc10} and Figure\,\ref{fig: cmap_cls-sfc10}).
In this section, we first consider sources associated with G25B1 and then G25B2 and B3 (G25B$'$; see Section\,\ref{sec: g25-sed}).

G25B1 is spatially coincident with $\HJP$ (see Figure\,\ref{fig: cmap_cls-sfc10}).
The X-ray observation found PSR\,J1838-0655 embedded in a PWN with an extent of $1.3'$ at the edge of $\HJP$\,\citep{2008ApJ.681.515G}.
The H.E.S.S. source is a TeV PWN powered by this PSR with a spin-down luminosity of $5.5 \times 10^{36}$\,erg\,s$^{-1}$.
The SED measured by the LAT smoothly connects to that measured by H.E.S.S., which suggests that photons of the LAT and H.E.S.S. data have the same origin (see Figures\,\ref{fig: sed_x2} and \ref{fig: sed_TeV}).
G25B1 has a photon index of about 1.5 (see Section\,\ref{sec: g25-sed}), which is consistent with the expected relativistic electron distribution of PWNe: $dN/dp \propto p^{-2}$.
G25B1 is most likely a PWN powered by PSR\,J1838$-$0655.
The $\gamma$-ray luminosity of G25B1 is calculated to be $(3.1 \pm 0.8) \times 10^{35}$\,$(d/6.6\,{\rm kpc})^2$\,erg\,s$^{-1}$ in 3--500\,GeV.
Here we adopt 6.6\,kpc for the distance to $\HJP$ following \cite{2008ApJ.681.515G} who assume that the source is associated with an adjacent massive star cluster.
The ratio of the $\gamma$-ray luminosity to a spin-down luminosity of the PSR is about 3\%. 
Combined with the TeV luminosity, the ratio of the total $\gamma$-ray luminosity to the spin-down energy is about 6\%.

As stated in Section\,\ref{sec: g25-sed}, the $\gamma$-ray size of the PWN probably ranges from $0 \fdg 2$ to $0 \fdg 4$,
although it is difficult to determine a precise size in the GeV band because of contamination from the G25B2 and B3 regions.
In addition, our analysis indicates that the spatial size in the TeV range might be larger than the previously-reported value ($0 \fdg 24 \times 0 \fdg 10$ in size),
as suggested by a preliminary report based on more accumulated H.E.S.S. data\,\citep{HESS1837.new}.
To estimate relativistic particles responsible for the $\gamma$ rays, we consider a leptonic model where the $\gamma$ rays are from IC scattering.
We adopt the simple assumption that the spectral distribution of the electrons has a cut-off power-law function: $dN/dp \propto p^{-s} \exp{(-p/p_{\rm cut})}$.
The target photon fields for IC scattering are the CMB, infrared, and starlight photons adopted from the \textsc{GALPROP} code\,\citep{Porter2008}.
The temperature and energy density of the CMB, infrared, and starlight photons are $2.3 \times 10^{-4}$\,eV and 0.26\,eV\,cm$^{-3}$, $3.6 \times 10^{-3}$\,eV and 1.2\,eV\,cm$^{-3}$, and 0.30\,eV and 3.0\,eV\,cm$^{-3}$ respectively.
We vary the normalization of the electron distribution, index $s$, and a cut-off momentum $p_{\rm cut}$ to reproduce the observed $\gamma$-ray data.
As shown in Figure\,\ref{fig: sed_TeV}, this simple model can explain the observed $\gamma$ rays.
Given the above discussion, here we use the $\gamma$-ray morphology measured for $\HJP$ in the LAT band.
Note that the choice of the LAT morphology does not significantly affect the resulting parameters.
The obtained total energy of the electrons ($>3$\,GeV) is $2.6 \times 10^{48}\,(d/6.6\,{\rm kpc})^2$\,erg with the index $s = 1.9$ and $p_{\rm cut} = 13$\,TeV\,$c^{-1}$.

Next we consider the source of the $\gamma$ rays from the G25B$'$ region.
The spectral shape of the region is similar to that of G25A. 
In addition, there is no PWN, SNR, or PSR with spin-down luminosity $>1\times10^{34}$\,erg\,s$^{-1}$ in the region. 
As for G25A, those celestial objects are unlikely to explain the $\gamma$ rays from the region.
On the other hand, Figure\,\ref{fig: co-all1} shows that molecular clouds at $v=45\mbox{--}65$\,km\,s$^{-1}$ overlap the regions.
However the molecular clouds are spatially much larger than the G25B$'$ region.
Although we cannot exclude the possibility that a part of the molecular clouds is illuminated by relativistic protons,
there is no strong support that the $\gamma$ rays come from the molecular clouds.

In summary, G25B1 associated with $\HJP$ is most likely a PWN.
On the other hand, G25B$'$ has no clear association like G25A.
Interestingly the region has similar $\gamma$-ray features as G25A (see Section\,\ref{sec: sfr} for details).
In that section, we will discuss the possibility that the $\gamma$ rays of the regions originate in an SFR.

\subsubsection{Source G25C}
\label{sec: d_g25c}

The spectrum of G25C ($\Gamma \sim 2.1$) is distinctly harder than the Galactic diffuse emission. 
This source is detected at high energies $\gtrsim 3$\,GeV and in this energy range its SED is not sensitive to uncertainties of the Galactic diffuse model (see Figure\,\ref{fig: sed_x3}).
Therefore we conclude that G25C is a discrete $\gamma$-ray object despite its relatively low statistics ($\sim 4\,\sigma$ detection) on the Galactic plane.
Although we classify it as a point source, G25C has $\TSe$ of 7, which might indicate spatial extension.
Actually in the residual map (Figure\,\ref{fig: cmap_cls-sfc10}), G25C appears as a spatially-extended structure rather than a clear point-like shape.
However much greater statistics would be needed to resolve a detailed spatial distribution for this source.

No identified source is associated with G25C. 
The hard photon index of 2.1 disfavors a PSR origin, since $\gamma$-ray PSRs usually have an energy cutoff of 1--10\,GeV\,\citep[see e.g.,][]{2013ApJS..208...17A}.
Interestingly the hard spectral shape of G25C is almost the same as for G25B$'$.
Given its spatial proximity and possible spatial extension, G25C may originate from the same celestial object that powers extended source G25B$'$.
In addition, possible TeV emission may be seen to the south of the G25B$'$ region in the H.E.S.S. image\,(Figure\,\ref{fig: cmap_cls-sfc10}).

\subsection{Counterparts to the X-Ray Emissions}
\label{sec: d-xray}

In Section\,\ref{sec: xray}, we reported the X-ray emissions from multiple point sources in the G25.18+0.26 region.
Here we discuss counterparts to these X-ray emissions.

Before identifying the sources, we compare our results with past X-ray studies in this direction.
{\em ASCA} GIS found an unidentified X-ray source AX\,J1836.3$-$0647 at this direction\,\citep{2001ApJS..134...77S}.
The observed flux of the source is $8.2 \times 10^{-13}$\,erg\,cm$^{-2}$\,s$^{-1}$ in a range of 0.7--10\,keV, which is 76\% of the total flux of sources in group (i) and (ii) obtained in our analysis. 
Both fluxes are similar and the difference probably comes from the fact that the {\em ASCA} source consists of multiple X-ray sources that extend over $\sim0 \fdg 2$ in diameter; it is difficult to determine precise spectral parameters of such a faint extended source with {\em ASCA}.
{\em Swift} XRT also observed this source for 8\,ks on 2007 March, 
but did not detect it; the upper limit of the flux is $2 \times 10^{-13}$\,erg\,cm$^{-2}$\,s$^{-1}$ in 0.3--10\,keV\,\citep{Degenaar:2012ke}.
The authors conclude that AX\,J1836.3$-$0647 is a strongly variable or transient source because the upper limit is much lower than the flux reported by {\em ASCA}.
However they derive the upper limit on the assumption that the {\em ASCA} source is a point source.
We consider that this incorrect assumption is the reason for the very low XRT upper limit on the source.
As stated above, the {\em ASCA} flux is almost the same as the total flux in our results, based on the \XMM\ data that was taken only six months after the {\em Swift} observation. This suggests that the total flux of the sources is almost stable.

~

~

\begin{figure*}[t] 
\begin{center}
\includegraphics[width=7.5in]{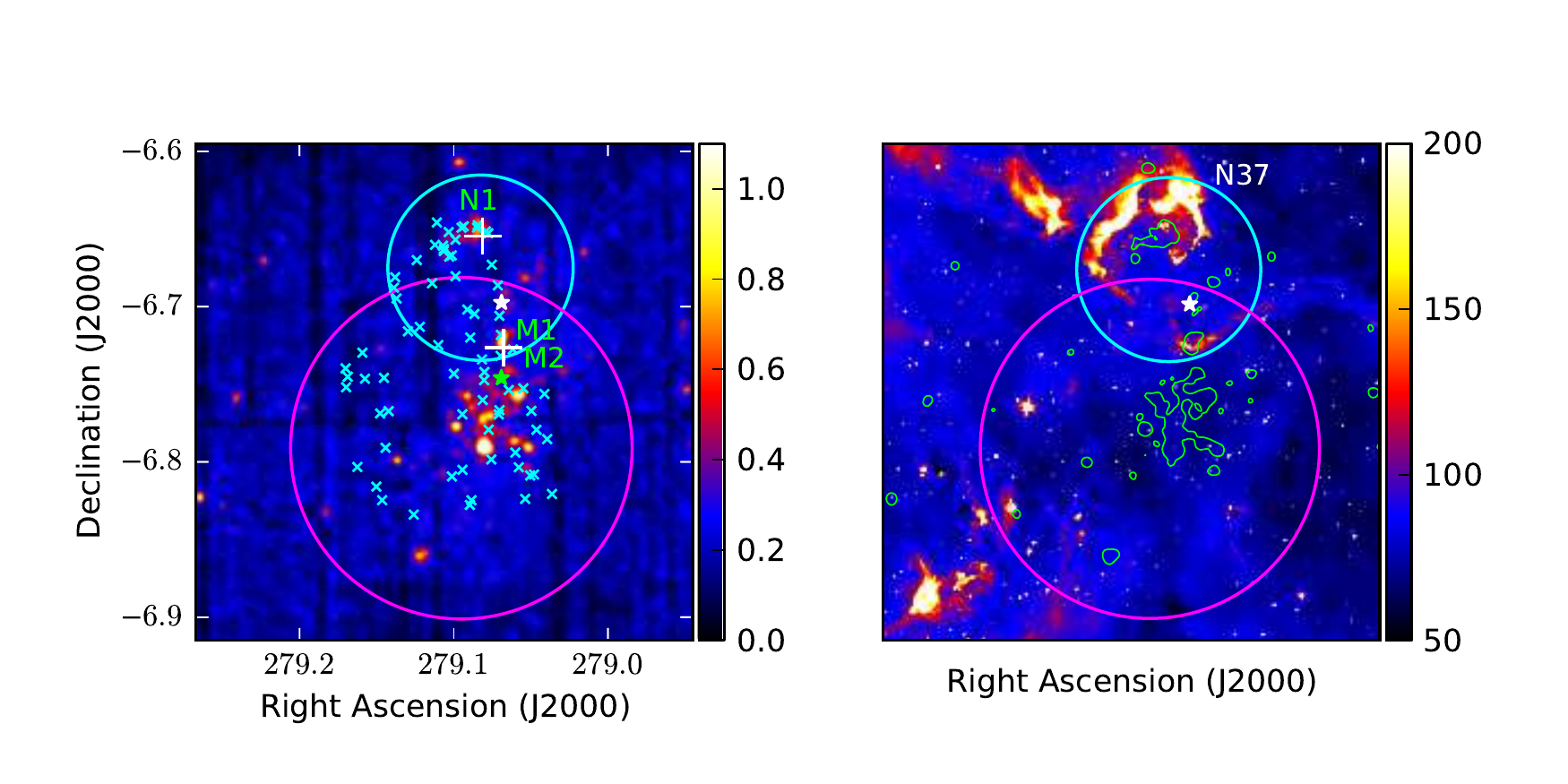}
\caption{
Left:
combined MOS1+MOS2+pn count map of the AX\,J1836.3$-$0647 field in the 0.2--12\,keV band.
The magenta and the cyan circles represent the regions of group (i) and (ii) respectively (see Section\,\ref{sec: xray}).
The two white crosses display positions of \HII\ regions G25.294+0.307 and G25.220+0.289.
The cyan Xs display all the likely massive stars of Alicante\,6 except for the two O-type (O7\,II and O7\,V) stars, which are designated by the white and the green stars respectively.
Right:
{\em Spitzer} IRAC 8-$\mu$m image of the same field as that of the left panel in units of MJy\,sr$^{-1}$.
The green contours represent the X-ray count map of the left panel.
The white star and the magenta and the cyan circles are the same as those in the left panel.
\label{fig: imgs-X-IR}}
\end{center}
\end{figure*}

\subsubsection{Sources in Group (ii)}
\label{sec: group2}

We separately treat the sources in group (ii) from those in group (i), 
because the group (ii) sources M1, M2, and N1 are spatially coincident with known \HII\ regions.
Figure\,\ref{fig: imgs-X-IR}\,(left) shows that source N1 is coincident with \HII\ region G25.294+0.307, which has $v_r$ of 39.6\,km\,s$^{-1}$\,\citep{Lockman89}.
The velocity corresponds to a kinematic distance of 2.8/12.6\,kpc.
The region appears surrounded by an IR bubble called N37\,(\cite{2006ApJ...649..759C}; see also the right panel in Figure\,\ref{fig: imgs-X-IR}).
This is generally understood to be a phenomenon due to winds of putative massive star(s) sweeping up the surrounding dust to create a bubble, ionize the region, and illuminate the bubble by heating the dust.
The \HII\ region and the bubble are probably associated with the young cluster Alicante\,6 reported by \cite{Marco:2011dpa}.
They find a significant population of massive stars located at a distance of $3.0^{+0.6}_{-0.4}$\,kpc by using optical and IR data.
The estimated distance is consistent with the closer kinematic distance of the \HII\ region.
They find that most massive stars of the cluster are of later types than O stars and only two stars are O-type (O7\,II and O7\,V) stars (see Figure\,\ref{fig: imgs-X-IR}).
They also find that a number of B0-1\,V stars concentrate in the cavity of the bubble.
Such spatial coincidence strongly suggests that the stars are physically associated with N37.
The authors also state that the O7\,II star located near N37 is a possible ionizing source to create the bubble.
Given the facts listed here, the \HII\ region G25.294+0.307 and the probable associated bubble N37 are most likely to be associated with the young cluster Alicante\,6.

Figure\,\ref{fig: imgs-X-IR} shows that source N1 appears to be spatially confined within the bubble N37.
Given the spatial association, source N1 is likely to be associated with the bubble.
If this is the case, the observed X-ray luminosity is calculated to be $1.4 \times 10^{32} (d/3.0\,{\rm kpc})^2$\,erg\,s$^{-1}$ in 0.5--8\,keV.
The X-ray image may suggest that the source N1 is extended and consists of unresolved multiple sources,
given that the PSF of the EPIC has a full width at half maximum of $\sim 6''$.\footnote{http://xmm.esac.esa.int/external/xmm\_user\_support/documentation/ uhb\_2.1/node14.html}

Sources M1 and M2 are coincident with the \HII\ region G25.220+0.289 recently found by \cite{Anderson:2011cu} as shown in Figure\,\ref{fig: imgs-X-IR}\,(left).
The \HII\ region has $v_r$ of 42.4\,km\,s$^{-1}$, which is similar to that of G25.294+0.307.
This indicates that the sources M1 and M2 are also associated with Alicante\,6.
Interestingly the O7\,II star of Alicante\,6 is located between the \HII\ regions G25.294+0.307 and G25.220+0.289.
If the star ionizes G25.294+0.307 as suggested by \cite{Marco:2011dpa}, G25.220+0.289 may be also ionized by this star.
Actually the 8\,${\mu}$m IR map shows a filament-like structure coincident with G25.220+0.289, which may be swept up and excited by the O star (Figure\,\ref{fig: imgs-X-IR}).
Given the observational evidence, the \HII\ region G25.220+0.289 may be associated with Alicante\,6.
The observed total X-ray luminosity is calculated to be $1.4 \times 10^{32} (d/3.0\,{\rm kpc})^2$\,erg\,s$^{-1}$ in the 0.5--8\,keV band.
We note that the sources M1 and M2 have no counterpart massive stars in Alicante\,6.
This seems inconsistent with our expectation that the sources M1 and M2 are massive stars based on the derived high X-ray luminosity.
This might be because molecular clouds obscure the sources in the optical wavelength.
Actually Table\,\ref{tbl: xmodel-2T-PQ} shows that $\NH$ obtained with the $2T$ model is $\sim 2.4 \times 10^{22}\,{\rm cm}^{-2}$, which corresponds to the optical extinction $A_V \sim 11$\,mag,
when we apply the relation $\NH = 2.2 \times 10^{21}\,A_V\,{\rm mag}^{-1}\,{\rm cm}^{-2}$\,\citep{Ryter1996}.
More X-ray statistics would be needed to determine reliable values of the spectral parameters for the interpretation.

\subsubsection{Sources in Group (i)}
\label{sec: group1}

The sources in group (i) clearly concentrate around the center of the magenta circle in Figure\,\ref{fig: imgs-xray}.
As displayed in Table\,\ref{tbl: xmodel-2T-PQ}, 
the $N_{\rm H}$ of sources P1 and Q$_{\rm sum}$ have similar values of $(1.0\mbox{--}1.4) \times 10^{22}$\,cm$^{-2}$ for a typical spectral shape of OB-association members, when we adopt the $2T$ model (see Section\,\ref{sec: xray}).
This suggests that the sources are located at the same distance: they are associated in physical space.
In the $1T$ model, $\NH$ is almost the same as that of the $2T$ model for source P1 and the low-temperature component ($kT \lesssim 1\,{\rm keV}$) is dominant in both models.
The choice of the models does not affect the interpretation of P1.
On the other hand, $\NH$ toward Q$_{\rm sum}$ in the $1T$ model is $0.4 \times 10^{22}$\,cm$^{-2}$, 
which is significantly different from $(1.1\mbox{--}1.4) \times 10^{22}$\,cm$^{-2}$ for the $2T$ model.
In the $1T$ model, the obtained temperature of 6--7\,keV is too hot for usual stellar association members.
In addition, such a low $\NH$ means that sources Q1--Q15 are most likely to be nearby (within a few kpc).
We can roughly estimate $A_V$ to be $\sim 2$\,mag based on the $\NH$ using the relation $\NH = 2.2 \times 10^{21}\,A_V\,{\rm mag}^{-1}\,{\rm cm}^{-2}$\,\citep{Ryter1996}.
Given such a low extinction and the close distance, most members of the association G25.18+0.26 should be detectable in the optical band.
This is inconsistent with the fact that no clear optical counterpart has been found for G25.18+0.26.
Therefore the $1T$ model is unlikely to represent physical features of Q$_{\rm sum}$.
 The spectrum of Q$_{\rm sum}$ also can be explained by the PL model with $\NH$ of $0.51 \times 10^{22}$ and $\Gamma = 1.90$.
Typical non-thermal extragalactic sources are represented by a PL model with $\Gamma$ of 1.5--2.5.
However most of the sources Q1--Q15 must not be extragalactic sources 
because the derived $N_{\rm H}$ is lower than the total column density of \HI\ in this direction ($\sim 1.7 \times 10^{22}\,{\rm cm}^{-2}$).\footnote{http://heasarc.nasa.gov/cgi-bin/Tools/w3nh/w3nh.pl}
Such a concentration of non-thermal sources is unlikely in our Galaxy.
Therefore we conclude that the $2T$ model best represents the physical parameters of sources P1 and Q1--Q15 and that these sources consist of a stellar association/cluster in our Galaxy.
Hereafter we call this object G25.18+0.26.

The young cluster Alicante\,6 is also located in this direction (see Section\,\ref{sec: group2}). 
However the sources in group (i) are unlikely to be associated with this object for the following reasons.
First if they are really associated, the absorption-corrected X-ray luminosities of the individual sources are in the range $(1\mbox{--}6) \times 10^{32}\,(d/3\,{\rm kpc})^2$\,erg\,s$^{-1}$ in the 0.5--8\,keV band.
To calculate the X-ray fluxes, we adopt the $2T$ model with abundance 1.0 in Table\,\ref{tbl: xmodel-2T-PQ}.
The fluxes of sources Q1--Q15 are calculated by fitting the normalizations with the best-fitting $2T$ model for source\,Q$_{\rm sum}$ with the spectral shape fixed.
The obtained high luminosities mean that most sources in group (i) should be massive stars.
As shown in Figure\,\ref{fig: imgs-X-IR} (left), however, the massive stars of Alicante\,6 do not coincide with the X-ray sources, indicating that they are not associated.
In addition, the morphology of the IR image suggests that the most luminous star is likely to be located near the center of the cyan circle (the group (ii) region; see Section\,\ref{sec: group2}).
However the X-ray image strongly indicates that the most massive stars concentrate around the center of the magenta circle, 
if we assume that sources in the groups (i) and (ii) are located at the same distance.
The inconsistency also supports our conclusion that the OB association/cluster G25.18+0.26 is not associated with Alicante\,6.

Given that there is no optical counterpart to G25.18+26, the OB association is likely to be farther away than Alicante\,6 ($d = 3\,{\rm kpc}$). 
$\NH$ of sources P1 and Q$_{\rm sum}$ are $(1.0\mbox{--}1.4) \times 10^{22}\,{\rm cm}^{-2}$,
which is relatively high given that the total Galactic \HI\ column density at this direction is about $1.7 \times 10^{22}\,{\rm cm}^{-2}$.
This suggests that G25.18+0.26 is at a distant location, which is consistent with our expectation.
However we should note that the values of $\NH$ corresponds to $A_V = 5\mbox{--}6$\,mag
when we adopt a relation of $\NH = 2.2 \times 10^{21}\,A_V\,{\rm mag}^{-1}\,{\rm cm}^{-2}$\,\citep{Ryter1996}.
In this case the counterparts to the X-ray sources are expected to be visible in the optical band, 
which is inconsistent with the observational evidence.
This may be explained by an uncertainty of the relation between $\NH$ and $A_V$.
Actually \cite{Vuong03} indicates $\NH = 1.6 \times 10^{21}\,A_V\,{\rm mag}^{-1}\,{\rm cm}^{-2}$.
When we adopt this relation, the converted $A_V$ is $6\mbox{--}9$\,mag.
When we adopt the highest estimated value of $A_V$ of $\sim 10$\,mag,
the OB association is expected to be invisible in the optical band.
Another possible explanation is that the actual $\NH$ values are higher than the obtained ones.
Our analysis shows that $\NH$ could go up to $1.8 \times 10^{22}\,{\rm cm}^{-2}$ (Section\,\ref{sec: xray}).
In addition, 
the value of $\NH$ (Q$_{\rm sum}$) is calculated for combined spectra of Q1--Q15 because of the limited statistics.
Some of the integrated sources might be foreground ones, which would make the value of $\NH$ smaller than the real values of the OB association G25.18+0.26.
More observations in the X-ray and at other wavelengths will determine the precise distance of G25.18+0.26.

\cite{Rahman:2013di} claim another candidate OB association, SFC\,10, in this direction using near-IR observations, which are less affected by obscuration than the optical band.
They select the candidate association based on an observed excess number density of IR sources compared with that of the surrounding region.
The reddening of the selected stars indicates that the association is distant ($\gtrsim 6$\,kpc). 
The claimed association has a larger size than that of this work but the most dense part is almost spatially coincident with the X-ray association (see Figure\,\ref{fig: maps_CO}).
This may indicate that the IR and X-ray association have the same origin. 
More studies are needed to confirm this indication.

The spectrum of R$_{\rm sum}$ indicates $\NH$ of $(8\mbox{--}9) \times 10^{22}$\,cm$^{-2}$ independent of the choice of model (see Section\,\ref{sec: xray}).
The column density is much higher than those of sources P1 and Q$_{\rm sum}$ ($\NH = 1.0\mbox{--}1.4 \times 10^{22}\,{\rm cm}^{-2}$) or the total Galactic \HI\ column density in this direction ($\sim 1.7 \times 10^{22}$\,cm$^{-2}$).
This means that sources R1--R4 are embedded in dense gas and obscured with $A_V \sim 40$\,mag using the relation $\NH = 2.2 \times 10^{21}\,A_V\,{\rm mag}^{-1}\,{\rm cm}^{-2}$.
One potential explanation for this obscuration is that these sources are members of G25.18+0.26;
young associations/clusters usually have such sources obscured by their parental molecular clouds.
Another interpretation is that the sources are extragalactic. To determine their associations, we would need more data to precisely study individual sources.

\subsection{SFR Scenario}
\label{sec: sfr}

As discussed in Section\,\ref{sec: origin}, there are no clear associations with G25A, B$'$, or C.
Interestingly the G25A and G25B$'$ regions show similar characteristics: elongated morphologies with similar surface brightnesses and hard energy spectra. 
Given their proximity, one celestial object plausibly could be responsible for the $\gamma$-ray emission of both regions.
Although G25C may have similar characteristics, we need more statistics to study it in detail.
Here we discuss $\gamma$-ray emission of G25A and G25B$'$ (hereafter G25).
Note that the exclusion of G25C does not change our conclusion in the discussion.

We propose a scenario that the observed $\gamma$-ray emission is due to a massive SFR.
The first evidence to support the scenario is that the $\gamma$-ray properties of G25 are similar to those of the Cygnus cocoon, 
the only firm case of $\gamma$-ray detection from an SFR in our Galaxy\,(\CC):
both $\gamma$-ray sources are spatially extended;
their energy spectra are described as a power law with hard photon indices of 2.1--2.2 without any significant spectral curvature at least up to $\sim 500$\,GeV;
because of their large sizes ($2^\circ$--$4^\circ$), the LAT is able to obtain SEDs of five sub-regions for each source 
and finds a uniform spectral shape ($\Gamma \simeq 2.1\mbox{--}2.2$) in the regions.
These similarities suggest that the $\gamma$-ray production process in G25 might be similar to that in the Cygnus cocoon, i.e., a massive SFR.

In this section, we will provide three more lines of observational evidence to support our proposition.
First, if a massive SFR exists in this region, an accompanying bubble structure is expected.
Although there is no clear evidence that a massive SFR is associated with this region,
\cite{Rahman:2010in} recently claim a candidate massive SFR in this direction based on radio and IR observational data.
In Section\,\ref{sec: bubble}, we investigate the surrounding gas distributions to confirm the bubble structure that they report.
Second, if the bubble mentioned in Section\,\ref{sec: bubble} constitutes an SFR, a massive OB association/cluster is expected in this region.
Actually the X-ray observation in the present work reveals that the OB association/cluster G25.18+0.26 resides in the region (see Section\,\ref{sec: d-xray}). 
In Section\,\ref{sec: connection}, 
we provide observational evidence in support of the possibility that the bubble is an SFR created by the OB association G25.18+0.26.
Finally, if the observed $\gamma$ rays come from the SFR proposed in Section\,\ref{sec: connection}, 
the $\gamma$ rays are expected to be largely confined in the bubble (i.e., a cavity delineated by the claimed bubble).
This is a morphological property revealed by the LAT observation of the SFR Cygnus\,cocoon.
In Section\,\ref{sec: gassoc} we examine relations between the spatial structures of the $\gamma$ rays and the claimed bubble,
to check whether the morphology follows our expectation.
Note that the mechanism of the confinement will be discussed in Section\,\ref{sec: accG25}.

\subsubsection{Bubble Structure}
\label{sec: bubble}

Toward the direction $(l, b) \sim (24 \fdg 9, 0 \fdg 1)$, \cite{Rahman:2010in} find copious 8\,$\mu{\rm m}$ emission, 
which has a bubble structure associated with \HII\ regions with line-of-sight velocities $v_r$ of $\sim$100\,km\,s$^{-1}$.
Their interpretation is that the emission is associated with a star-forming complex created by hidden massive OB association(s).
They estimate the distance to the bubble as 6.1\,kpc based on the median velocity ($\sim 100$\,km\,s$^{-1}$) of the \HII\ regions.

\begin{figure}[ht] 
\begin{center}
\includegraphics[width=3.5in]{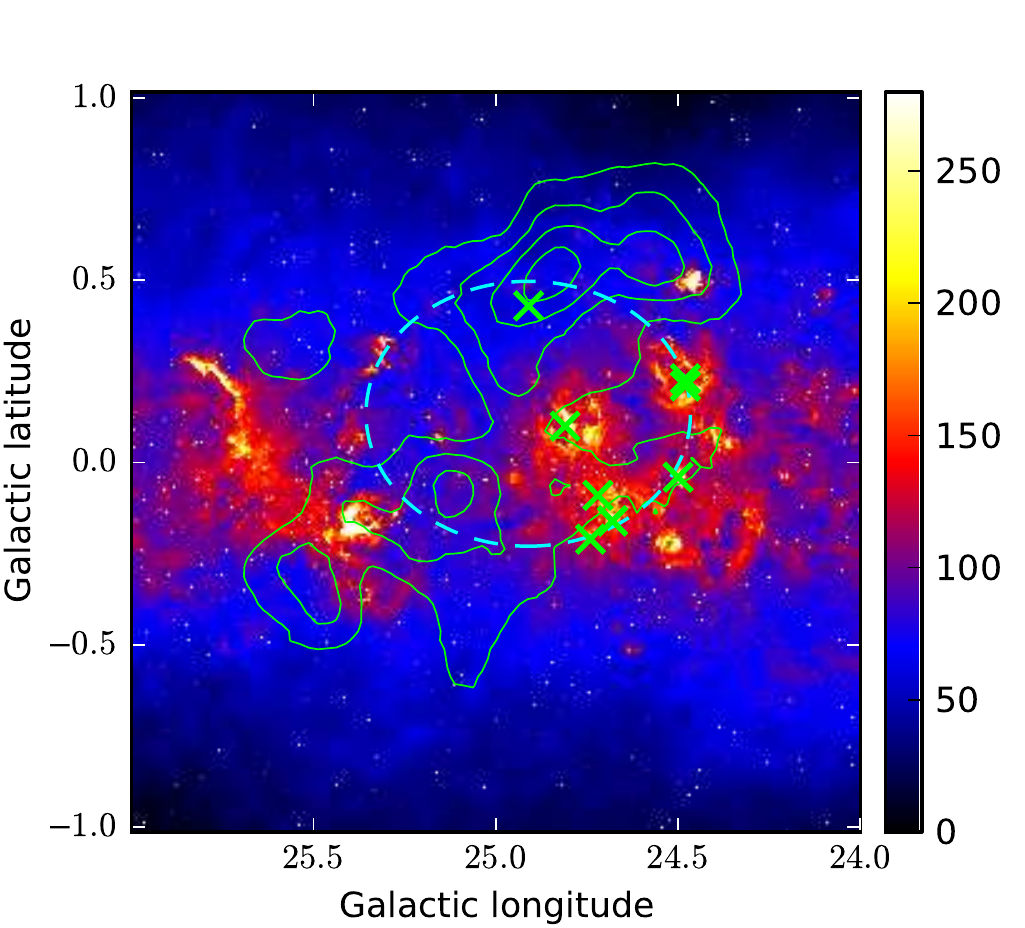}
\caption{\small
{\em Spitzer} IRAC 8-$\mu$m image around the G25 region in units of MJy\,sr$^{-1}$.
The cyan dashed-line ellipse displays the bubble claimed by \cite{Rahman:2010in}.
The green xs are the \HII\ regions which are considered to be associated with the bubble (see text).
The green contours show the residual map of the LAT data above 3\,GeV as shown in Figure\,\ref{fig: cmap_cls-sfc10}.
\label{fig: maps_IR}}
\end{center}
\end{figure}

\begin{figure*}[h!] 
\begin{center}
\includegraphics[width=6.9in]{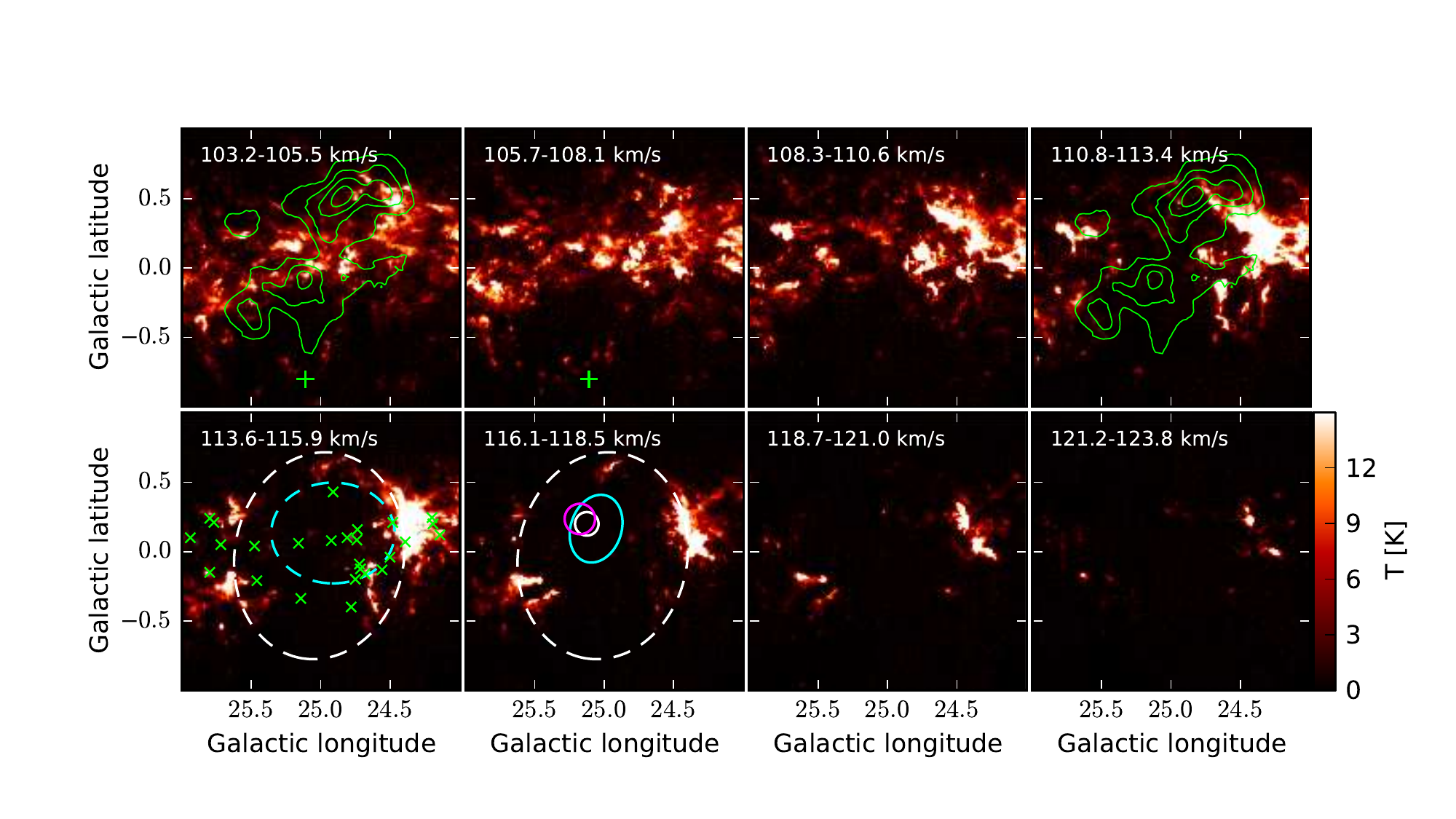}
\caption{\small
Integrated $^{13}{\rm CO}\,J=1\mbox{--}0$ emission every 2\,km\,s$^{-1}$ within the range 103--124\,km\,s$^{-1}$. 
The green contours show the residual map of the LAT data above 3\,GeV from Figure\,\ref{fig: cmap_cls-sfc10}.
The green cross shows the position of G25C. 
The cyan and white dashed-line ellipses represent the bubbles suggested by  \cite{Rahman:2010in} and this work respectively.
The green Xs are positions of \HII\ regions with $v_r$ above 100\,km\,s$^{-1}$ (see text).
The cyan solid-line ellipse delineates the OB association claimed by \cite{Rahman:2013di}, 
while the white solid-line circle represents a region where the number density of IR sources is especially high within the OB association (see Section\,\ref{sec: d-xray}).
The magenta circle is the same as that in Figure\,\ref{fig: imgs-xray}.
\label{fig: maps_CO}}
\end{center}
\end{figure*}

Figure\,\ref{fig: maps_IR} shows that G25 is spatially coincident with the reported bubble, although a factor of $\sim1.8$ larger.
To study the bubble structure in detail, we investigate the $^{13}$CO map for the velocity range around $v_r = 100$\,km\,s$^{-1}$ in this direction (Figure\,\ref{fig: maps_CO}). 
Figure\,\ref{fig: maps_CO} clearly shows a bright arch-like structure around $l \sim 24 \fdg 5$ for $v_r$ in the range 110--120\,km\,s$^{-1}$.
The structure can also be seen in the IR map and composes the western part of the claimed bubble (hereafter cardinal directions are in Galactic coordinates unless otherwise mentioned).
We also overlay \HII\ regions with $v_r >100$\,km\,s$^{-1}$ around this region. 
The positions and velocities of the \HII\ regions are taken from the Green Bank Telescope \HII\ Region Discovery Survey\,\citep{Anderson:2011cu} 
and the Boston University Catalog\footnote{Available at http://www.bu.edu/iar/files/script-files/research/hii\_regions/index.html}.
The latter compiles previous studies of 
ultra-compact \HII\ regions\,\citep{Araya02,Watson03,Sewilo04},
classical \HII\ regions\,\citep{Lockman89}, 
and \HII\ regions that are truly diffuse\,\citep{Lockman96}.
Some of the \HII\ regions are located on the arch (Figure\,\ref{fig: maps_CO}). 
These can be naturally interpreted as swept-up gas illuminated by putative massive stars.
These results support an interpretation that the arch is a part of the bubble created by a massive OB association.

Based on the $^{13}$CO data, the arch structure is expected to exist at least from 110 to 120\,km\,s$^{-1}$ and probably beyond 124\,km\,s$^{-1}$.
Such large dispersion of the velocity is difficult to be explained only by the motion of the Galactic arm,
suggesting that the dispersion is probably explained by local motion of the clouds.
The mean velocity of the gas distribution is $\sim 115$\,km\,s$^{-1}$, which is higher than the velocity at the tangent point (113\,km\,s$^{-1}$) for the direction $(l, b) = (25^{\circ}, 0^{\circ})$.
Figure\,\ref{fig: maps_CO} shows the existence of clouds above the velocity at the tangent point, which means that these clouds locally move away from us.
This motion may be caused by an expansion of the bubble.
Since the arch-like molecular clouds are located near the tangent point and a part of them exists above the velocity at the tangent point,
here we adopt the distance to the tangent point ($= 7.7$\,kpc) as the distance to the clouds.
The estimated distance of 7.7\,kpc is greater than the distance estimated by \cite{Rahman:2010in} (6.1\,kpc).
The authors derive the distance from the mean velocities of eight \HII\ regions which they assume are associated with the bubble.
However one of the \HII\ regions has $v_r$ of 84.8\,km\,s$^{-1}$. 
As discussed above, a possible bubble is expected to exist around $v_r$ of 110\,km\,s$^{-1}$.
Even considering the dispersion of $v_r$ for the bubble, the associated objects are most likely to exist in the velocity range of 95--125\,km\,s$^{-1}$.
Therefore the \HII\ region with $v_r$ of 84.8\,km\,s$^{-1}$ is unlikely to be associated with the bubble. 
When we exclude this object, the mean velocity of the other seven \HII\ regions is 111\,km\,s$^{-1}$, which is very close to the velocity adopted in this paper (113\,km\,s$^{-1}$).
Therefore we adopt 7.7\,kpc for the distance.


Using the $^{13}$CO map, we confirmed the western shell structure of the claimed bubble.
On the other hand, we cannot find any clear shell structures for the other parts of boundaries of the bubble (see Figure\,\ref{fig: maps_CO}).
\cite{Rahman:2010in} delineate the boundary based on the morphology of the IR image and positions of the \HII\ regions, which they consider associate with the bubble.
Except for the clear western shell structure, however, no bright shell structure is found in the IR map either.
Therefore the other boundaries may be located at different positions from the assumed ones.
Actually we find that an arch-like structure extends from the western part to the northern one for $v_r$ in the range of 114--119\,km\,s$^{-1}$ (see Figure\,\ref{fig: maps_CO}). The arch exists outside the boundary of the candidate bubble.
At the southern boundary, we find the \HII\ regions with $v_r > 100$\,km\,s$^{-1}$ around $b = -0\fdg5$, which is outside the claimed bubble. This may suggest that the southern boundary of the bubble is located at or beyond these \HII\ regions.
These findings indicate the possibility that the bubble is more extended than that defined in \cite{Rahman:2010in}:
we delineate a possible boundary of the bubble that is $1 \fdg 5 \times 1 \fdg 2$ in size as shown in Figure\,\ref{fig: maps_CO}.
Hereafter we call this structure ``the G25 bubble".
Note that the eastern boundary passes thorough molecular clouds at $l \sim 25 \fdg 5$ beyond the velocity at the tangent point (113\,km\,s$^{-1}$).
These high-velocity clouds may be pushed away and compressed by a putative massive OB association/cluster.
Figure\,\ref{fig: co-all2} shows that these high-velocity clouds concentrate around $l \sim 25^{\circ}$, which supports this possibility.

\begin{figure*}[h!] 
\begin{center}
\includegraphics[width=6.9in]{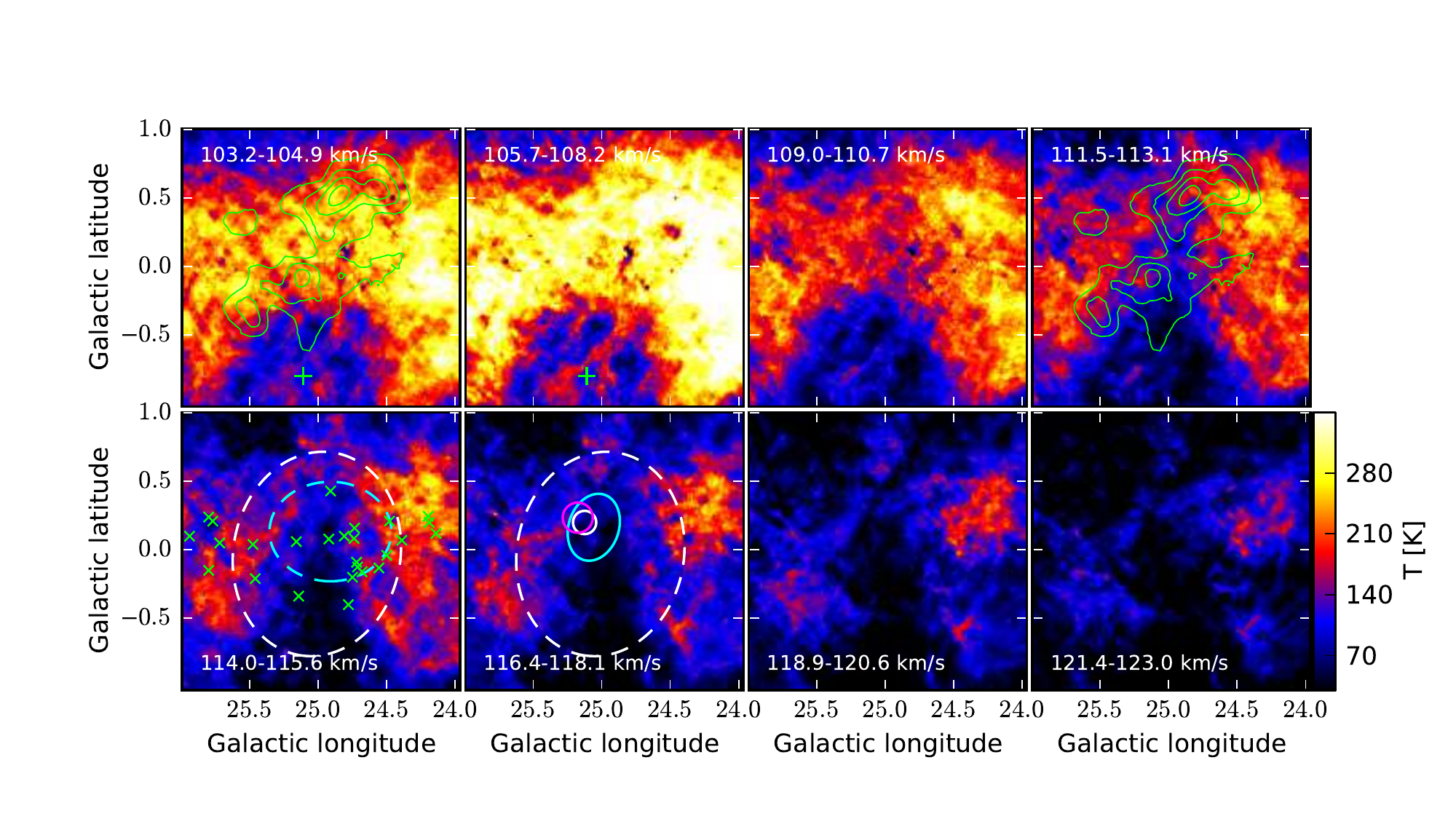}
\caption{\small
Integrated \HI\ gas every 2\,km\,s$^{-1}$ within the range 103--124\,km\,s$^{-1}$.
The contours and marks are the same as those in Figure\,\ref{fig: maps_CO}.
\label{fig: maps_HI}}
\end{center}
\end{figure*}

The bubble with a size of $1 \fdg 5 \times 1 \fdg 2$ has a physical size of $210 \times 170$\,pc at 7.7\,kpc.
This size is comparable to the thickness of a dense part of the Galactic plane, 
which means that ambient gas and clouds in the north and south are generally expected to be more sparse than those in west and east on the Galactic plane.
Figures\,\ref{fig: maps_CO} and \ref{fig: maps_HI} show that
the intensities of the CO and the \HI\ emissions at the northern and the southern boundaries are indeed less than those in the western and the eastern boundaries.
Given the non-uniform gas distributions, the bubble is expected to be elongated in the Galactic-latitude direction
if we assume that the bubble is created by a powerful stellar object on the Galactic plane.
The morphology of the possible boundary is consistent with this expectation.

We also investigate the spatial distribution of \HI\ gas using the VLA Galactic Plane Survey (VGPS)\,\citep{VGPS1}.
Figure\,\ref{fig: maps_HI} shows that a spatial distribution of \HI\ is very similar to that of $^{13}$CO above 110\,km\,s$^{-1}$ in Figure\,\ref{fig: maps_CO}. 
At lower velocities the spatial distribution of \HI\ gas is not similar to that of the molecular clouds:  we cannot find any clear spatial correlation between them for the other velocities at this direction.
If we assume a powering source such as a massive OB association or/and an SNR around $(l, b) \sim$ $(25^{\circ}, 0^{\circ})$, 
then the wind of the source would sweep the surrounding molecular clouds and \HI\ gas away; 
such pressure would make a bubble whose spatial structures are similar regardless of their initial distributions.
Therefore the clear spatial correlation between the molecular clouds and the \HI\ gas can be naturally understood 
if they are shells of the bubble created by an OB association or/and an SNR.

\subsubsection{Does OB Association G25.18+0.26 Create a Bubble?}
\label{sec: connection}

Intense free-free emission is detected from and around this region\,\citep{2010ApJ...709..424M}.
Since free-free emission is expected to be mainly due to reprocessed ionizing photons emitted by young massive stars,
the intense emission suggests that massive OB association(s) reside in the region.
Therefore an OB association/cluster is the most prominent candidate for powering the G25 bubble (Section\,\ref{sec: bubble}),
although no such celestial object has previously been found within the bubble at the distance around 7.7\,kpc.
A candidate for such massive OB associations/clusters is the newly-found OB association/cluster G25.18+0.26 (see Section\,\ref{sec: d-xray}). 
In the following, we provide observational evidence to support this possibility.
In the case that G25.18+0.26 creates the bubble, the radius of G25.18+0.26 ($\sim 0 \fdg 1$) will be physically large, 13\,($d$/7.7\,kpc)\,pc.
The relatively large size indicates that the object is an OB association rather than an OB cluster\,\citep[see e.g.,][]{PortegiesZwart:2010kc}. 
Hereafter we call the system of the G25 bubble created by the OB association G25.18+0.26 ``SFR G25".

\begin{figure*}[t!]
\begin{center}
\includegraphics[width=6in]{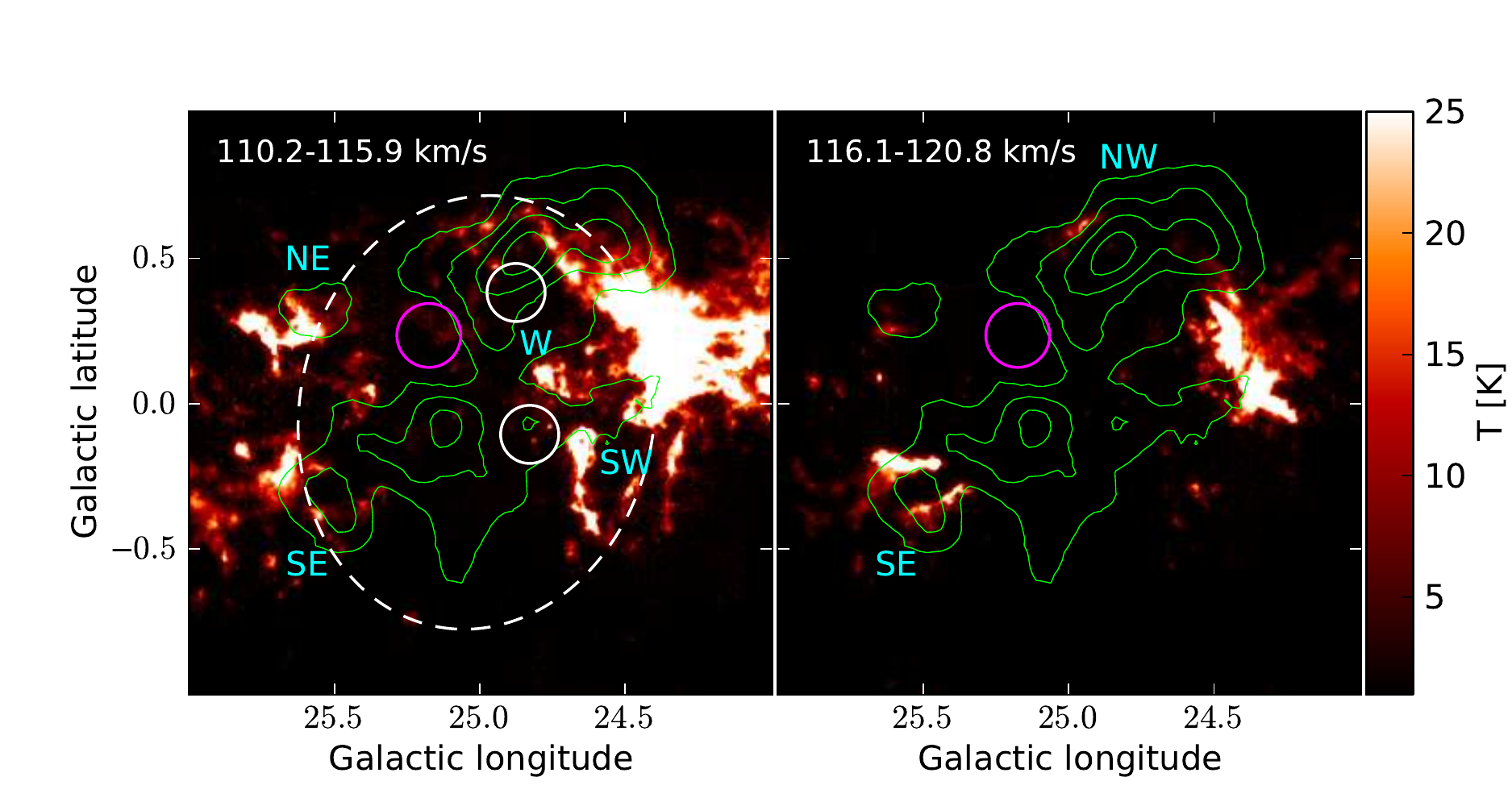}
\caption{\small
Integrated $^{13}{\rm CO}\,J=1\mbox{--}0$ emission integrated over 110.2--115.9\,km\,s$^{-1}$ (left) and 116.1--120.8\,km\,s$^{-1}$ (right). 
The magenta circle, the white dotted ellipse, and the overlaid contours are the same as those in Figure\,\ref{fig: maps_CO}.
The white circles represent the selected positions for estimating the column density of \HI\ and molecular gas (see Appendix\,\ref{sec: tgt}).
\label{fig: maps_CO-cls}}
\end{center}
\end{figure*}

Since an OB association pushes the surrounding gas away to create a bubble, it should be located inside the bubble.
G25.18+0.26 meets this criterion as shown in Figure\,\ref{fig: maps_CO-cls}.
$\NH$ of sources P1 and Q$_{\rm sum}$ are $(1.0\mbox{--}1.4) \times 10^{22}\,{\rm cm}^{-2}$ (see Table\,\ref{tbl: xmodel-2T-PQ}),
which is relatively high given the fact that the total Galactic \HI\ column density at this direction is about $1.7 \times 10^{22}\,{\rm cm}^{-2}$.
This suggests that G25.18+0.26 has a distant location. Our assumed distance of 7.7\,kpc is consistent with this expectation.
However we note that the value of $\NH$ seems to be small for 7.7\,kpc, given that the association is located in the inner Galaxy ($l \sim 25^\circ$).
If no foreground molecular clouds obscure the association, the obtained $\NH$ may be consistent with $d = 7.7\,{\rm kpc}$.
The issue of the $\NH$ value and the distance is discussed in Section\,\ref{sec: group1}.
Here we assume that G25.18+0.26 is located at 7.7\,kpc.

The physical size of the G25 bubble is large, $210 \times 170$\,pc. To create such a large bubble, the responsible OB association should be massive.
To check whether the OB association G25.18+0.26 is massive, here we roughly estimate a total mass from its X-ray luminosity.
The X-ray luminosity functions (XLFs) of the OB associations/clusters are known to have similar shapes\,\citep[e.g.,][]{2010ApJ...713..871W};
the similarities could be understood in terms of a universal shape of the initial mass function for OB associations and an empirical relation between the mass and the X-ray luminosity of a star.
Therefore we expect that the X-ray luminosity of the OB association is roughly proportional to its mass (see also Appendix\,\ref{sec: validation}).

We estimate the X-ray luminosity of G25, assuming that sources P1 and Q1--Q15 are members of G25.18+0.26.
Note that R1--R4 are not included since their association with G25 is uncertain (see Section\,\ref{sec: group1}).
The total luminosity of the sources is estimated to be $L_{\rm G25.18}$ of $2.8 \times 10^{33}\,(d/7.7\,{\rm kpc})^2$\,erg\,s$^{-1}$ in the 2--8\,keV band 
using the total absorption-corrected flux of sources P1 and Q$_{\rm sum}$ based on the $2T$ model with the solar abundance (see Section\,\ref{sec: xray}).
The luminosity is integrated above 2\,keV to reduce the dependency on $N_{\rm H}$.
The luminosities of the individual sources are in the range $(0.7\mbox{--}3) \times 10^{32}\,(d/7.7\,{\rm kpc})^2$\,erg\,s$^{-1}$ (see Section\,\ref{sec: d-xray}):
the estimated luminosity is integrated for X-ray sources with luminosities above $7 \times 10^{31}$\,erg\,s$^{-1}$ in the 2--8\,keV band.
Next we estimate the X-ray luminosity of a known OB association for a comparison with that of G25.18+0.26.
Here we adopt Cyg\,OB2 because it is one of the most massive OB associations ($\sim 3 \times 10^4 \Ms$; \cite{2010ApJ...713..871W}) and located at a distance of only $\sim1.45$\,kpc\,\citep{2003ApJ...597..957H}.
In Appendix\,\ref{sec: XcygOB2}, we estimate the corresponding luminosity $L_{\rm Cyg}$ as $6.3 \times 10^{33}$\,erg\,s$^{-1}$.

Using the derived luminosity, we roughly estimate a total mass of G25.18+0.26 to be $\sim (L_{\rm G25.18}/L_{\rm Cyg}) \times M_{\rm Cyg} = 1 \times 10^4\,\Ms$,
where $L_{\rm G25.18}$ and $L_{\rm Cyg}$ are the 2--8\,keV luminosities of G25.18+0.26 and Cyg\,OB2 respectively, and $M_{\rm Cyg}$ is the mass of Cyg\,OB2 of $3 \times 10^4\,\Ms$. 
Note that we may miss some X-ray sources of G25.18+0.26 above $7 \times 10^{31}$\,erg\,s$^{-1}$ due to the sensitivity limit of the \XMM\ observation.
In addition at least some of sources R1--R4 may be members of the association.
Given these considerations, we probably underestimate the total luminosity of G25.18+0.26. However such effect would be a factor of $\sim 2$ and not crucial for this rough estimation.
In any case, the derived mass is comparable to that of Cyg OB2, which means that G25.18+0.26 is one of the most massive OB associations in our Galaxy if it is located at 7.7\,kpc.

\subsubsection{Gamma-ray Association with the G25 Bubble}
\label{sec: gassoc}


As shown in Figure\,\ref{fig: maps_CO-cls}, 
the $\gamma$-ray emission of G25 is spatially well matched with the G25 bubble suggested based on the molecular-cloud map in Section\,\ref{sec: bubble}.
In the following, we compare spatial structures between the $\gamma$-ray emissions and molecular clouds.
In Figure\,\ref{fig: maps_CO-cls} (left), the G25 emission appears to extend over the boundary of the G25 bubble at the northwest.
Interestingly the molecular clouds with the highest velocities (the right panel) have a spatial break at the northwestern boundary and a bright part of the $\gamma$-ray emission appears to extend through the break.
This can be interpreted as a ``breakout" of the bubble: the $\gamma$ rays may come from the outflow of high-energy particles.
In the western and southwestern parts the $\gamma$ rays and molecular clouds are anti-correlated: 
the $\gamma$ rays appear to extend in cavities between dense parts of the molecular clouds.
In the southeastern and northeastern parts, bright $\gamma$-ray emission comes from just inside the molecular-cloud boundaries but not within the clouds.

We can summarize the relations between the spatial structures of $\gamma$-ray emission and molecular clouds as below:
(i) the $\gamma$ rays are anti-correlated with the molecular shells and clumps so that the $\gamma$ rays appear to be confined in cavities delineated by dense shells of the G25 bubble 
(except for the northwestern boundary where a breakout may occur); 
(ii) the bright $\gamma$-ray emission comes from regions near (but not on) the dense molecular-cloud shell or clump (i.e., the western and eastern boundaries);
(iii) no strong $\gamma$-ray emission is detected near the sparse molecular-cloud boundaries (i.e., the northeastern and southern boundaries).
The spatial properties show that the $\gamma$-ray emission is confined in the G25 bubble,
which is consistent with our proposition that the $\gamma$ rays come from SFR G25.
These spatial properties will be compared with those of the Cygnus cocoon SFR in Section\,\ref{sec: cygx}.

\subsection{High-energy Particles in SFR G25}
\label{sec: model}

\begin{figure*}[t!] 
\begin{center}
\includegraphics[width=5in]{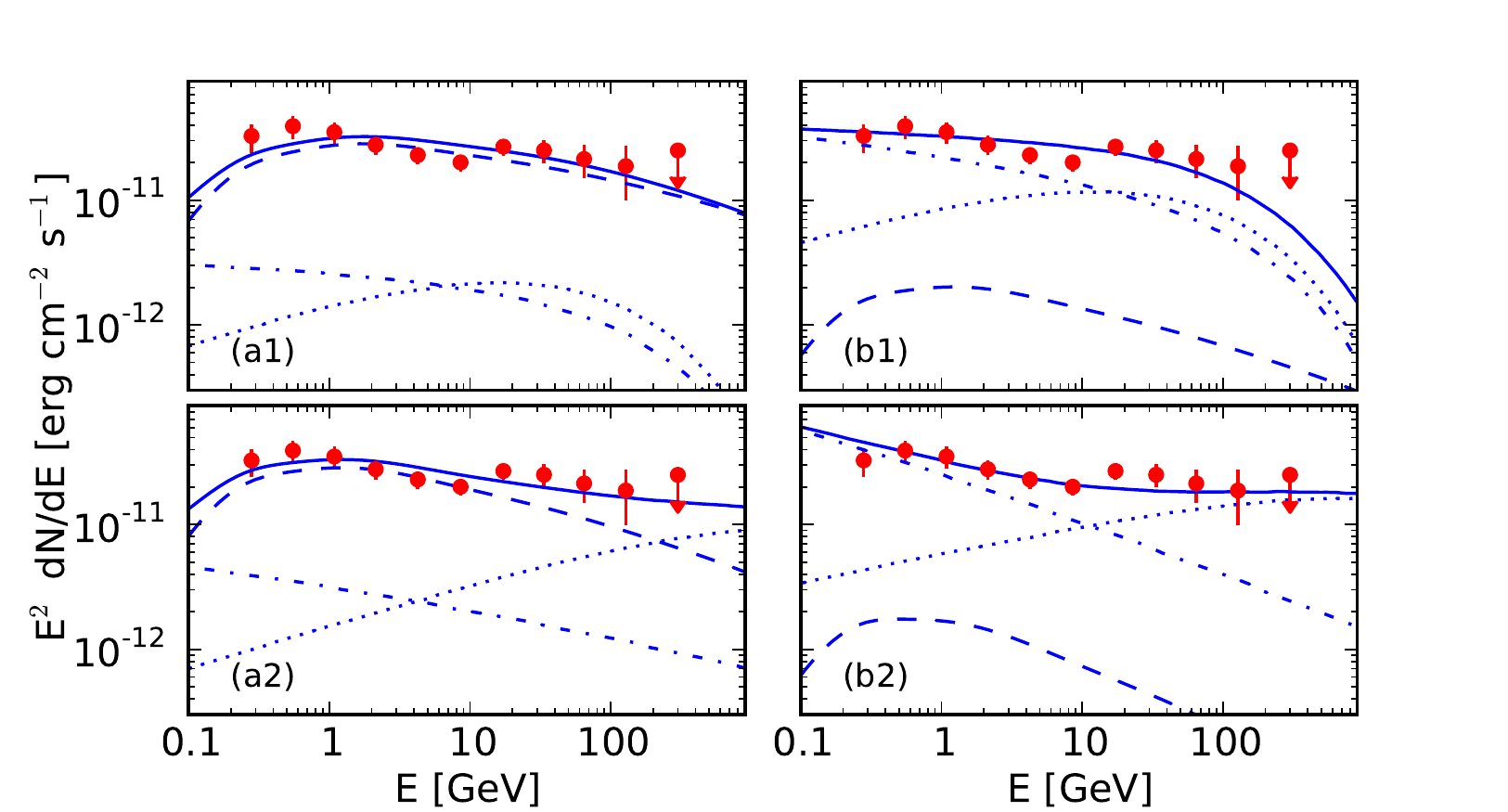}
\caption{\small
SEDs of G25A with model curves for the four cases in Table\,\ref{tbl: model}.
The observed SEDs are represented by red points and error bars (see Section\,\ref{sec: g25-sed}).  
The errors are set by adding in quadrature statistical and systematic errors.
The $\gamma$-ray emission is modeled with a combination of 
$\pi^0$-decay (dashed line), Bremsstrahlung (dot-dashed line), and IC scattering (dotted line).
The total model curve is represented by a solid line.
\label{fig: model_G25A}}
\end{center}
\end{figure*}

\begin{figure*}[t!] 
\begin{center}
\includegraphics[width=5in]{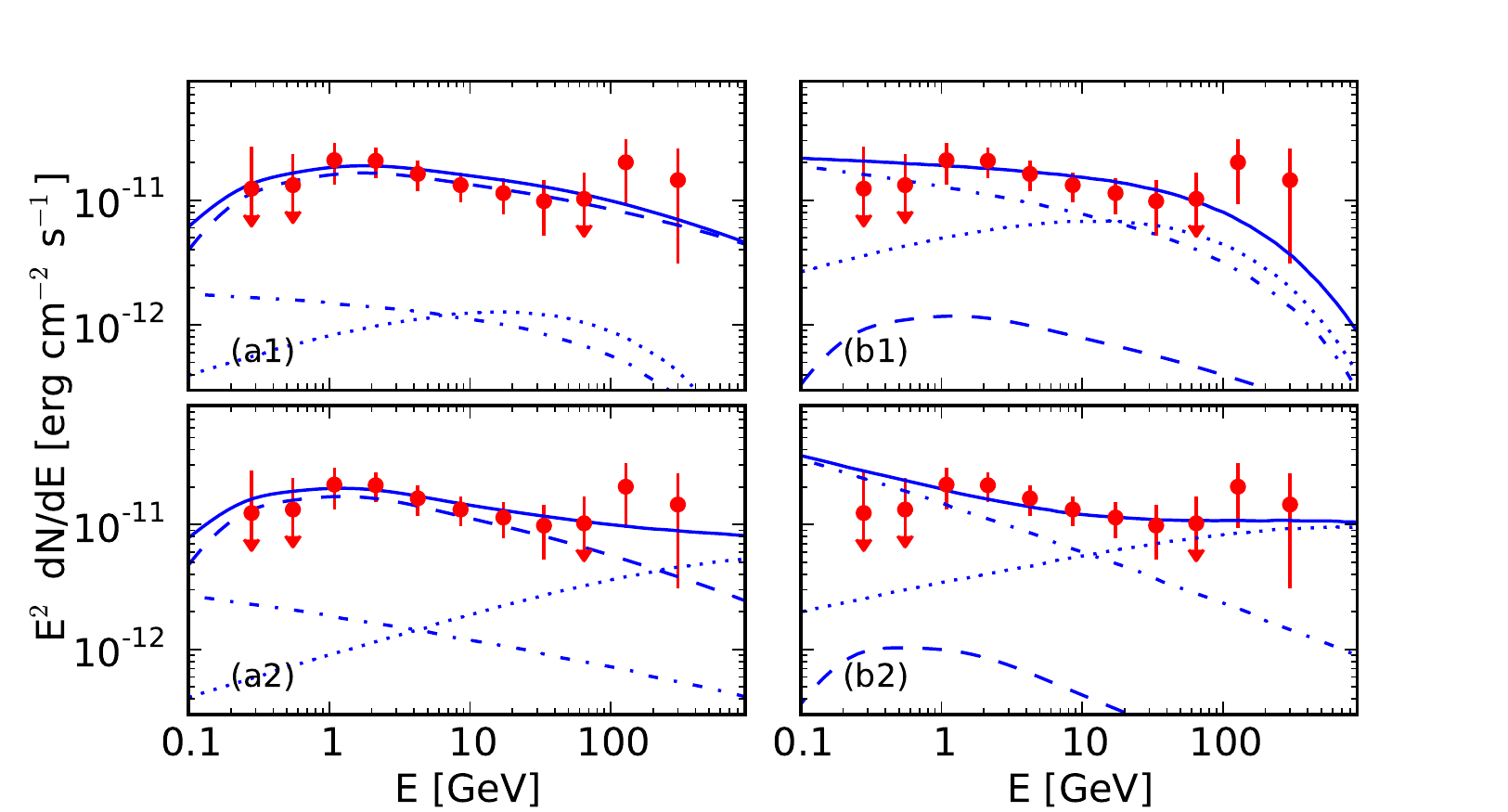}
\caption{\small
Same as Figure\,\ref{fig: model_G25A} but for G25B$'$.
\label{fig: model_G25B}}
\end{center}
\end{figure*}

Here we discuss spectral distributions of the high-energy particles radiating the observed $\gamma$ rays.
We adopt the SFR scenario of Section\,\ref{sec: sfr} and focus on the $\gamma$ rays of the G25A and G25B$'$ regions.
Model spectral distributions of the relativistic particles are adjusted to reproduce the observed $\gamma$ rays.
We consider three kinds of emission mechanisms: the $\pi^0$-decay emission due to high-energy protons, 
and the Bremsstrahlung and the Inverse Compton (IC) scattering processes by high-energy electrons. 
To calculate the $\gamma$-ray emission via these mechanisms, 
we adopt the target gas density ($n = 20\,{\rm cm}^{-3}$) and the target photons estimated in Appendix\,\ref{sec: tgt}.

The spectral distributions of the relativistic particles are assumed to be cut-off power-law functions of the form $dN_{p,e}/dp \propto p^{-s} {\rm exp}\,(-p/p_{{\rm cut}(p,e)})$.
The spectral index of $s$ is assumed to be the same between the relativistic protons and electrons, 
while the cutoff momentum of $p_{{\rm cut}(p,e)}$ can change independently.
To predict the $\gamma$-ray spectrum, we consider two values for the number ratio of the relativistic electrons to protons ($K_{ep}$):  0.01 and 1. 
The value $K_{ep} = 0.01$ is similar to what is locally observed at GeV energies\,\citep[e.g.,][]{2009NJPh...11j5021B}.
The distributions of target gas and photons in Appendix\,\ref{sec: tgt} are adopted for the model calculations.
We reproduce the observed $\gamma$-ray spectra of G25A and G25B$'$ by adjusting spectral parameters of the relativistic particles: 
the normalization of the relativistic protons, $s$, $p_{{\rm cut}(p)}$, $p_{{\rm cut}(e)}$.
The resulting parameters and spectra are shown in Table~\ref{tbl: model} and Figures\,\ref{fig: model_G25A} and \,\ref{fig: model_G25B}.

In the case of $K_{ep} = 0.01$ (which we refer to as the hadronic scenario), the resulting $\gamma$ rays are dominated by the $\pi^0$-decay emission due to the relativistic protons.
The spectral shape of the $\pi^0$-decay emission is almost the same as that of the parent protons.
Since we do not find any significant spectral curvature in the $\gamma$-ray SEDs of G25A and G25B$'$ (see Section\,\ref{sec: g25-sed}),
we set $p_{{\rm cut}(p)}$ at 100\,TeV\,$c^{-1}$; the $\gamma$-ray spectral shape via $\pi^0$-decay emission has no significant spectral curvature in the 1--500\,GeV range.
On the other hand, the photon index of the IC component due to the relativistic electrons is hard, $\sim (s+1)/2 = 1.6$ so that it might contribute to the SED at high energies ($\gtrsim 10$\,GeV).
Here we consider two cases where $p_{{\rm cut}(e)}$ are 1.0 and 100\,TeV\,$c^{-1}$.
The $p_{{\rm cut}(e)}$ of 1.0\,TeV\,$c^{-1}$ corresponds to the break momentum due to the synchrotron radiation loss for a magnetic field $B = 10$\,$\mu$G for 0.1\,Myr.
A magnetic field of the order of 10--20\,$\mu{\rm G}$ is expected to be a reasonable value in the bubble created by the massive OB association\,\citep[see e.g.,][]{Parizot:2004bp}.
Note that the time of 0.1\,Myr is about an order of magnitude less than the expected age of SFR G25 ($\sim {\rm Myr}$; see Section\,\ref{sec: acc}).
For the case $p_{{\rm cut}(e)} = 100$\,TeV\,$c^{-1}$, the IC SED shows no significant spectral curvature in the LAT band.
As shown in Figure\,\ref{fig: model_G25A}, both cases can reproduce the observed $\gamma$-ray spectrum.

For $K_{ep} = 1$ (the leptonic scenario), the resulting $\gamma$ rays are primarily due to Bremsstrahlung and IC scattering processes by high-energy electrons.
As for the case of $K_{ep} = 0.01$, we consider two values for $p_{{\rm cut}(e)}$, 1.0 and 100\,TeV\,$c^{-1}$.
For both values the predicted $\gamma$ rays can also explain the observed spectrum.
Note that the value of $p_{{\rm cut}(p)}$ does not affect the total reproduced $\gamma$ rays.

We assume that the energetic particles uniformly fill the bubble
with the volume $V_{\rm A} = 2.7 \times 10^{61}\,(d/7.7\,{\rm kpc})^3\,{\rm cm}^3$ and $V_{{\rm B}'} = 1.7 \times 10^{61}\,(d/7.7\,{\rm kpc})^3\,{\rm cm}^3$ for G25A and G25B$'$ respectively.
The volume $V_{\rm A}$ is derived as $(4\pi/3) {r_{\rm A}^3}$, where $r_{\rm A}$ is the geometric mean of the semi-minor and the semi-major axes of G25A (see Table\,\ref{tbl: mulpsc-sfc10});
the volume $V_{{\rm B}'}$ is derived as $V_{\rm B} - \pi {r_{\rm B1}^2} l_{\rm B}$, where $V_{\rm B}$ is obtained for G25B in the same manner as for G25A, $r_{\rm B1}$ is the radius of G25B1, 
and $l_{\rm B} = (\pi/2) r_{\rm B}$ is an average line-of-sight length of G25B. 
Under this assumption, the energy densities in Table\,\ref{tbl: model} are calculated to be $U_{p,e} = W_{p,e}/V$, where $V$ is $V_{\rm A}$ or $V_{{\rm B}'}$ for G25A or G25B$'$ respectively.
Note that here we assume that the high-energy particles uniformly fill each region (i.e., the filling factor = 1.0).
However the value may be less than 1.0 given that some parts of the regions have no (or very weak) $\gamma$-ray emission (see Figure\,\ref{fig: cmap_cls-sfc10}).
In that case, the actual $U_{p,e}$ will be higher than the estimated values in Table\,\ref{tbl: model}.

\begin{deluxetable}{lcccccc}
\tabletypesize{\scriptsize}
\tablecaption{Parameters of Multi-Wavelength Models \label{tbl: model}}
\tablewidth{0pt}
\tablehead{
\colhead{Model} & \colhead{$K_{ep}$} & $s$   & \colhead{$p_{{\rm cut}(e)}$}  & \colhead{$p_{{\rm cut}(p)}$} & \colhead{(a)$W_p$/(b)$W_e$ } & \colhead{(a)$U_p$/(b)$U_e$} \\ 
\colhead{}          & \colhead{}                & \colhead{}  & \colhead{(TeV $c^{-1}$)}  & \colhead{(TeV $c^{-1}$)}   & \colhead{($10^{50}$ erg)}            & \colhead{(eV cm$^{-3}$)}
}
\startdata
G25A \\
~~(a1) & 0.01 & 2.2 & 1.0 & 100 & 2.7 & 6.2 \\
~~(a2) & 0.01 & 2.3 & 100 & 100 & 2.5 & 5.8 \\
~~(b1) &      1 & 2.3 & 1.0  & 100 & 0.19 & 0.44 \\
~~(b2) &      1 & 2.5 & 100 & 100 & 0.19 & 0.44 \\
~\\
G25B$'$ \\
~~(a1) & 0.01 & 2.2 & 1.0 & 100 & 1.6 & 5.9 \\
~~(a2) & 0.01 & 2.3 & 100 & 100 & 1.5 & 5.5 \\
~~(b1) &      1 & 2.3 & 1.0  & 100 & 0.11 & 0.40 \\
~~(b2) &      1 & 2.5 & 100 & 100 & 0.11 & 0.40 \\
\enddata
\tablecomments{The total kinetic energies ($E_{\rm kin}$) of radiating protons ($W_p$) and electrons ($W_e$) are calculated for $E_{\rm kin} > 1\,{\rm GeV}$.
The target gas density ($n = 20\,{\rm cm}^{-3}$) and photon field estimated in Appendix\,\ref{sec: tgt} are adopted.}
\end{deluxetable}

\subsection{Acceleration in SFRs}
\label{sec: acc}

In Section\,\ref{sec: sfr}, we proposed and discussed the SFR scenario where
G25A and G25B$'$ are associated with the G25 bubble created by a massive OB association; 
and the OB association is G25.18+0.26, which is found in this work.
Here in Section\,\ref{sec: cygx}, we check that the SFR scenario for G25 is consistent with the case of the Cygnus cocoon, 
the only prior firm case of $\gamma$-ray detection from a SFR in our Galaxy (\CC).
To do this, we compare physical parameters and spatial structures between the G25 and the Cygnus cocoon regions.
The existence of many similarities will support the SFR scenario.
In Section\,\ref{sec: accG25}, based on the SFR scenario, we investigate possible acceleration mechanisms in SFR G25 
using the observed $\gamma$-ray spectral and spatial properties.
Given that G25 is likely to be $\gamma$-ray emissions from an SFR,
we can expect that other young massive SFRs accelerate particles as SFR G25 does.
We discuss some implications in Section\,\ref{sec: accSFR}.

\subsubsection{Comparison with the Cygnus Cocoon}
\label{sec: cygx}

We summarize physical parameters of G25 and the Cygnus cocoon in Table\,\ref{tbl: sfr}.
Note that $L_{\gamma}$ is integrated over the 1--100\,GeV band for comparison with the luminosity of the Cygnus cocoon.
The $\gamma$-ray luminosities for 0.2--500\,GeV are $(1.38 \pm 0.06) \times 10^{36}\,(d/7.7\,{\rm kpc})^2\,{\rm erg}\,{\rm s}^{-1}$ 
and $(0.80 \pm 0.06) \times 10^{36}\,(d/7.7\,{\rm kpc})^2\,{\rm erg}\,{\rm s}^{-1}$ for G25A and G25B$'$ respectively.
In this section, we consider only the hadronic scenario ($K_{ep} = 0.01$).
As stated in Section\,\ref{sec: connection},  
the most massive O stars of the OB association are likely to be still on the main sequence and drive powerful stellar winds into the bubble.
Since the lifetime of such O stars is about 3\,Myr\,\citep[e.g.,][]{MStar}, G25.18+0.26 is expected to be a young association of $\lesssim 3\,{\rm Myr}$ age.
The stellar-wind power of G25.18+0.26 is estimated from that of Cyg\,OB2 based on the assumption that the power is proportional to the total mass of each OB association,
and the fact that both G25.18+16 and Cyg\,OB2 are massive and have similarly young ages.
In the table, we also display corresponding parameters of the Cygnus cocoon.
Although two OB associations, Cyg\,OB2 and NGC\,6910, are coincident with the region, 
here we consider only Cyg\,OB2 because it is about 200 times more powerful (see \CC).

Table\,\ref{tbl: sfr} shows many similarities between the G25 and the Cygnus cocoon regions. 
Each is large (100--200\,pc) and hosts a young ($\lesssim {\rm several}\,{\rm Myr}$) massive ($\gtrsim 10^4\,\Ms$) OB association. 
Each bubble is probably created by the corresponding OB association.
Their sizes are compatible with the simple estimate of \cite{WR77}.

We also compare the spatial properties of G25 and the Cygnus cocoon.
In Section\,\ref{sec: gassoc}, we find the three relations between the spatial structures of the G25 $\gamma$ rays and the surrounding molecular clouds.
In the following, we will investigate the corresponding spatial relations for the Cygnus cocoon region.
Note that we use the 8-${\mu}$m map in Figure\,1 of \CC\ to trace the boundary of the bubble (or the cavity), 
since the CO map in \CC\ has a relatively low spatial resolution.
Here we interpret bright IR shells in \CC\ as dense shells of gas.
As shown in Figure\,3 of \CC, the detected $\gamma$ rays can be divided into three spatial regions: 
main central, eastern, and southwestern regions.

First, as discussed in Section\,\ref{sec: gassoc},
the $\gamma$ rays of G25 are generally anti-correlated with the molecular shells and clumps so that the $\gamma$ rays appear to be confined in the candidate bubble.
The Cygnus cocoon shares this feature:
as stated in \CC, the entire $\gamma$-ray emitting region is surrounded by the bright IR shells and confined within the cavity in the Cygnus cocoon region.

The next feature in common is that the bright $\gamma$-ray emission exists near (but not toward) the dense molecular-cloud shell or clump.
The $\gamma$ rays in the central region of the Cygnus cocoon follow this tendency, since they exist adjacent to the bright western shell. 
On the other hand, the $\gamma$ rays in the other two regions appear not to follow the relation, since they are  ``on" the bright shell.
However this does not necessarily mean that $\gamma$ rays are emitted from the dense shell.
For example, even if $\gamma$ rays are emitted from a cavity surrounded by dense shells (like those in the central region),
the $\gamma$-ray emission could be observed on the bright shells if the shells are along the line of sight of the $\gamma$-ray emitting region.
Actually the spatial size and peak position of the $\gamma$-ray emitting region are different from those of the IR shells in both cases, 
which may indicate that the $\gamma$ rays are irrelevant to the shell region.
For now we cannot conclude whether the $\gamma$ rays in the two regions follow this relation.

\begin{deluxetable}{lcc}
\tabletypesize{\scriptsize}
\tabletypesize{\footnotesize}
\tablecaption{Physical Parameters of the G25 Region and the Cygnus Cocoon Region \label{tbl: sfr}}
\tablewidth{0pt}
\tablehead{
\colhead{} &  \colhead{G25\tablenotemark{a}} & \colhead{Cygnus cocoon\tablenotemark{b}}
}
\startdata
High-energy properties \\
~Photon index & 2.1 & 2.2\\
~Diameter (pc) & $\sim 180$ & $\sim 100$ \\
~$L_{\gamma}$  ($10^{36}$\,erg\,s$^{-1}$) & 1.3 & 0.09 \\
~$W_p$ ($10^{50}$\,erg) & 4.3 $(20\,{\rm cm}^{-3}/n)$ & $0.13\,(60\,{\rm cm}^{-3}/n)$ \\
~$U_p$ (eV\,cm$^{-3}$) & 6.1 $(20\,{\rm cm}^{-3}/n)$& 1.6 $(60\,{\rm cm}^{-3}/n)$\\
~$nU_p$ ($10^{2}$\,eV\,cm$^{-6}$) & 1.2 & 1.0\\
Coincident OB association \\
~Name & G25.18+0.26 & Cyg\,OB2 \\
~Mass ($10^4\,\Ms$) & $\sim 1$ & 3\tablenotemark{c} \\
~Age (Myr) &  $\lesssim 3$ & 1--7\tablenotemark{c} \\
~Stellar-wind power ($10^{38}$\,erg\,s$^{-1}$) & $\sim 1$ & 2--3 \\
\enddata
\tablecomments{The parameters of the G25 bubble are determined based on a 7.7\,kpc distance.
$L_{\gamma}$ is calculated for 1--100\,GeV. $W_p$ and $U_p$ are calculated for $E_{\rm kin} > 1\,{\rm GeV}$.}
\tablerefs{
(a) this work; (b) \CC\ unless otherwise mentioned; (c) \cite{2010ApJ...713..871W}
}
\end{deluxetable}

The last relation is that no strong $\gamma$-ray emission is detected near the sparse molecular-cloud boundaries.
In the Cygnus cocoon region, the LAT does not detect $\gamma$ rays from the southern region, which no bright shell surrounds,
even though the region is adjacent to a plausible powering source Cyg\,OB2.
We also point out another spatial relation: both $\gamma$-ray sources appear to lack a core for the coincident young massive OB association; the $\gamma$ rays shine not on but around the association.

While the G25 and the Cygnus cocoon regions have many similarities, as stated above, 
values of $L_{\gamma}$, $W_p$, and $U_p$ are different between the SFRs (see Table\,\ref{tbl: sfr}).
The $\gamma$-ray luminosity of G25 is 14 times larger than that of the Cygnus cocoon.
Given that the luminosity is proportional to $nW_p = nU_pV$, 
the difference is probably due to the volume ($V$) of the sources; the estimated volume of G25 is 9 times larger than that of the Cygnus cocoon\,(\CC).
Both $L_{\gamma}$ and $V$ of G25 are about an order of magnitude larger than for the Cygnus cocoon.
Actually $nU_p$ of G25 and the Cygnus cocoon are almost the same as shown in Table\,\ref{tbl: sfr}.
Although the values of $nU_p$ are almost the same, the values of $U_p$ are different between the regions (see Table\,\ref{tbl: sfr}).
The difference is due to the different estimated target gas densities.
The estimated density of the Cygnus cocoon ($n \approx 60\,{\rm cm}^{-3}$; \CC) is higher than that of G25 ($n = 20\,{\rm cm}^{-3}$).
This is probably because the estimated gas density for the Cygnus cocoon includes shells of the dense gas.
The total mass of the gas is dominated by H$_2$ and \HI\ gas, 
which are expected to be swept up to the boundaries of the bubble and compose dense shells.
As described above, based on the morphology of the $\gamma$-ray emitting region and the surrounding gas, 
we find that the $\gamma$-ray emission may come from cavities created by the OB association but not from the dense shells.
If this is true for the Cygnus cocoon, the value of $n$ should be lower than $60\,{\rm cm}^{-3}$.
For example, if we assume $n$ of G25 ($20\,{\rm cm}^{-3}$) for the Cygnus cocoon,
then $U_p$ is estimated to be $\sim 5\,{\rm eV}\,{\rm cm}^{-3}$, which is similar to 6\,eV\,${\rm cm}^{-3}$ for G25.

\subsubsection{The Acceleration Process in SFR G25}
\label{sec: accG25}

\paragraph{Acceleration Mechanism}

The observed bright $\gamma$-ray emission provides evidence that the accelerated particles are distributed in the SFR.
Given the similarities of the spectral and spatial properties between G25 and the Cygnus cocoon region,
the particles in both regions are expected to be accelerated by the same kind of powering source via the same acceleration process. 
The powering source for G25 is likely to be stellar winds of the OB association G25.18+0.26,
since there is no other powerful object such as a PWN or SNR within the G25 bubble around the assumed distance of 7.7\,kpc.
This is consistent with the plausible scenario that the OB association Cyg\,OB2 is a powering object for the Cygnus cocoon region\,(\CC).
As shown in Table\,\ref{tbl: sfr}, $W_p$ of G25 and the Cygnus cocoon are $4 \times 10^{50}\,{\rm erg}$ and $1 \times 10^{49}\,{\rm erg}$ respectively.
The total energy of the stellar winds of G25.18+0.26 can be roughly estimated as $\sim 3 \times 10^{51}\,{\rm erg}$ 
on the assumption that the stellar-wind power of $1 \times 10^{38}\,{\rm erg}\,{\rm s}^{-1}$ is continuously supplied for 1\,{\rm Myr}.
The energy available is great enough to provide the estimated $W_p$ for both sources.
In this section, we consider only the hadronic scenario.

There are two main candidate sites for particle acceleration powered by the stellar winds:
(1) the particles are accelerated through DSA at the wind boundary and/or in the turbulent wind itself in OB associations\,\citep[e.g.,][]{CCJ83,2008MNRAS.385.1105B};
(2) the particles are trapped in the bubble and gain energy by stochastic acceleration due to magnetic turbulence\,\citep[e.g.,][]{Bykov01,Parizot:2004bp,2010A&A...510A.101F}.
In case (1), 
G25 is interpreted as $\gamma$-ray emission by runaway particles from the original acceleration site of the OB association G25.18+0.26.
In this scenario, we assume that the OB association continuously accelerates particles for $\sim 1{\rm Myr}$.
The diffusion coefficient of the particles is considered to be energy dependent; the higher-energy particles diffuse faster.
Given that the LAT $\gamma$-ray emission does not come from the dense shells, the accelerated particles of $\lesssim 1$\,TeV are still confined within the bubble.
In this case, the spectral shapes of the observed $\gamma$ rays will be spatially different within the bubble: 
a harder spectrum is expected further from the association.
We note that the LAT observations show that G25A and G25B$'$ have the same spectral shape of $\Gamma \simeq 2.1$,
though they are located at different distances from G25.18+0.26 ($\sim 50$\,pc and $\sim 80$\,pc respectively). 
In addition, even the divided sections of G25A and G25B$'$ have an uniform spectral shape (see Table\,\ref{tbl: sed-sfc10}).
In case (2), 
G25 is interpreted as $\gamma$-ray emission by relativistic particles that experience {\it in-situ} acceleration by magnetic turbulence in the bubble.
A part of the stellar winds is considered to be converted to secondary shocks and magnetic turbulence via wind-wind interactions and shock-clump interactions inside the bubble.
The particles gain energy via stochastic acceleration during their stay in the bubble because of the magnetic turbulence that repeatedly scatters them.
Since the acceleration is expected to occur within the whole bubble region, 
the spectral shapes are expected to be more-or-less uniform within the bubble, which is consistent with our results.

\paragraph{Spatial Properties of the Acceleration Site}

As stated in Section\,\ref{sec: cygx}, the values of $nU_p$ are similar between G25 and the Cygnus cocoon.
In Table\,\ref{tbl: sfr}, the values of $n$ are estimated as 20\,cm$^{-3}$ and 60\,cm$^{-3}$ for G25 and the Cygnus cocoon, respectively.
$W_p$ ($= U_pV$) is mainly determined by $V$: 
the energy content of the SFRs is proportional to the volume of the cavities created by the OB associations.
Actually $L_{\gamma}$ of G25 is 1.3\% of the stellar-wind power, while the corresponding value is 0.04\% for the Cygnus cocoon (Table\,\ref{tbl: sfr}); 
the $\gamma$-ray luminosity is not dependent on the stellar-wind power of the OB association.
This may suggest that the energy content of the accelerated particles does not strongly depend on the stellar-wind power.
Rather, the energy content seems to depend on the volume of cavities where the accelerated particles can be trapped.

The spatial structures of the $\gamma$-ray emission in G25 also provide hints about how the acceleration process operates.
Given the spatial property (i) in Section\,\ref{sec: gassoc}, the relativistic particles seem to be confined within the bubble.
Such an efficient confinement is expected to be achieved by slow diffusion.
If the bubble is filled with magnetic field turbulence, 
the diffusion coefficient ($D$) would be smaller than that of the interstellar medium.
For example, if we assume $D = 10^{27}\,{\rm cm}^{-2}\,{\rm s}^{-1}$ for a 1\,TeV particle\,\citep{Parizot:2004bp}, 
the corresponding diffusion length will be $\sim 2\sqrt{Dt} = 115\,{\rm pc}$ for $t = 1\,{\rm Myr}$.
This is larger than the radius of the bubble.
To confine the particles with $\lesssim 1\,{\rm TeV}$ energy, the diffusion coefficient and/or the trapping time should be smaller than the assumed values.
Even in this case, the higher-energy particles are expected to already escape the bubble.
On the other hand, when we focus on the western and southwestern parts,
$\gamma$ rays are present, although not associated with the gas clumps and the dense shell (see Section\,\ref{sec: gassoc} and Figure\,\ref{fig: maps_CO-cls}).
This phenomenon may suggest that 
the particles are confined because they cannot penetrate the dense-gas regions (clumps and shells) rather than because of the slow diffusion.
The breakout region apparent in the northwest may also support this hypothesis; 
a part of the $\gamma$-ray emission emerges from the bubble via the region where the dense shell does not exist.
Deep TeV observations of this region will give us a clue to how the high-energy particles are trapped within the bubble.

The spatial properties (ii) and (iii) in Section\,\ref{sec: gassoc} are also worth mentioning here.
The $\gamma$-ray emission comes not from an entire region but only parts of the G25 bubble.
We can explain the properties in two ways, given that the $\gamma$-ray emission is proportional to $nW_p$.
First, the gas density in the $\gamma$-ray emitting region may be higher.
In this scenario, the particles are accelerated and fill the entire bubble
but the LAT is able to detect $\gamma$ rays only near the dense shell where the target gas density is enhanced.
Second, the $\gamma$-ray emitting regions are ones where the relativistic particles exist.
In this case, the particles are mainly accelerated and trapped within the parts of the bubble where the $\gamma$ rays are observed.
This may happen, for example, if the magnetic turbulence is enhanced around the dense gas 
by the reflected shocks generated via interactions of the shocks and the dense gas\,\citep[e.g.,][]{Parizot:2004bp}.
For now it is difficult to conclude which scenario is most plausible.

The Cygnus cocoon region has spatial characteristics similar to those of G25 described above (Section\,\ref{sec: cygx}). 
This may indicate that such properties are tightly connected to the acceleration and the emission processes of the SFRs:
for the acceleration in SFRs, a powerful OB association is required but the morphology and size of the SFR created by the OB association may also play an important role.
The density of accelerated particles within the bubble appears to be much larger than 
that in the surrounding molecular clouds as is manifested by the apparent anti-correlation between the gamma-ray 
emission and the $^{13}$CO map. 
The diffusion coefficients within the bubble and the molecular clouds are expected to be quite different, possibly resulting in the large degree of spatial variation of the CR density. 
More sophisticated theoretical models are needed to draw definitive conclusions. This is beyond the scope of the present paper.

\subsubsection{Acceleration in Young Massive SFRs}
\label{sec: accSFR}

Given this discussion, G25 is likely to be the second case of $\gamma$-ray detection from an SFR in the Galaxy.
In addition, we find many similarities of the physical parameters and spatial structures between G25 and the Cygnus cocoon, 
which suggests that the same acceleration mechanism operates in both.
This would indicate that the particle acceleration in the Cygnus cocoon is not a special case and that
other young massive SFRs have capabilities to accelerate particles via the same mechanism.
However, a recent work by \cite{MG16} has shown that the \HII\ regions powered by young stellar clusters 
such as the Rosetta and Orion Nebulae are not detected by the {\it Fermi}-LAT. 
This could be explained by the different bubble sizes; the G25 bubble and Cygnus cocoon 
are larger than the \HII\ regions by more than an order of magnitude. 
The timescale of diffusive escape of accelerated particles from the \HII\ regions is 
shorter than the age of the SFR, and consequently most of the accelerated particles should already 
have left the \HII\ regions. Further detections of other SFRs with different sizes will 
allow us to test such a scenario.

It is interesting to note that 
massive SFRs could constitute a substantial fraction of unassociated {\it Fermi} sources.
In the 2FGL catalog, $\sim30$\% of the 1873 sources have no clear associations\,\citep{2012ApJS..199...31N}.
The proportion of the unassociated sources increases to $\sim60$\% of 400 sources near the Galactic plane ($| b | < 5^\circ$).
The corresponding number of unassociated sources on the Galactic plane in the 1FHL catalog is also non-negligible ($\sim20$\% of 76 sources).
Note that the number of securely identified sources is lower.
These facts would suggest that there are new $\gamma$-ray source classes to be found in the Galaxy.
Our Galaxy is estimated to contain $\gtrsim 80$ associations/clusters with large masses of $10^{4\mbox{-}5}$\,$\Ms$ and ages of $\lesssim 25$\,Myr\,\citep{IVD10}.
Therefore SFRs created by such objects are prominent candidates for the unassociated {\it Fermi} sources.

Stochastic acceleration could accelerate particles to $>{\rm TeV}$ energies in superbubbles (SBs).
SBs are created by the collective power of stellar winds of massive stars and SNRs in OB associations/clusters\,\citep[e.g.,][]{1992MNRAS.255..269B,1998ApJ...509L..33H}.
SBs are considered to be filled with the enhanced CR density, secondary shocks, and strong magnetic turbulence caused by collective shocks of SNRs and stellar winds.
Since such an environment is significantly different from that of the interstellar field, 
the particle acceleration process in SBs is theoretically expected to be different from that of isolated SNRs in the interstellar medium\,\citep[e.g.,][]{Bykov01,Parizot:2004bp,2010A&A...510A.101F}.
Most SNRs are born in OB associations/clusters, which means that the supernova shocks generally propagate in SBs.
Since SNRs are believed to be accelerating sites for the Galactic CRs, the acceleration process in SBs is a key to understanding the Galactic CR acceleration.

This work combined with \CC\ observationally support the theoretical prediction 
that bubbles created by OB associations are sites where particles are accelerated to high energies 
via the stochastic acceleration powered by collective shocks\,\citep{BAMssr}.
The observed SFRs can be interpreted as SBs in an early phase when the stellar-wind power dominates.
If stochastic acceleration also operates in many other SBs even in the later stage where when the SNR power dominates ,
the acceleration process of the Galactic CRs should be different from the process for isolated SNRs.
Actually we find that $U_p$ of G25 is $6\,{\rm eV}\,{\rm cm}^{-3}$ (see Table\,\ref{tbl: sfr}), which is $\sim 6$ times higher than the locally observed value for CRs.
This may indicate that supernova shocks in SBs will generally propagate in a sea of significantly higher $U_p$ than the local value.
Note that $U_p$ of the Cygnus cocoon in Table\,\ref{sec: sfr} is comparable to the local value, but the value may be greater and be similar to that of G25 (see Section\,\ref{sec: cygx}).
In addition, given the uncertainties, $U_p$ may be still greater, ranging up to $100\,{\rm eV}\,{\rm cm}^{-3}$.

\section{Conclusion}
\label{sec: conclusion}

We report that a study of LAT observations of extended emission is likely to be the second case of a $\gamma$-ray detection from an SFR in our Galaxy.
We analyze extended $\gamma$-ray emission in the G25.0+0.0 region using $\sim 57$-months LAT data in the energy range 0.2--500\,GeV.
The emission is divided into two extended regions (G25A and G25B) and one point-like source (G25C).
We divide G25A into three sections and do not find any significant variation of the spectral shapes among them.
We also divide G25B into three sections and find that one section (G25B1) has a hard spectrum,
while the other sections (G25B$'$) have softer spectra.
The spectrum of G25B1 is well represented by a power law with 
photon index $\Gamma = 1.53 \pm 0.15$ and integrated 3--500\,GeV flux of $(1.0 \pm 0.2) \times 10^{-9}$\,photon\,cm$^{-2}$\,s$^{-1}$.
We do not find any significant spectral curvature of G25A, G25B$'$, or G25C; their SEDs are well represented by power laws.
The resulting parameters of G25A are $\Gamma = 2.14 \pm 0.02$ and the integrated 0.2--500\,GeV flux is $(1.13 \pm 0.05) \times 10^{-7}$\,photon\,cm$^{-2}$\,s$^{-1}$;
those of G25B$'$ are $\Gamma = 2.11 \pm 0.04$ and flux $(6.0 \pm 0.6) \times 10^{-8}$\,photon\,cm$^{-2}$\,s$^{-1}$;
those of G25C are $\Gamma = 1.8 \pm 0.2$ and the flux of $(2.8 \pm 1.8) \times 10^{-9}$\,photon\,cm$^{-2}$\,s$^{-1}$.
There is no indication of variability for any of the sources for the period spanned by the observations.

We do not find any sources at other wavelengths clearly associated with G25A, G25B$'$, or G25C.
Both the G25A and G25B$'$ regions show similar characteristics: elongated morphologies with similar surface brightness and hard energy spectra. 
Given their proximities, it is plausible that one celestial object is responsible for their $\gamma$-ray emission.
On the other hand, the hard spectrum of G25B1 suggests that its $\gamma$ rays originate from a different object than the other regions.
Actually G25B1 is spatially coincident with $\HJP$, a candidate PWN powered by PSR\,J1838$-$0655. 
In addition, the LAT SED is smoothly connected to that of the H.E.S.S. data.
The LAT spectral shape is the expected one from typical relativistic electron distribution of PWNe: $dN/dp \propto p^{-2}$.
Given the evidence, G25B1 is the most likely to be produced by the PWN.

We study the AX\,J1836.3$-$0647 field, which is located in the G25 region, using a $\sim17$\,ks  \XMM\ observation (0.5--12\,keV).
We find that an unidentified X-ray source AX\,J1836.3$-$0647 is composed of a cluster of $\sim 20$ X-ray sources and study their spectra.
Given their thermal spectra and spatial concentration, 
we conclude that AX\,J1836.3$-$0647 is an OB association/cluster; we call this new object G25.18+0.26.

In this work, we propose that
G25A and G25B$'$ are associated with an SFR created by a massive OB association; 
and the OB association is G25.18+0.26 found by this work. 
Four lines of observational evidence support this scenario.
First, 
the $\gamma$-ray properties of G25 resemble those of the Cygnus cocoon region, 
the only firm case of $\gamma$-ray detection from an SFR in our Galaxy:
the $\gamma$-ray sources in both regions are spatially extended;
their energy spectra are described as power laws with $\Gamma$ of 2.1--2.2 without any significant spectral curvature at least up to a few hundred GeV;
the LAT finds no significant spectral variation in either region.
Second, 
recent radio and IR observations have revealed a candidate massive SFR in the direction of G25.
We also confirm a corresponding bubble-like structure of the molecular and the \HI\ gas (G25 bubble) at 7.7\,kpc, 
which may be created by a putative powering source.
Third, 
we find that G25.18+0.26 resides in the G25 bubble.
The unabsorbed X-ray luminosity of the object at the assumed distance (7.7\,kpc) 
is comparable to that of Cygnus\,OB2, one of the most massive OB associations in the Galaxy.
This suggests that G25.18+0.26 is also a massive OB association ($\sim 2 \times 10^4\,\Ms$).
In addition, with near-IR data, a candidate massive OB association has been claimed in this direction.
Finally,
the observed $\gamma$ rays look confined in the G25 bubble. This property is also shared with the Cygnus cocoon.

We examine the observational results based on the SFR scenario.
We construct models to reproduce the observed $\gamma$ rays of G25A and G25B$'$. 
Either the hadronic ($K_{ep} = 0.01$) or the leptonic ($K_{ep} = 1$) scenarios can explain the data, 
where $K_{ep}$ is a number ratio of the relativistic electrons to protons.
In either case, the momentum cutoff in the radiative-particle distributions is required to be $\gtrsim 1\,{\rm TeV}$.

The relativistic particles appear to be confined and have a uniform spectral shape within the G25 bubble.
This suggests that the particles are trapped and stochastically accelerated by magnetic turbulence in the bubble.
The acceleration power is likely to be provided through stellar winds of the young massive SFR.
In addition, the bright $\gamma$-ray emissions appear to exist only near (but not on) the dense shell or clump.
The phenomenon is interpreted in two ways.
The first is that the gas density in the $\gamma$-ray emitting region is higher than that elsewhere in the bubble:
the particles are accelerated and fill the entire bubble
but the LAT is able to detect $\gamma$ rays from regions of the enhanced gas density.
In the second interpretation, the $\gamma$-ray emitting regions are interpreted as where the relativistic particles exist:
the particles are mainly accelerated and trapped within the $\gamma$-ray emitting regions.

We compare physical parameters of G25 and the Cygnus cocoon and find that 
the energy content of the accelerated particles does not strongly depend on the stellar-wind power.
Rather, the energy content seems to depend on the volume of cavities where the accelerated particles are trapped.
We also find that the spatial properties of G25 are similar to those of the Cygnus cocoon region.
Such properties may be tightly connected to the acceleration and the emission processes in the SFRs.
This could explain why no $\gamma$-ray emission has been detected from other known young massive SFRs up to now.
The accelerated particles appear to be confined within the bubble;
perhaps because the particles are unable to penetrate the dense shells and clumps.
This seems inconsistent with the expectation that the accelerated protons penetrate into the dense gas and emit $\gamma$ rays via interactions with it.
The uncertain mechanism of such an unusual confinement may be key to understanding the detailed acceleration process at SFRs.

G25 has no spectral curvature in the 0.2--500\,GeV range; the source is expected to be detectable by the current generation of TeV telescopes, 
although no TeV detection has been reported yet except for the G25B1 region.
The G25A region is an especially good target, since the $\gamma$-ray surface brightness is high and no other $\gamma$ source seems to contaminate the region.
Spectral and spatial information in the TeV band will enable us to study this interesting source in more detail.
The detailed properties of the newly-found OB association G25.18+0.26 are still unknown.
This is mainly because the limited \XMM\ statistics prevent us from studying the individual members of the object.
More observations by X-ray telescopes such as \XMM\ and \Chandra\ will reveal its detailed features.

\begin{acknowledgements}

We thank Dr.\,S.\,W.\,Digel, Dr.\,L.\,Tibaldo, Dr.\,T.\,Jogler, and Dr.\,J.\,W.\,Hewitt for helpful discussion.
We also appreciate the referee for his/her valuable comments, which improved our paper.
We thank Dr. K.\,Kosack for kindly providing the TeV data of HESS\,J1841$-$055.
We thank Canadian Astronomy Data Centre, which is operated by the Herzberg Institute of Astrophysics, National Research Council of Canada.
The \textit{Fermi} LAT Collaboration acknowledges generous ongoing support
from a number of agencies and institutes that have supported both the
development and the operation of the LAT as well as scientific data analysis.
These include the National Aeronautics and Space Administration and the
Department of Energy in the United States, the Commissariat \`a l'Energie Atomique
and the Centre National de la Recherche Scientifique / Institut National de Physique
Nucl\'eaire et de Physique des Particules in France, the Agenzia Spaziale Italiana
and the Istituto Nazionale di Fisica Nucleare in Italy, the Ministry of Education,
Culture, Sports, Science and Technology (MEXT), High Energy Accelerator Research
Organization (KEK) and Japan Aerospace Exploration Agency (JAXA) in Japan, and
the K.~A.~Wallenberg Foundation, the Swedish Research Council and the
Swedish National Space Board in Sweden.

Additional support for science analysis during the operations phase is gratefully
acknowledged from the Istituto Nazionale di Astrofisica in Italy and the Centre National d'\'Etudes Spatiales in France.
\end{acknowledgements}

\appendix

\section{G27 Region}
\label{sec: g27}

\begin{figure*}[t] 
\begin{center}
\includegraphics[width=7.5in]{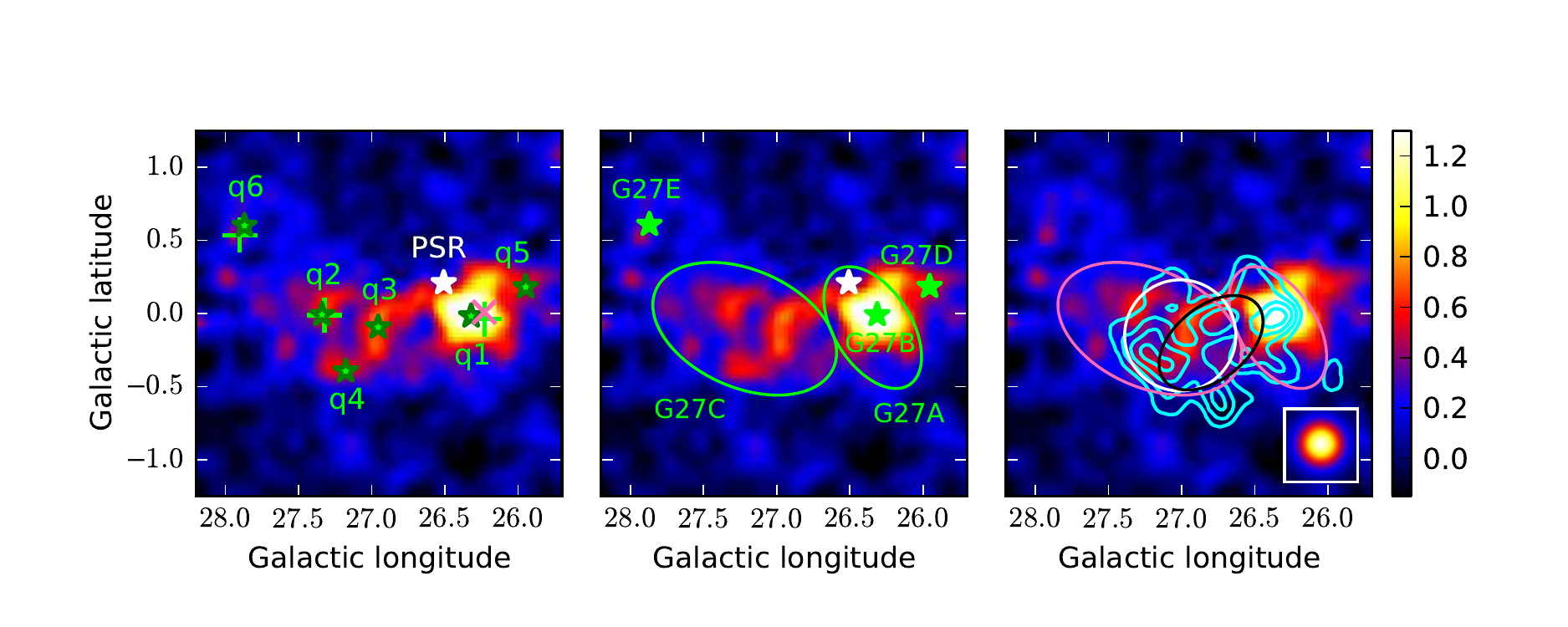}
\caption{\small
Left: {\it Fermi}-LAT background-subtracted map of the G27 region above 3\,GeV (in units of counts per pixel). The pixel size is $0 \fdg 025$. Smoothing with a Gaussian kernel of $\sigma = 0 \fdg 15$ is applied. Note that we subtract the $\gamma$ rays of $\LP$ using its best-fit values, since the pulsar is so bright that it dominates the region in this energy range (see Figure\,\ref{fig: cmap_wide}).
The position of $\LP$ is represented by the white star.
The green stars represent the point sources described in Table~\ref{tbl: mulpsc-g27}.
The green crosses and the magenta X represent positions of the 2FGL and 1FHL sources respectively. 1FHL\,1839.1$-$0557 is the only 1FHL source in this region.
Middle: same map as the left panel. Green ellipses delineate the G27A and G27C regions. The point sources G27B, D, and E are represented by the green stars. The position of $\LP$ is represented by the white star.
Right: same map as the left panel. 
$\HJ$ and its corresponding TeV image are represented by the black ellipse and the cyan contours\,\citep{Aharonian:2008dk}.
The contour levels are 30, 45, 60, and 75 counts per pixel, where the pixel size is 0 \fdg 0039.
The white circle represents the best-fit Gaussian morphology ($1\,\sigma$) from \citet{2013ApJ...773...77A}.
We also show the G27A and C regions (magenta ellipses) for comparison.
The simulated point source in the inset is the same as that in Figure\,\ref{fig: cmap_cls-sfc10}.
\label{fig: cmap_cls-g27}}
\end{center}
\end{figure*}

The G27 region contains six point sources in a $\sim 1 \fdg 5$ square area as described in Section~\ref{sec: pd}. The high concentration of point sources might indicate that the observed $\gamma$-ray emission is extended as with G25. To check this possibility, we estimate the extension of the observed $\gamma$-ray emission using the same procedure as in G25 (see Section\,\ref{sec: g25-sp}).
Note that as the first step of the procedure we check for the existence of extended emission around the bright \LP, which is located in the G27 region, 
since the 2PC catalog reports unidentified extended off-peak emission around this pulsar\,\citep{2013ApJS..208...17A}. 
We construct an input model by replacing sources in G27 from $model2$ (see Section\,\ref{sec: pd}) for the multiple sources in Table~\ref{tbl: mulpsc-sfc10}.
Figure~\ref{fig: cmap_cls-g27} shows background-subtracted maps of G27 using the best-fit values of this model; the subtracted background model contains background sources outside the G27 region and the contributions from the Galactic diffuse emission and isotropic diffuse background.

Here we provide notes for each step in the procedure.
As stated above, we first check for the existence of extended emission around $\LP$ by adding an elliptical shape around this pulsar.
As shown in Table\,\ref{tbl: sp-g27}, there is significant extended emission ($\TSe =48$), which is named ``G27A".
Note that the elliptical region contains $q1$, but $q1$ still has a significant TS of 147, so that we keep this point source in the model.
Next we fit $q1$ using an elliptical shape and obtain no significant extension ($\TSe = 2$). 
We keep this source solely as a point source and rename it ``G27B".
Then we fit $q2$ using an elliptical shape. The resulting size is so large ($\sim 1^\circ$) that the ellipse includes $q3$ and $q4$. 
The TS of each of these sources decreases to less than 25 so that we remove both and refit the elliptical shape. We call this source ``G27C", which is significantly extended ($\TSe = 257$; see Table\,\ref{tbl: sp-g27}). We also compare the maximum likelihoods of models of the elliptical shape (G27C) and three point sources ($q2$, $q3$, and $q4$). 
The resulting value is $-2\ln (L_{\rm 3psc}/L_{\rm G27C}) = 41$.
This means that the elliptical shape explains the observed $\gamma$-ray emission better than the three point sources. We adopt the elliptical shape, given that it has higher likelihood even though it has 5 degrees of freedom fewer than the model with the three point sources does.
We fit $q6$ using an elliptical shape and obtain $\TSe = 0$, which means that the source is not significantly extended. 
We keep $q5$ as a point source and rename it ``G27D".
With the same procedure, we find that the remaining source $q6$ is also a point source; we rename it ``G27E".

\begin{deluxetable*}{lccccccrrc}
\tabletypesize{\scriptsize}
\tabletypesize{\footnotesize}
\tablecaption{Components of G27 \label{tbl: sp-g27}}
\tablewidth{0pt}
\tablehead{
\colhead{Name} & \colhead{Spatial} & Center position & Positional & \colhead{Semi-major} & \colhead{Semi-minor} & \colhead{Angle\tablenotemark{b}} & $\TSe$ & TS & Sources\\
\colhead{} & \colhead{type} &  \colhead{($l$, $b$)}& error\tablenotemark{a} & axis (deg) & axis (deg) & (deg) &&& \colhead{ in Table\,\ref{tbl: mulpsc-g27} }
}
\startdata
G27A & ellipse & (26\fdg34, -0\fdg10) & 0\fdg03 & $0.47\pm0.03$ & $0.26\pm0.03$ & $-58\pm10$ & 48 & 236 & -- \\
G27B & point & (26\fdg31, -0\fdg01) & 0\fdg02 & -- &--&--& 2 & 34 & $q1$ \\
G27C & ellipse & (27\fdg22, -0\fdg10) & 0\fdg03 & $0.66\pm0.04$ & $0.40\pm0.03$ & $-24\pm4$ & 257 & 273 & $q2$, $q3$, $q4$ \\
G27D & point & (25\fdg96, \,0\fdg18) & 0\fdg02 & -- &--&--& 0 & 28 & $q5$ \\
G27E & point & (27\fdg87, \,0\fdg61) & 0\fdg04 & -- &--&--& 0 & 29 & $q6$ \\
\enddata
\tablenotetext{a}{The uncertainty of the center position at 68\% confidence level}
\tablenotetext{b}{Measured counter-clockwise from the Galactic longitude axis to the major axis}
\end{deluxetable*}

We test for spectral curvature for the sources described in Table\,\ref{tbl: sp-g27} for the 0.2--500\,GeV energy range.
Based on this test, we adopt an appropriate spectral function and obtain the best-fit spectral parameters using {\sf gtlike}.
The results are summarized in Table\,\ref{tbl: sed-g27}.
G27B and D have ${\rm TS_{Cut}} > 9$ and ${\rm TS_{BPL}} > 12$, which means a significant spectral curvature ($\gtrsim 3\,\sigma$).
Since a broken power law has one more free parameter than a cut-off power law, here we adopt a cut-off power law.
Note that the choice of a cut-off or broken power law does not change the results for our main target region, G25.
We also measure the spectral parameters of $\LP$.
The parameters in the 2PC catalog are $\Gamma = 1.6 \pm 0.1$, $E_{\rm cut} = 4.1 \pm 0.4\,{\rm GeV}$, and the integrated 0.1--100\,GeV flux of $(2.2 \pm 0.2) \times 10^{-7}$\,photon\,cm$^{-2}$\,s$^{-1}$. 
This flux corresponds to a 0.2--500\,GeV flux of $1.3 \times 10^{-7}$\,photon\,cm$^{-2}$\,s$^{-1}$.
We confirm that the values in the 2PC catalog are similar to those obtained in this analysis (Table\,\ref{tbl: sed-g27}).
We also note that TS of G27A in the 0.2--500\,GeV range is smaller than in 3--500\,GeV.
This means that coupling of data of G27A and B occurs in the 0.2--500\,GeV range probably due to worse angular resolution at lower energy.
More study is needed to measure the precise SED of each source. 
Here we do not study them further because decoupling the SEDs of the background sources 
does not have a significant effect on our main target region, G25.

\begin{deluxetable*}{llcccrrr}
\tabletypesize{\scriptsize}
\tabletypesize{\footnotesize}
\tablecaption{SEDs of G27 Components Above 0.2\,GeV \label{tbl: sed-g27}}
\tablewidth{0pt}
\tablehead{
\colhead{Name} &  \colhead{Spectral}  & \colhead{Flux\tablenotemark{a}}                       & \colhead{Photon} &  Cutoff  &  ${\rm TS_{Cut}}$ & ${\rm TS_{BPL}}$ & \colhead{TS} \\
\colhead{}           & \colhead{Function}  & \colhead{($10^{-8}$\,ph~cm$^{-2}$~s$^{-1}$)} & \colhead{Index} &   (GeV) & &
}
\startdata
G27A & PL & $1.0 \pm 0.3$ & $1.71 \pm 0.09$ & -- & 0 & 3 & 60\\
G27B & PLCUT& $8.7 \pm 0.8$ & $1.82 \pm 0.07$ & $4.5 \pm 0.6$ & 49 & 56 & 715 \\
G27C & PL & $5.7 \pm 0.5$ & $1.98 \pm 0.03$ & -- & 0 & 0 & 413\\
G27D & PLCUT & $1.9 \pm 0.6$ & $2.0 \pm 0.2$ & $15 \pm 10$ & 10 & 17 & 60\\
G27E & PL & $3.9 \pm 0.6$ & $2.58 \pm 0.08$ & -- & 3 & 0 & 103\\
PSR J1838-0537 & PLCUT & $12.1 \pm 0.8$ & $1.85 \pm 0.05$ & $4.9 \pm 0.4$ & -- & -- & 1325\\
\enddata
\tablenotetext{a}{The flux is integrated over the 0.2--500\,GeV energy range}
\end{deluxetable*}

Figure\,\ref{fig: cmap_cls-g27} (right) shows that G27A, B, and C are spatially coincident with an extended TeV source, $\HJ$\,\citep{Aharonian:2008dk}.
Actually, \citet{2013ApJ...773...77A} report extended LAT emission associated with $\HJ$ using $>10$\,GeV data.
Thanks to a longer and lower-energy dataset, we reveal that the observed $\gamma$-ray emissions are more elongated than previously reported.
The TeV source appears to have a multiple-peaked morphology, which might indicate the $\gamma$-ray emissions come from multiple origins.
In this paper, we treat this source as a background source so that we do not perform a further analysis of this complex region.
A complete study will be reported by a forthcoming paper\,($Fermi$-LAT and H.E.S.S. collaborations in preparation).
We also note that a further detailed analysis of this region will have a negligible effect on the results obtained for the G25 region.

\section{X-Ray Luminosity of Young OB Associations/Clusters}
\label{sec: validation}

As stated in Section\,\ref{sec: connection}, the X-ray luminosity of an OB association is expected to be roughly proportional to its mass.
We  note that the similarities of the shape of the XLF are not observationally assured at high luminosities ($\gtrsim 10^{32}$\,erg\,s$^{-1}$), 
because only a small number of stars have been detected at such high luminosities.
However such high luminosities almost entirely arise from very massive stars, which are expected to be born in massive OB associations/clusters.
Therefore we expect the assumption of similarity to be roughly valid even at the high luminosities.
Here we check the expectation for some OB associations/clusters.

The Orion nebula cluster (ONC) has a total mass of $1.9 \times 10^3\,\Ms$\,\citep{2010ApJ...713..871W}.
For this relatively low massive cluster, only two members meet the criteria of $> 7 \times 10^{31}$\,erg\,s$^{-1}$ in 2--8\,keV (see Section\,\ref{sec: connection}) 
and the total luminosity of the two is $4 \times 10^{32}$\,erg\,s$^{-1}$\,\citep{2005ApJS..160..379F}.
The luminosity of the ONC is $\approx 6\%$ of Cyg OB2 (see Appendix\,\ref{sec: XcygOB2}); the ratio is almost the same as that of their mass ($\approx 6$\%).
We also check four more relatively less-massive ($\sim 300$--600\,$\Ms$) clusters, Cep\,OB3b, NGC\,2264, RCW\,36, and NGC\,2244\,\citep[][and references therein]{N2244_mass,CepOb3_mass,OBa_mass}.
The corresponding luminosities of all the sources are estimated to be $\sim 1\times10^{32}$\,erg\,s$^{-1}$\,\citep{CepOb3_Xray,OBac_Xray},
$\sim 30$\% of that of ONC, except for NGC\,2264.
The luminosity of NGC\,2264 cannot be obtained because it has no X-ray sources with luminosity above the threshold ($7 \times 10^{31}$\,erg\,s$^{-1}$).
We also investigate a young massive OB association Westerlund\,2 ($\sim 1 \times 10^4\,\Ms$; \citet{Wd2_mass}).
Five members meet the criterion of $> 7 \times 10^{31}$\,erg\,s$^{-1}$ and
the corresponding X-ray luminosity is estimated to be $4 \times 10^{33}$\,($d$/5\,kpc)$^2$\,erg\,s$^{-1}$, $\sim 60$\% of that of Cygnus\,OB2\,\citep{Wd2_Xray}.
The distance to Westerlund\,2 is still debated; estimates of the distance span the range 2.8--8\,kpc\,\citep[e.g.,][]{Wd2_mass,Wd2_d1,Wd2_d2}.
In any case, we confirm that this massive association has a high X-ray luminosity ($> 10^{33}$\,erg\,s$^{-1}$).
All in all, these results are roughly consistent with our expectation: 
more massive associations/clusters are more X-ray luminous even above $\sim 10^{32}$\,erg\,s$^{-1}$.

\section{X-Ray Luminosity of Cygnus OB2}
\label{sec: XcygOB2}

Here we estimate the X-ray luminosity of the OB association Cyg\,OB2 to compare with that of G25.18+0.26 in Section\,\ref{sec: connection}.
To do that, we use $\sim 1700$ X-ray sources in Cyg OB2 cataloged by \cite{2009ApJS..184...84W} based on two {\em Chandra} observations in 2004 January and July. 
The completeness limit of the catalog is $\sim 10^{30} (d/1.45\,{\rm kpc})^2$\,erg\,s$^{-1}$ in the 0.5--8\,keV range.
The first and second observations cover the central and northeastern regions of Cyg\,OB2 respectively with almost no overlap.
Each observation has a $17' \times 17'$ field of view.

Since the observations do not cover the entire region of Cyg\,OB2, we need to extrapolate to estimate the total luminosity of X-ray sources outside the observed regions.
To estimate the total luminosity of Cyg\,OB2, we make reasonable approximations:
the spatial distribution of Cyg\,OB2 is spherically symmetric;
the first observation is centered on Cyg\,OB2;
the second observation is exactly adjacent to the first one;
and each observation field is a circle with an effective radius of $r_e = 17'/\sqrt{\pi}$.
We do not consider X-ray sources in the region $r > 3\,r_e$, where $r$ is the distance from the center.
This is because the number density of members rapidly drops off from the center so that the sources at $r > 3\,r_e$ make negligible contributions to the total luminosity of the association.
Let us define the total number of sources in the central and northeastern regions as $N_1$ and $N_2$ respectively.
Under the noted approximations, the number density at $r_e < r \leq 3\,r_e$ is roughly the same as that of the northeastern region ($n_2 = N_2/\pi r_{e}^{2}$).
Therefore we can estimate the total number in Cyg\,OB2 as $N \approx N_1 + n_2 \times \pi ((3r_e)^2 - r_e^2) = N_1 + 8\,N_2$.
Since the shape of the XLF is almost the same in both regions\,\citep{2010ApJ...713..871W}, the total number of sources is proportional to the total luminosity.
Then we can estimate the total luminosity of the association as $L = N/(N_1 + N_2) \times L_{1+2} = 11/4 \times L_{1+2}$, 
where $L_{1+2}$ is the observational X-ray luminosity.
To derive the last term, we use a relation of $N_2 \approx 1/3 N_1$ based on the observational fact that the number of X-ray sources in the central region is about three times larger than that in the northeastern region\,\citep{2010ApJ...713..871W}.

From the catalog in \cite{2009ApJS..184...84W}, 
we obtain $L_{1+2} = 2.3 \times 10^{33}\,{\rm erg}\,{\rm s}^{-1}$ in 2--8\,keV for sources with luminosities above $7 \times 10^{31}$\,erg\,s$^{-1}$ assuming a distance of 1.45\,kpc.
Finally we derive the estimated total luminosity $L_{\rm Cyg}$ of $6.3 \times 10^{33}$\,erg\,s$^{-1}$ for Cyg\,OB2.
The integrated number of sources is also estimated to be $11/4 \times 7 \approx 19$.
For reference, we also provide the estimated luminosity of $5.2 \times 10^{33}$\,erg\,s$^{-1}$ in 2--8\,keV
for sources with a luminosity above $1 \times 10^{32}$\,erg\,s$^{-1}$.
The estimated integrated number of sources is $\approx 8$. 
This indicates that the total luminosity is dominated by the most luminous stars so that the choice of the threshold luminosity 
does not significantly affect our estimate.

\section{Target Gas and Photons for Modeling}
\label{sec: tgt}

Here we evaluate the target gas density and the target photons for the modeling in Section\,\ref{sec: model}.
The values are estimated based on the SFR scenario (see Section\,\ref{sec: sfr}), in which
the $\gamma$ rays are radiated from cavities delineated by shells of the G25 bubble but not from the shells themselves.
To evaluate the target gas density, we estimate densities of each gas phase (\HII, \HI, and H$_2$) 
and then sum the estimated values.
We also estimate the target photon field.

First we estimate the number density of \HII\ based on the free-free emission. 
\cite{2010ApJ...709..424M} report intense free-free emission within a diameter of $3 \fdg 92 \times 1 \fdg 66$ at $(l, b) = (24 \fdg 5, 0^{\circ})$, which includes the region of the G25 bubble.
The flux of the free-free emission is estimated to be 1377\,Jy at 90\,GHz.
Here we assume that the G25 bubble (created by the massive OB association G25.18+0.26) 
is responsible for all the free-free emission within its region,
given that free-free emission is mainly due to reprocessed ionizing photons emitted by young massive stars.
For the estimation of the flux, we assume that the surface brightness ($S^{\rm ff}$) is uniform in the region;
under this assumption $S^{\rm ff}$ is 269\,Jy\,deg$^{-2}$ at 90\,GHz.
Then we evaluate the emissivity of the bubble as $F^{\rm ff}_{\rm B} = S^{\rm ff} \times {\Omega}_{\rm B} = 381\,{\rm Jy}$, 
where ${\Omega}_{\rm B}$ is the solid angle of the bubble with a size of $1\fdg5 \times 1\fdg2$.
The corresponding luminosity at 90\,GHz ($L^{\rm ff}_{\rm \nu}$) is $2.7 \times 10^{25}$\,($d$/7.7\,kpc)$^2$\,erg\,s$^{-1}$\,Hz$^{-1}$.
The free-free emission can be described as $L^{\rm ff}_{\rm \nu} \approx \int \epsilon_0 n_{e}^2 dV$, 
where $\epsilon_0$ is $2.7 \times 10^{-39}$\,erg\,cm$^{3}$\,s$^{-1}$\,Hz$^{-1}$, $n_e$ is an electron density, and $V$ is a volume of the emission region.
Here we follow the approximation for the 90\,GHz emission in \cite{2010ApJ...709..424M}.
The volume of the bubble is $V \approx (4\pi/3)\,r_{\rm m}^3 = 9.0 \times 10^{61} (d/7.7\,{\rm kpc})^3\,{\rm cm}^3$, 
where $r_{\rm m}$ is the geometric mean of the semi-minor and semi-major axes ($=90$\,pc).
Assuming uniform emission within the bubble, we can estimate the electron density in the bubble using the relation 
$n_e \approx \sqrt{L^{\rm ff}_{\rm \nu}/(\epsilon_0 V)} = 10.5\,(d/7.7\,{\rm kpc})^{-0.5}$\,cm$^{-3}$.
Finally the ionized hydrogen gas density in the bubble is estimated as $n(\mbox{\HII}) \approx n_e = 10.5\,{\rm cm}^{-3}$.

Next we estimate the number density of \HI. 
To do that, we first evaluate the column density of $N( \mbox{\HI})$ for the G25A and G25B$'$ sources.
The \HI\ column density is evaluated under the assumption of the optically thin limit\,\citep{1990ARA&A..28..215D}:
$N (\mbox{\HI}) \simeq$ $1.82 \times 10^{18} \int T_{\rm b} (\mbox{\HI}; v_r)\,dv_r$\,cm$^{-2}$,
where $T_{\rm b} (\mbox{\HI}; v_r)$ is the brightness temperature of the observed 21\,cm line at $v_r$.
The gas within the bubble is expected to have random motions especially under our assumption that the bubble is powered by the massive OB association G25.18+0.26.
Therefore we need to integrate some range of $v_r$ around $v_r = 113\,{\rm km}\,{\rm s}^{-1}$, the corresponding velocity for $d$ of 7.7\,kpc.
However as shown in Figure\,\ref{fig: maps_HI}, there must be some gas which is not associated with the bubble in this complex region around the velocity.
In order not to overestimate the gas density in the bubble, we integrate the \HI\ gas above the velocity at the tangent point of 113\,km\,s$^{-1}$.
Under the assumption that the bubble is located at the tangent point, most gas above $v_r = 113$\,km\,s$^{-1}$ is expected to be associated with the bubble.
Since the integrated gas is expected to consist of about a half of all \HI\ gas in the bubble if the gas randomly moves,
the total column density can be estimated by doubling the obtained value.
Here we adopt the range $113\,{\rm km}\,{\rm s}^{-1} \leq v_r < 125\,{\rm km}\,{\rm s}^{-1}$ for the integration.
Parts of the G25A and G25B$'$ regions are used to estimate the number density of \HI\ of each region (see Figure\,\ref{fig: maps_CO-cls}).
This is because we try to estimate the gas density and avoid confusion from the dense shell; as discussed in Section\,\ref{sec: gassoc}
the observed $\gamma$ rays appear associated with the inside of the shell and not the shell itself.
We assume that the \HI\ gas of the obtained density is uniformly distributed within each source.
Using the methods described above, we estimate $N(\mbox{\HI}) = 1.1 \times 10^{21}$ and $0.77 \times 10^{21}$\,cm$^{-2}$ for G25A and G25B$'$.
The length of the line of sight through G25A and G25B$'$ is estimated as $L_{\rm A,B'} = 2 \times r_{\rm A,B'} = 121, 114\,{\rm pc}$, 
where $r_{\rm A,B'}$ is the geometric mean of the semi-minor and the semi-major axis for each source.
Finally we obtain the estimated $n(\mbox{\HI}) = N/L$ of 2.9 and 2.2\,cm$^{-3}$ for G25A and G25B$'$.

Thirdly we estimate the number density of molecular gas (H$_2$).
As shown in Figure\,\ref{fig: maps_CO}, no strong $^{13}$CO emission is found within the G25 bubble.
In the previous sections, we use the data from GRS ($^{13}$CO\,$J = 1\mbox{--}0$) because of the good angular resolution.
The GRS traces the morphology of dense H$_2$ clouds so that GRS is an appropriate choice.
On the other hand, to trace sparse, lower-density molecular gas, we should use $^{12}$CO\,$J = 1\mbox{--}0$ line emission.
Here we take data for $^{12}$CO\,$J = 1\mbox{--}0$ emission from \cite{2001ApJ...547..792D}.
We evaluate the H$_2$ column density of $N({\rm H}_2) \simeq X_{\rm CO} \int T_{\rm b}^{\rm CO}(v_r) {\rm d}v\,{\rm cm}^{-2}$, 
where $X_{\rm CO}$ is $1.8 \times 10^{20}\,({\rm km}\,{\rm s}^{-1})^{-1}$\,\citep{2013XCO}
and $T_{\rm b}^{\rm CO}(v_r)$ is the brightness temperature of the observed $^{12}$CO\,$J = 1\mbox{--}0$ line emission at $v_r$.
We apply the same method as $n(\mbox{\HI})$ to obtain $n({\rm H}_2)$ (see above); the extracted regions are shown in Figure\,\ref{fig: maps_CO-cls}.
Finally we evaluate $n({\rm H}_2)$ of 2.5\,cm$^{-3}$ and 3.0\,cm$^{-3}$ for G25A and G25B$'$ respectively.

Finally total gas densities are estimated to be $n_{{\rm A,B}'} = n(\mbox{\HII}) + n(\mbox{\HI}) + 2 \times n({\rm H}_2) = 18, 19\,{\rm cm}^{-3}$.
Given the similar values of $n_{\rm A}$ and $n_{{\rm B}'}$, we adopt a total gas density $n$ of 20\,cm$^{-3}$ for modeling both sources in Section\,\ref{sec: model}.
Here we do not consider dark gas (neutral gas not traced by the emission line surveys considered above),
since the contribution of dark gas would be less than the \HII\ gas on our assumption that the G25 bubble is almost fully ionized by the OB association G25.18+0.26.
We note that even if the dark gas density is comparable with the \HII\ gas density, such an increase in density does not make a significant difference in the model results of Section\,\ref{sec: model}.

For the distributions of the target photons, we consider contributions from the stellar light of G25.18+0.26, the dust emission in the G25 bubble, and cosmic microwave background.
To estimate the stellar light, we need to estimate the number of massive stars in G25.18+0.26.
In Section\,\ref{sec: connection}, the total mass of G25.18+026 is estimated to be about half of that of Cyg\,OB2, which indicates that the number of massive stars in G25.18+0.26 is also about half of that of Cyg\,OB2.
Therefore we estimate the stellar light of G25.18+0.26 as half that of Cyg\,OB2; the estimated stellar luminosity of Cyg\,OB2 is obtained from \CC.
The obtained stellar light field is represented as six black-body components.
The average energy density of the stellar light is determined by the energy density at a typical distance from the center of G25.18+0.26 to each source: 54\,pc for G25A and 81\,pc for G25B$'$.
It is difficult to estimate the dust emission in the bubble in this complex region.
Here we assume that environment of the G25 bubble is similar to that of the Cygnus cocoon, a cavity created by Cyg\,OB2,
and adopt the dust emission of the Cygnus cocoon estimated in \CC\ as that of the G25 bubble.

\bibliographystyle{apj}
\bibliography{bibs_G25_v2}

\end{document}